\documentclass[12pt,a4paper]{article}
\usepackage{amssymb,amsmath}
\include{epsf}

\newcommand{\eqn}[1]{(\ref{#1})}

\newcommand{\app}{\section}

\newbox\ncintdbox \newbox\ncinttbox
\setbox0=\hbox{$-$} \setbox2=\hbox{$\displaystyle\int$}
\setbox\ncintdbox=\hbox{\rlap{\hbox to
\wd2{\hskip-.125em\box2\relax \hfil}}\box0\kern.1em}
\setbox0=\hbox{$\vcenter{\hrule width 4pt}$}
\setbox2=\hbox{$\textstyle\int$}
\setbox\ncinttbox=\hbox{\rlap{\hbox to
\wd2{\hskip-.175em\box2\relax \hfil}}\box0\kern.1em}
\newcommand{\ncint}{\mathop{\mathchoice{\copy\ncintdbox}%
{\copy\ncinttbox}{\copy\ncinttbox}{\copy\ncinttbox}}\nolimits}

\newbox\nncintdbox \newbox\nncinttbox
\setbox0=\hbox{$-_n$} \setbox2=\hbox{$\displaystyle\int$}
\setbox\nncintdbox=\hbox{\rlap{\hbox to
\wd2{\hskip-.125em\box2\relax \hfil}}\box0\kern.1em}
\setbox0=\hbox{$\vcenter{\hrule width 4pt}$}
\setbox2=\hbox{$\textstyle\int$}
\setbox\nncinttbox=\hbox{\rlap{\hbox to
\wd2{\hskip-.175em\box2\relax \hfil}}\box0\kern.1em}

\newcommand{\mbf}[1]{{\boldsymbol {#1} }}
\newcommand{\complex}{{\bb C}} 
\newcommand{\zed}{{\bb Z}} 
\newcommand{\mat}{{\bb M}} 
\newcommand{\id}{{1\!\!1}} 

\newcommand{\SSigma}{{\mbf \Sigma}}
\def\ii{{\,{\rm i}\,}}
\def\dd{{\rm d}}
\def\DD{{\rm D}}
\def\P{{\sf P}}
\def\B{{\sf B}}
\def\a{{\sf a}}
\def\b{{\sf b}}
\def\A{{\sf A}}
\def\U{{\sf U}}
\def\V{{\sf V}}
\def\K{{\sf K}}
\def\T{{\sf T}}
\def\X{{\sf X}}
\def\Y{{\sf Y}}
\def\G{{\sf G}}
\def\mPhi{{\mbf\Phi}}
\def\mPi{{\mbf\Pi}}
\def\mphi{{\mbf\phi}}
\def\mpsi{{\mbf\psi}}
\def\mnu{{\mbf\nu}}
\def\mtau{{\mbf\sigma}}
\def\mnabla{{\mbf\nabla}}
\newcommand{\torus}{{\mathbb T}}
\newcommand{\circles}{{\mathbb S}}
\newcommand{\heisen}{{{\cal E}_{p,q}}}

\font\mybb=msbm10 at 12pt
\def\bb#1{\hbox{\mybb#1}}

\newcommand{\Tr}[1]{\:{\rm Tr}\,#1}
\def\e{{\,\rm e}\,}

\newcommand{\nn}{\nonumber \\}
\def\be{\begin{equation}}
\def\ee{\end{equation}}
\def\beq{\begin{equation}}
\def\eeq{\end{equation}}
\newcommand{\del}{\partial}
\def\bea{\begin{eqnarray}}
\def\eea{\end{eqnarray}}
\def\bd{\begin{displaymath}}
\def\ed{\end{displaymath}}

\newcommand{\atheta}{{{\cal A}_\theta}}
\newcommand{\aalpha}{{{\cal A}_\alpha}}
\newcommand{\athetan}{{{\cal A}_n}}
\newcommand{\bthetan}{{{\cal B}_n}}

\makeatletter
\newdimen\normalarrayskip              
\newdimen\minarrayskip                 
\normalarrayskip\baselineskip
\minarrayskip\jot
\newif\ifold             \oldtrue            \def\new{\oldfalse}
\newcommand{\half}{{\textstyle{1\over 2}}}

\newcommand{\ca}{{\cal A}}
\newcommand{\cb}{{\cal B}}
\newcommand{\cc}{{\cal C}}

\newcommand{\ce}{{\cal E}}

\newcommand{\ch}{{\cal H}}

\newcommand{\cp}{{\cal P}}
\newcommand{\cs}{{\cal S}}

\newcommand{\IC}{{\mathbb C}}

\newcommand{\IM}{{\mathbb M}}
\newcommand{\IN}{{\mathbb N}}

\newcommand{\IP}{{\mathbb P}}

\newcommand{\IR}{{\mathbb R}}
\newcommand{\IS}{{\mathbb S}}
\newcommand{\IT}{{\mathbb T}}
\newcommand{\IZ}{{\mathbb Z}}
\newcommand{\bw}{{\overline{w}}}
\def\bar#1{\overline{#1}}
\def\norm#1{{\Vert#1\Vert}}
\def\abs#1{{\vert#1\vert}}
\def\raw{\rightarrow}

\def\wh{\widehat}
\def\bra#1{\left\langle #1\right|}
\def\ket#1{\left| #1\right\rangle}
\def\hs#1#2{\left\langle #1,#2\right\rangle}

\begin{document}
\begin{titlepage}
\begin{flushright}

\baselineskip=12pt
DSF--36--03\\ HWM--03--30\\ EMPG--03--22\\ hep--th/0401072\\
\hfill{ }\\ January 2004
\end{flushright}

\vspace{0.5 cm}

\begin{center}

\baselineskip=24pt

{\Large\bf Matrix Quantum Mechanics and\\ Soliton Regularization of\\
    Noncommutative Field Theory}

\baselineskip=14pt

\vspace{1cm}

{{\bf Giovanni Landi}$^{a,b,}$\footnote{\baselineskip=12pt Email: {\tt
landi@univ.trieste.it}}, {\bf Fedele
Lizzi}$^{b,c,d,}$\footnote{\baselineskip=12pt Email: {\tt
    fedele.lizzi@na.infn.it}} and {\bf
Richard J.~Szabo}$^{d,}$\footnote{\baselineskip=12pt Email: {\tt
R.J.Szabo@ma.hw.ac.uk}}}
\\[6mm]
$^a$ {\it Dipartimento di Scienze Matematiche, Universit\`a di
     Trieste\\ Via A. Valerio 12/1, I-34127 Trieste, Italia}
\\[6mm]
$^b$ {\it INFN, Sezione di Napoli, Napoli, Italia}
\\[6mm]
$^c$ {\it Dipartimento di Scienze Fisiche, Universit\`{a} di
Napoli {\sl Federico II}\\ Monte S.~Angelo, Via Cintia, 80126
Napoli, Italia}
\\[6mm]
$^d$ {\it Department of Mathematics, Heriot-Watt University\\
Scott Russell Building, Riccarton, Edinburgh EH14 4AS, U.K.}
\\[10mm]

\end{center}

\vskip 1 cm

\begin{abstract}

\baselineskip=12pt

We construct an approximation to field theories on the
noncommutative torus based on soliton projections and partial isometries
which together form a matrix algebra of functions on the sum of
two circles. The matrix quantum mechanics is applied to the
perturbative dynamics of scalar field theory, to tachyon dynamics
in string field theory, and to the Hamiltonian dynamics of
noncommutative gauge theory in two dimensions. We also describe
the adiabatic dynamics of solitons on the noncommutative torus and
compare various classes of noncommutative solitons on the torus and
the plane.

\end{abstract}

\end{titlepage}

\newpage

\setcounter{page}{2}

{\baselineskip=12pt
    \tableofcontents}

\newpage

\section{Introduction and Summary}
\setcounter{equation}{0}

Among the principal characteristics of noncommutative
spaces~\cite{co94,Landi,Madorebook,Ticos}, whichever way we may
choose to define them, is the fact that the concept of locality
becomes evanescent and disappears altogether. Noncommutativity
typically introduces a length scale below which it is no longer
possible to resolve ``points'' in the space. If a noncommutative
space cannot be described by the local fields defined on it, it is
still possible to use those fields, which technically live in a
noncommutative $C^*$-algebra, to describe some geometric
properties of the space. In some instances, for example when the
noncommutative spaces are deformations of ordinary ones, it may
still be possible to ``see'' the points underlying the algebra,
and the noncommutativity is typically described by the
nonvanishing commutator of coordinates. This description may be
appealing for the connections which can be made with ordinary
geometry, but it does hide some novel characteristics of
noncommutative geometry which can have important physical
implications and provide useful calculational tools. For instance,
there exist solitonic solutions in noncommutative geometry which
have no counterparts in commutative geometry~\cite{GMS}. By
solitons we mean nonvanishing finite energy extrema of the action
functional of a given field theory, and in the examples to be
considered in this paper they are described by projections or
partial isometries of the underlying noncommutative algebra. In
the following we will in fact use the words solitons and
projections/partial isometries synonymously.

One of the main physical interests in noncommutative geometry is the
fact that it arises naturally in string theory, and in particular the
noncommutative torus~\cite{ri81,co80} describes naturally the stringy
modifications to classical spacetime~\cite{CDS,LLSNC,SW}
(see~\cite{KS1,DN1,Sz1} for reviews). In the context of open string
field theory, the algebraic structure of noncommutative geometry
allows a particularly simple construction of both stable and unstable
D-branes in terms of projections and partial
isometries~\cite{DMR1,HKLM1,Witten1}. Also related to
this operatorial nature is the fact that noncommutative field theories
can be regulated and studied by means of matrix
models~\cite{CDS,AIIKKT1,AMNS1,AMNS2,MRW,BGI,LVZ,GW,LSZ,Stein1}. In the case
of field theories on the noncommutative torus, the matrix
regularization yields field theories on the fuzzy torus and is
intimately related to the lattice regularization of noncommutative
field theories~\cite{AMNS1,AMNS2}.

Although the matrix model formalisms have many computational
advantages, they have several pitfalls as well. Foremost among
these are the complicated double scaling limits required to
reproduce the original continuum dynamics. The complicated nature
is related to the mathematical fact that the algebra of functions
on a manifold can never be the exact inductive limit of finite
dimensional algebras, and examples abound for which the finite
approximations fail to capture relevant physical aspects or
produce phenomena which are unphysical artifacts of the matrix
regularization. Technically, we may say that no algebra of
functions can be an \emph{approximately finite (AF)
algebra}~\cite{lls}. In this paper we will show how to overcome
this problem by exploiting one of the aforementioned novel
characteristics of noncommutative field theories, namely the
presence of projection operators (or projections for brevity). As
we have mentioned, they play an important role in the effective
field theories of D-branes in that they are finite energy extrema
of the potential energy, or \emph{solitons}.

In what follows, after a review of the Elliott-Evans construction of
the sequence of algebras approximating the noncommutative
torus~\cite{ee}, we will construct viable field theories based on
it. The interest in this construction is many-fold. The approximate
algebras are generated by projections and partial isometries which
together generate the direct sum of two copies of the algebra of
matrix-valued functions on a circle, and therefore the approximation
to a noncommutative field theory is effectively a \emph{matrix quantum
   mechanics} which can be solved exactly in some cases. Unlike the usual
lattice
approximations, the noncommutative torus is the inductive limit of
the sequence of algebras in the strong rigorous sense. From a
computational point of view, this means that the continuum limit is
much simpler. It is important to realize though that it is {\it not}
simply the 't~Hooft planar limit of the matrix model, and the notion
of planarity in the matrix quantum mechanics coincides with that of
the original noncommutative field theory~\cite{MVRS1}.

We will show that the field theory corresponding to the
soliton approximation can be used, as a quantum mechanics, in a quantitatively
useful manner for field theoretic calculations. For example, we will
analyse in detail the dynamics of a noncommutative scalar field theory
and show that ultraviolet-infrared (UV/IR) mixing~\cite{MVRS1} is
cured by the approximation (but of course reappears in the limit). We
also show that the approximation already captures quantitative aspects
of tachyon condensation in string field theory, and further
demonstrate how the exact solution of gauge theory on the
noncommutative torus~\cite{PS1} is captured by the Hamiltonian
dynamics of the matrix quantum mechanics. The approximation presented
in this paper thereby has the opportunity to capture important
nonperturbative aspects of noncommutative field theories.

We will also study the adiabatic dynamics of projections according to a
$\sigma$-model action defined on soliton moduli space. We will find
that the extrema of the action are solitons which satisfy a certain
self-duality or
anti-self-duality condition. The typical soliton of this
kind, the Boca projection~\cite{bo0}, is the torus equivalent of the
GMS solitons on the noncommutative
plane~\cite{GMS}. The field configurations correspond to smooth
``bump'' functions which are localized within the scale of
noncommutativity, and they are very different from the projections
which generate the matrix algebras. The latter projections generalize
the Power-Rieffel projections~\cite{ri81}, and the corresponding
fields wind around the torus thereby exhibiting a more non-local
structure. In the context of tachyon condensation on the two-dimensional
noncommutative torus, the Boca projection has been employed
in~\cite{MM1,GHS1,KS2,KMT1,HP1} and the Powers-Rieffel projection
in~\cite{BKMT1,SS1,MM1}. From the dynamically obtained Boca
projection we will then use the matrix regularization on the
noncommutative torus to induce approximations also of field theories
on the noncommutative plane.

\subsubsection*{Outline}

The structure of the remainder of this paper is as follows. In
section~\ref{se:solitons} we introduce the main characteristics describing
field
theories on the noncommutative torus, their connection with tachyon
condensation, and the sequence of projections which will form the diagonal part
of the
matrix approximation. In section~\ref{se:masu} we describe in detail
the construction of the matrix subalgebras and the way the
approximation is realized. In section~\ref{NCFTMQM}, which is
the crux of the paper, we describe how to construct the
matrix quantum mechanics equivalent (in the limit) of a generic
noncommutative field theory. In section~\ref{se:applications} we
present three examples of the formalism, involving the perturbative
dynamics of $\phi^4$ scalar field theory on the noncommutative torus,
the construction of D-branes as decay products in tachyon
condensation, and a Hamiltonian analysis of noncommutative Yang-Mills
theory in two dimensions. In the final section~\ref{se:moduli}, we
describe the relationships between the solitons used for the matrix
approximation and the
Boca projection, the toroidal generalization of the GMS lump
configurations, which leads to the planar version of the matrix model
regularization. There we also describe the dynamics of solitons on the
noncommutative torus through a $\sigma$-model defined on
their configuration space. Some technical details are presented in
five appendices at the end of the paper. Some aspects of the present paper
have been announced in~\cite{trailer}.

\section{Solitons on the Noncommutative Torus \label{se:solitons}}
\setcounter{equation}{0}

In this section we will review the construction of solitonic field
configurations on the two-dimensional noncommutative torus, primarily
to introduce the physical notions, the notation and the definitions
which will be used throughout this paper. We will begin with a review
of the geometry of the noncommutative torus, emphasizing those
ingredients which are important for the construction of noncommutative
field theories. We shall then briefly review the construction of
D-branes as solitons in the effective field theory of open strings, as
this will set the main physical motivation for most of our subsequent
analysis. Then we will describe an important set of projections for
the noncommutative torus.

\subsection{Field Theories on the Noncommutative Torus\label{FTNT}}

Consider an ordinary square two-torus $\IT^2$ with coordinate functions
$U =\e^{2 \pi\ii x}$ and $V =\e^{2 \pi\ii y}$, where $x,y\in[0,1]$. By
Fourier expansion the algebra $C^\infty(\IT^2)$ of complex-valued
smooth functions on the torus is made up of all power series of the form
\be\label{2ta}
a =\sum_{(m,r) \in \IZ^2} a_{m,r}~U^m\,V^r~,
\end{equation}
with $\{a_{m,r}\}\in S(\IZ^2)$ a complex-valued Schwartz function
on $\IZ^2$. This means that the sequence of complex numbers $\{a_{m,r} \in
\IC~|~ (m,r) \in\IZ^2 \}$ decreases rapidly at ``infinity'', i.e. for
any $k\in\IN_0$ one has bounded semi-norms
\be\label{nctsn}
\norm{a}_k = \sup_{(m,r)\in \IZ^2} ~|a_{m,r}|\,\big(1+\abs{m}+\abs{r}
\big)^k < \infty ~.
\end{equation}
Let us now fix a real number $\theta$. The
algebra $\atheta =C^\infty(\IT^2_\theta)$ of smooth functions on the
noncommutative torus is the associative algebra made up of all elements
of the form \eqn{2ta}, but now the two generators $U$ and $V$ satisfy
\be\label{nct}
V\,U =\e^{2\pi\ii \theta}~U\,V~.
\end{equation}
The algebra $\atheta$ can be made into a $*$-algebra by defining a
$*$-involution~$\dag$ by
\be\label{nctsta}
U^\dag := U^{-1} ~, ~~~ V^\dag := V^{-1}~.
\end{equation}
{}From \eqn{nctsn} with $k=0$ one gets a $C^*$-norm and the corresponding
closure of $\atheta$ in this norm is the universal $C^*$-algebra $A_\theta$
generated by two unitaries with the relation \eqn{nct}; $\atheta$ is
dense in $A_\theta$ and is thus a pre-$C^*$-algebra.

In the following we shall use the one-to-one correspondence
between elements of the noncommutative torus algebra $\atheta$ and the
commutative torus algebra $C^\infty(\IT^2)$ given by the Weyl map
$\Omega$ and its inverse, the Wigner
map. As is usual for a Weyl map, there are operator ordering
ambiguities, and so we will take the prescription
\be
\Omega\left(\,\sum_{(m,r) \in \IZ^2} f_{m,r}~\e^{2\pi\ii(m\,x + r\,y)}
\right):=\sum_{(m,r) \in \IZ^2} f_{m,r}~
\e^{\pi\ii m\,r\,\theta}~U^m\,V^r \ .
\label{weymap}\end{equation}
This choice (called Weyl or symmetric ordering) maps real-valued
functions on $\torus^2$ into Hermitian elements of $\atheta$. The
inverse map is given by
\be
\Omega^{-1}\left(\,\sum_{(m,r) \in \IZ^2} a_{m,r}~U^m\,V^r\right)=
\sum_{(m,r) \in \IZ^2} a_{m,r}~\e^{-\pi\ii m\,r\,\theta}~
\e^{2\pi\ii(m\,x +r\,y)} \ .
\label{wigmap}\end{equation}

Clearly, the map $\Omega:C^\infty(\torus^2)\to\atheta$ is not an algebra
homomorphism; it can be used to deform the commutative product on
the algebra $C^\infty(\IT^2)$ into a noncommutative star-product by defining
\be
f \star g := \Omega^{-1}\big(\Omega(f)\,\Omega(g) \big) ~, ~~~
f,g \in C^\infty\left(\IT^2\right) ~.
\end{equation}
A straightforward computation gives
\be
f \star g= \sum_{(r_1,r_2) \in \IZ^2} (f \star g)_{r_1,r_2}
\,\e^{2\pi\ii(r_1 x + r_2y)} \, ,
\end{equation}
with the coefficients of the expansion of the star-product given by a
twisted convolution
\be
(f \star g)_{r_1,r_2} = \sum_{(s_1,s_2) \in \IZ^2} f_{s_1,s_2} \,
g_{r_1-s_1,r_2-s_2}~\e^{\pi\ii(r_1 s_2 - r_2 s_1)\,\theta}
\label{twco}
\end{equation}
which reduces to the usual Fourier convolution product in the limit
$\theta=0$. Up to isomorphism, the deformed product depends only on
the cohomology class in the group cohomology ${\sf H}^2(\IZ^2,U(1))$ of
the $U(1)$-valued two-cocycle on $\IZ^2$ given by
\be
\lambda (\boldsymbol{r},\boldsymbol{s}) :=
\e^{\pi\ii(r_1 s_2 - r_2 s_1)\,\theta}
\label{2grcocy}
\end{equation}
with $\boldsymbol{r}=(r_1,r_2),\boldsymbol{s}=(s_1,s_2)\in\IZ^2$.

Heuristically, the noncommutative structure (\ref{nct}) of the torus
is the exponential of the Heisenberg commutation relation
$[y,x]=\ii\theta/2\pi$. Acting on functions of $x$ alone, the
operator $U$ is represented as multiplication by $\e^{2\pi\ii x}$ while
conjugation by $V$ yields the shift $x \mapsto x+\theta$,
\bea
\Omega^{-1}\Big(U\,\Omega\big(f(x)\big) \Big)&=&\e^{2\pi\ii x}\,f(x) \ ,
\nn\Omega^{-1}\Big(V\,\Omega\big(f(x)\big)\,V^{-1} \Big)&=&f(x+\theta) ~.
\label{xweywig}
\end{eqnarray}
Analogously, on functions of $y$ alone we have
\bea
\Omega^{-1}\Big(U\,\Omega\big(g(y)\big)\,U^{-1} \Big)&=&g(y-\theta) \ , \nn
\Omega^{-1}\Big(V\,\Omega\big(g(y)\big)\Big)&=&\e^{2\pi\ii y}\,g(y) ~.
\label{yweywig}
\end{eqnarray}

{}From \eqn{nct} it follows that $\atheta$ is commutative if and only
if $\theta$ is an integer, and one identifies $\ca_0$ with the algebra
$C^\infty(\IT^2)$. Also, for any $n\in\IZ$ there is an isomorphism
$\atheta\cong\ca_{\theta+n}$ since \eqn{nct} does not change under integer
shifts $\theta \mapsto \theta + n$. Thus we may restrict the noncommutativity
parameter to the interval $0\leq\theta<1$. Furthermore, since
$U V =\e^{- 2\pi\ii \theta}\,V U =\e^{2\pi\ii (1-\theta)}\,V U$, the
correspondence $V \mapsto U, U
\mapsto V$ yields an isomorphism $\atheta \cong\ca_{1-\theta}$. These are
the only possible isomorphisms and the interval $\theta\in[0, \frac{1}{2}]$
parametrizes a family of non-isomorphic algebras.

When the deformation parameter $\theta$ is a rational number, the
corresponding algebra is related to the commutative torus algebra
$C^\infty(\IT^2)$, i.e. $\atheta$ is Morita equivalent to it in
this case~\cite{ri81}. Let $\theta=p/q$, with $p$ and $q$ integers
which we take to be relatively prime with $q>0$. Then $\ca_{p/q}$
is isomorphic to the algebra of all smooth sections of an algebra
bundle $\cb_{p/q} \raw \IT^2$ whose typical fiber is the algebra
$\IM_q(\IC)$ of $q\times q$ complex matrices. Moreover, there is a
smooth vector bundle $E_{p/q} \raw \IT^2$ with typical fiber
$\IC^q$ such that $\cb_{p/q}$ is the endomorphism bundle ${\rm
End}(E_{p/q})$. With $\omega=\e^{2\pi\ii p/q}$, one introduces the
$q\times q$ clock and shift matrices
\be
\cc_q =\left({\begin{array}{lllll}
1& & & & \\ &\omega& & & \\
& &\omega^2& & \\& &
&\ddots& \\ & & & &
\omega^{q-1}
\end{array}}\right)~~~~~~,~~~~~~
\cs_q =\left({\begin{array}{lllll}
0&1& & &0\\ &0&1& & \\
& &\ddots&\ddots& \\
& & &\ddots&1\\ 1& & & &0\end{array}}\right) \ ,
\label{torfuz}\end{equation}
which are unitary and traceless (since $\sum_{k=0}^{q-1} \omega^k =
0$), satisfy
\be
(\cc_q)^q = (\cs_q)^q = \id_q \ ,
\label{clockshiftid}\end{equation}
and obey the commutation relation
\be
\cs_q\,\cc_q = \omega ~\cc_q\,\cs_q \ .
\label{clockshiftcommrel}\end{equation}
Since $p$ and $q$ are relatively prime, the matrices \eqn{torfuz} generate
the finite dimensional algebra
$\IM_q(\IC)$: they
generate a $C^*$-subalgebra which commutes only with multiples of
the identity matrix $\id_q$, and thus it has to be the full matrix
algebra. Were $p$ and $q$ not coprime the generated algebra would be a proper
    subalgebra of $\IM_q(\IC)$.
The matrix algebra generated by $\cc_q$ and $\cs_q$ is also
referred to as the fuzzy torus.

The algebra $\ca_{p/q}$ has a ``huge'' center ${\cal C}(\ca_{p/q})$
which is generated by the elements $U^q$ and $V^q$, and one identifies
${\cal C}(\ca_{p/q})$ with the commutative algebra $C^\infty(\IT^2)$ 
of smooth functions
on an ordinary torus $\IT^2$ which is `wrapped' $q$ times onto itself.
There is a surjective algebra homomorphism
\be\label{calsur}
\pi_q\,:\,\ca_{p/q}~\longrightarrow~\IM_q(\IC)
\end{equation}
given by
\be\label{calsur1}
\pi_q\left(\,\sum_{(m,r)\in\IZ^2} a_{m,r}~U^m\,V^r\right) =
\sum_{(m,r)\in\IZ^2}a_{m,r}~ (\cc_q)^m ~(\cs_q)^r~.
\end{equation}
Under this homomorphism the whole center ${\cal C}(\ca_{p/q})$ is
mapped to $\IC$.

Henceforth we will assume that $\theta$ is an irrational
number unless otherwise explicitly stated. On $\atheta$ there is a
unique normalized, positive definite trace which we shall denote by the symbol
$\ncint :
\atheta \raw \IC$ and which is given by
\be
\ncint\,\sum_{(m,r) \in \IZ^2} a_{m,r}~U^m\,V^r~:=~
a_{0,0}~=~\int\limits_{\IT^2}
\dd x~\dd y~\Omega^{-1}\left(\,\sum_{(m,r)\in\IZ^2}
   a_{m,r}~U^m\,V^r\right)(x,y) \ . \label{ncintadefa00}
\end{equation}
Then, for any $a,b\in\atheta$, one readily checks the properties
\beq
\ncint a\,b = \ncint b\,a ~, ~~~
\ncint \id = 1 ~, ~~~
\ncint a^\dag\,a > 0 ~, ~~a\not= 0~,
\end{equation}
with $\ncint a^\dag a = 0$ if and only if $a=0$ (i.e. the trace is faithful).
This trace is invariant under the natural action of the commutative torus
$\IT^2$ on $\atheta$ whose infinitesimal form is determined by two
commuting linear derivations $\del_1, \del_2$ acting by
\bea
&& \del_1U= 2\pi\ii U~, ~~~\del_1V=0 \ , \nn
&& \del_2U=0 ~, ~~~\del_2V= 2\pi\ii V ~.
\label{t2act}
\end{eqnarray}
Invariance is just the statement that $\ncint \del_\mu a=
0$, $\mu = 1, 2$ for any $a\in\atheta$.

The algebra $\atheta$ can be represented faithfully as operators
acting on a separable Hilbert space $\cal H$, which is the GNS
representation space $\ch = L^2(\atheta\,,\,\ncint\,)$ defined as the
completion of $\atheta$ itself in the Hilbert norm
\beq\label{opnor}
\norm{a}_{\rm GNS}:= \left(\,\ncint a^\dag\,a\right)^{1/2}
\end{equation}
with $a\in\atheta$. Since the trace is faithful, the map
$\atheta\ni a \mapsto \wh{a} \in\ch$ is injective and the faithful
GNS representation $\pi:\atheta \raw \cb(\ch)$ is simply given by
\be
\pi(a) \wh{b}= \wh{a\,b}
\end{equation}
for any $a,b\in\atheta$. The vector $1 = \wh{\id}$ of $\ch$ is cyclic
(i.e. $\pi(\atheta) 1$ is dense in $\ch$) and separating (i.e. $\pi(a) 1 = 0$
implies $a=0$) so that the Tomita involution
is just $J(\wh{a}) = \wh{a^\dag}$ for any $\wh{a}\in\ch$.
It is worth mentioning that the $C^*$-algebra norm on $\atheta$
given in \eqn{nctsn} with $k=0$ coincides with the operator norm in
\eqn{opnor} when $\atheta$ is represented on the
Hilbert space $\cal H$, and also with the $L^\infty$-norm in the
Wigner representation. For ease of notation, in what follows we will
not distinguish between elements of the algebra $\atheta$ and their
corresponding operators in the GNS representation.

\subsection{D-Branes as Noncommutative Solitons\label{DBraneNCSoliton}}

Let us now briefly recall how D-branes arise as soliton configurations
which are described as projection operators or partial isometries in the
algebra $\atheta$ of the noncommutative torus. We are interested in
systems of unstable D-branes in a closed Type~II superstring
background of the form ${\cal M}\times\torus^2$. The particular
configurations comprise D9-branes in Type~IIA
string theory and D9--$\overline{\rm D9}$ pairs in Type~IIB string
theory. As it is by now well-known, the effect of turning on a
non-degenerate $B$-field along $\torus^{2}$ leads to an effective
description of the dynamics of these systems in terms of
noncommutative geometry~\cite{SW}. Integrating out all massive string modes in
the weakly-coupled open string field theory yields a
low-energy effective action that describes a noncommutative field
theory of the open string tachyon field $T$ and the open string gauge
connection $\nabla$. The classical equations of motion admit
interesting soliton solutions~\cite{DMR1,HKLM1,Witten1} which lead to
an open string field theory description of D-branes in terms of
tachyon condensation as follows~\cite{Sen1}.

In the Type~IIA case (or alternatively the bosonic string), the
tachyon field $T$ on the D9-branes is
Hermitian and adjoint-valued, and the tachyon potential is of the form
$V(T^2-\id)$, whose global minimum at $T=\pm\,\id$ is identified as the
closed string vacuum containing no perturbative open string
excitations. Solving the classical field equations is in general
tantamount to seeking slowly-varying tachyonic configurations,
i.e. $[\nabla,T]=0$, which extremize the tachyon potential,
i.e. $T\,V'(T^2-\id)=0$. One thereby finds solutions in terms of
projection operators $\P\in\atheta$ as
\be
T=\id-\P \ , ~~ \P^2=\P=\P^\dag \ .
\label{IIATproj}
\end{equation}
A projection operator $\P_{(k)}$ of rank $k$ induces a $U(k)$ gauge symmetry on
the lower dimensional unstable D-brane (with worldvolume $\cal M$),
whose dynamical degrees of freedom are operators on
$\ker(T)\to\ker(T)$. Since the projections are intimately related to the
K-theory of the algebra $\atheta$, this construction also
illustrates the relation between D-branes and K-theory.

In the Type~IIB case, the tachyon field on the
D9--$\overline{\rm D9}$ pairs is complex, and the tachyon potential is
of the form
\be
V\left(T\,,\,T^\dag\right)=U\left(T^\dag\,
   T-\id\right)+U\left(T\,T^\dag-\id\right) \ ,
\label{tachpotIIB}\end{equation}
in order to respect the symmetry given by the action of the operator 
$(-1)^{F_{\rm
      L}}$ which corresponds to interchanging the branes and
anti-branes ($F_{\rm L}$ is the left-moving worldsheet fermion number
operator). Now the field equations imply that $T$ must satisfy the
defining equation of a partial isometry
\be
T\,T^\dag\,T=T \ .
\label{IIBTpartialisom}\end{equation}
The net brane charge is ${\rm index}(T)$ and, assuming for simplicity
that ${\rm coker}(T)=0$, the dynamical degrees of freedom on the
lower-dimensional BPS D-brane again arise from operators on
$\ker(T)\to\ker(T)$.

In both the IIA and IIB situations, the tensions and effective actions
of these soliton solutions match {\it exactly} with those of the lower
dimensional D-branes~\cite{AGMS1,HKL1,KS2}. In this way the projections and
partial
isometries of $\atheta$ generate {\it exact} D-brane solutions of the
equations of motion, with the correct value of the tension. In
constructing these D-brane projections, it is convenient to use not
just a single projection operator in (\ref{IIATproj}), but rather a
complete set of mutually orthogonal projections $\P^i$ with~\cite{KS2}
\be
\P^i\,\P^j=\delta_{ij}\,\P^j \ , ~~ \sum_i\P^i=\id \ .
\label{projcollection}\end{equation}
Appropriate combinations of
the projection operators $\P^i$ determine solutions of the
Yang-Mills equations on $\atheta$~\cite{PS1}. In the
following we will construct a natural system of projections and
partial isometries which determine matrix regularizations of these
sorts of noncommutative field theories.

\subsection{A Sequence of Projections\label{ProjSeq}}

The archetype of all projections on
the noncommutative two-torus is the Powers-Rieffel
projection~\cite{ri81}. To construct it, we first observe that
there is an injective algebra homomorphism
\bea
\rho\,:\,C^\infty\left(\IS^1\right)&\longrightarrow&\atheta ~, \nn
f(x) = \sum_{m\in\IZ}f_{m}~\e^{2 \pi \ii m\,x}&\longmapsto&\rho(f) =
\sum_{m\in\IZ}f_{m}~U^m ~,
\label{xsubalg}
\end{eqnarray}
and by using the commutation relations \eqn{nct} it follows, in
particular, that if $f(x)$ is mapped to $\rho(f)$, then $V \rho(f)
V^{-1}$ is the image of the shifted function $f(x+\theta)$.
The map \eqn{xsubalg} is just the Weyl map \eqn{weymap} restricted to
functions of the variable $x$ alone with the corresponding properties
\eqn{xweywig}.

One now looks for projections of the form
\be
\P_\theta=V^{-1}\,\rho(g)+\rho(f)+\rho(g)\,V \ .
\label{Ptheta}\end{equation}
In order that
(\ref{Ptheta}) define a projection operator, the functions $f,g \in
C^\infty(\IS^1)$ must satisfy some conditions. First of all, they are
real-valued and in addition obey
\bea
g(x)\,g(x+\theta) &=& 0 ~,\nn
\big(f(x)+f(x+\theta)\big)\,g(x)&=&g(x) \ , \nn
g(x)+g(x-\theta)&=&\sqrt{f(x)-f(x)^2} \ ,
\label{fgPthetaconds}
\end{eqnarray}
with $0\leq f\leq1$. These conditions are satisfied by putting
\bea
f(x)&=&\left\{\new
\begin{array}{ccrcl}
\mbox{\rm smoothly increasing from 0 to 1} & ~ & 0 & \leq~x~\leq & 1-\theta \\
1 & ~ &  1 - \theta & \leq~x~\leq & \theta \\
1 - f(x-\theta) & ~ & \theta & \leq~x~\leq &  1 \\
\end{array}
\right. \ , \nn&&{~~~~}^{~~}_{~~}\nn
g(x)&=&\left\{\new
\begin{array}{ccrcl}
0 & ~ & 0 &\leq~x~\leq & \theta  \\
\sqrt{f(x)-f(x)^2} & ~ & \theta & \leq~x~\leq & 1
\end{array}\right. \ .
\label{bumptheta}
\end{eqnarray}
There are myriads of other choices
possible for these bump functions, and later on we will use a
particular one which is useful for our generalizations.

It is straighforward to check that the rank (i.e. trace) of
$\P_\theta$ is just $\theta$. From (\ref{Ptheta}) and the expressions
in \eqn{bumptheta} one finds
\be
\ncint \P_\theta = f_0 = \int\limits_0^1\dd x~f(x)= \theta ~.
\end{equation}
Furthermore, the monopole charge (i.e.\ first Chern number) of
$\P_\theta$ is $1$. Given any projection $\P$, its Chern number is
given by~\cite{co80}
\be\label{topcha}
c_1(\P) = - \frac{1}{2\pi \ii}\,\ncint \P\,\left( \del_1\P\,\del_2\P -
\del_2\P\,\del_1\P\right)~.
\end{equation}
This quantity always computes the index of a Fredholm operator, and hence is
always an integer. For the projection $\P_\theta$ one finds
\be
c_1(\P_\theta)=-6\,\int\limits_0^1\dd x~g(x)^2\,f'(x)= 1~,
\end{equation}
where the last equality follows from the explicit choice
\eqn{bumptheta} for the function $f$.

When $\theta$ is an irrational number, together
with the trivial projection $\id$, the projection $\P_\theta$ generates
the $\K_0$ group. The trace on $\atheta$ gives a map
\bea
\ncint\,:\,\K_0(\atheta)&\longrightarrow&\zed+\zed\,\theta \ , \nn
r\,[\id] + m\,[\P_\theta]&\longmapsto&r\,\ncint \id + m\,\ncint \P_\theta
= r + m\,\theta
\label{K0Atheta}
\end{eqnarray}
which is an isomorphism of ordered groups~\cite{pva}. The positive
cone is the collection of (equivalence classes of) projections
with positive trace,
\be
\K_0^+(\atheta)=\big\{(r,m)\in\zed^2~\big|~r+m\,\theta\geq0
\big\} \ .
\label{K0+Atheta}\end{equation}
The projection $\P_\theta$ thereby leads to a complete set of projections for
the
entire lattice of D$p$--D$(p-2)$ brane charges.

For completeness and later use, let us also add at this point a few
remarks about the group
$\K_1(\atheta)$. This group is made up of equivalence classes of
homotopic unitary elements in $\atheta$. It is easy to see that all
powers $U^m\,V^r$ are mutually non-homotopic. If $U^m\,V^r$ and
$U^{m'}\,V^{r'}$ are homotopic, then so are
$\e^{-2\pi\ii(m-m')r'\theta}\,U^{m-m'}\,V^{r-r'}$ and $\id$. But there cannot
be a continuous path of unitaries from $U^m\,V^r$ to $\id$ since
$\ncint U^m\,V^r = 0$ for $(m,r)\not=(0,0)$, whereas $\ncint
\id=1$. Passing to the matrix algebra
    $\IM_N(\atheta):=\atheta\otimes\IM_N(\complex)$ does not improve the
    situation since the same argument works with $\ncint$ replaced by
    $\ncint\otimes \Tr$, where $\Tr$ is the usual $N\times N$ matrix
    trace. Thus
\be
\K_1(\atheta) = \IZ[U] \oplus \IZ[V]~. \label{K1Atheta}
\end{equation}

For our purposes we will find it more useful to define two
generalized families of projections $\{\P_n\}_{n\geq1}$ and
$\{\P_n'\}_{n\geq1}$ which are related to the
even and odd order
approximants of the noncommutativity parameter
\be
\theta=\lim_{n\to\infty}\,\theta_n \ , ~~
\theta_n:=\frac{p_n}{q_n} \ . \label{thetalim}\end{equation} Any
irrational number $\theta$ can be treated as a limit
(\ref{thetalim}) of rational numbers $\theta_n$ in a definite way
by using continued fraction expansions. The approximants of
$\theta$, as well as the limiting process in (\ref{thetalim}), are
described in appendix~\ref{appa}, where we also fix some number
theoretic notation. For each $n\in\IN$, following the Elliott-Evans
construction~\cite{ee}, we define two Powers-Rieffel type projections
by
\bea
\P_n & = & V^{-q_{2n-1}}\,\rho(g_n)+\rho(f_n)+\rho(g_n)\,V^{q_{2n-1}} \ ,
\label{projdef} \\
\P_n' & = &U^{q_{2n}}\,\rho'(g'_n)+\rho'(f'_n)+\rho'(g'_n)\,U^{-q_{2n}} \ ,
\label{projdefprime}
\end{eqnarray}
where $\rho'$ is the ``dual'' of the representation \eqn{xsubalg},
\bea
\rho'\,:\,C^\infty\left(\IS^1\right)&\longrightarrow&\atheta ~, \nn
g(y) = \sum_{r\in\IZ}g_{r}~\e^{2 \pi \ii r\,y}&\longmapsto&\rho'(g) =
\sum_{r\in\IZ}g_{r}~V^r ~,
\label{ysubalg}
\end{eqnarray}
and now $U \rho'(g) U^{-1}$ is the image of the shifted function
$g(y-\theta)$. Again, the map \eqn{ysubalg} is just the Weyl map \eqn{weymap}
restricted to functions of
the variable $y$ alone with the corresponding properties \eqn{yweywig}.

The importance of the projections \eqn{projdef} and \eqn{projdefprime} is that
they provide the building blocks for the construction~\cite{ee} of a
sequence of subalgebras $\athetan\subset\atheta$ which converge to the
full algebra $\atheta$ of the noncommutative torus. We shall describe this
construction at length in section~\ref{se:masu}. Each of these
subalgebras is a sum of two algebras of matrix-valued functions on a circle.
Heuristically, the picture which will emerge is that of two
``solitonic fuzzy tori'' which wrap around two circles. Any
field on the noncommutative torus will thereby admit a regularization
by two sets of matrix-valued soliton configurations, each of which is
a function on a circle.

In the remainder of this section we will describe the properties
of the projections $\P_n$ and $\P_n'$. Since for the time being we
will work at a fixed approximation level $n$, to simplify notation
we will suppress the subscript $n$ on the functions $f$, $g$, $f'$
and $g'$ and the subscripts $2n$ and $2n-1$ on the integers $p$
and $q$. To distinguish $q_{2n}$ from $q_{2n-1}$ we will denote
the former integer by $q$ and the latter one by $q'$, and similarly for
$p$. Subscripts will be reintroduced whenever we discuss the limiting
process explicitly.

Let us then look for a projection of the form $\P_n =
V^{-q'}\,\rho(g)+\rho(f)+\rho(g)\,V^{q'}$. As for the Powers-Rieffel
projection \eqn{Ptheta}, the real-valued functions $f$ and $g $ must
now satisfy the conditions
\bea
g(x)\,g(x+q'\,\theta) &=& 0 ~,\nn
\big(f(x)+f(x+q'\,\theta)\big)\,g(x)&=&g(x) \ , \nn
g(x)+g(x-q'\,\theta)&=&\sqrt{f(x)-f(x)^2} \ ,
\label{fgPnconds}
\end{eqnarray}
with $0\leq f\leq1$. We shall also require $f$ to have trace
\be
\beta =p'-q'\,\theta
\label{betatrace}\end{equation}
so that $\P_n$ is of rank $\beta$, and fix $g$ in such a manner that its
$\K_0$-class is $(p',-q')$.
These numbers $\beta$ also come in a sequence
    $\{\beta_{2n}\}$ which is defined in appendix~\ref{appa},
    eq.~(\ref{defbetaeven}).

\begin{figure}[htb]
\epsfxsize=3.5 in
\bigskip
\centerline{\epsffile{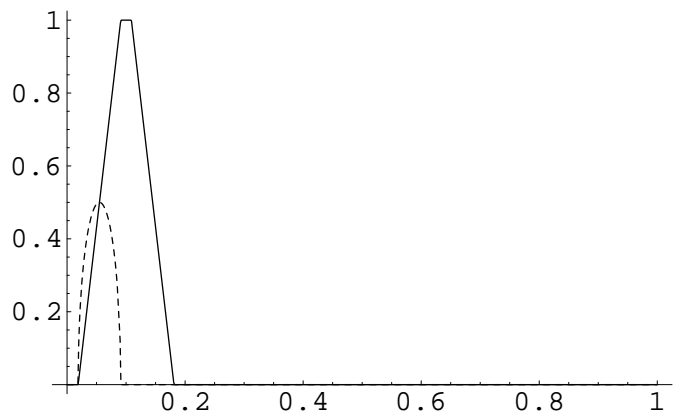}} \caption{\baselineskip=12pt
{\it Profiles of the bump functions $f$ (solid line) and $g$ (dashed
    line) used to construct the projection $\P_n$. The noncommutativity
    parameter is taken to be the inverse of the golden mean,
    $\theta=\frac2{\sqrt5+1}$, while the approximants are chosen as
    $\theta_{2n}=\frac35$ and $\theta_{2n-1}=\frac58$.}}
\bigskip
\label{zawinul12}\end{figure}

As before, the functions $f$ and $g$ are ``bump'' functions which now
differ from zero only in small intervals. Viewed as continuous functions,
they are given by\footnote{\baselineskip=12pt We should really give a
     ``smoothened'' version of these bump functions. This can always be
     accomplished without any difficulty \cite{bo} and we will
implicitly assume that it  has been done whenever necessary.}
\bea
f(x)&=&\left\{\new
\begin{array}{ccrcl}
0 & ~ & 0 &
    \leq~x~\leq &  \frac{1}{2q} - \delta \\
\frac{1}{\delta-\sigma}\,\left(x-\frac {1}{2q}+\delta\right)
& ~ & \frac{1}{2q} - \delta & \leq~x~\leq &  \frac{1}{2q} - \sigma
\\
1 & ~ & \frac{1}{2q} - \sigma & \leq~x~\leq &  \frac{1}{2q} + \sigma
\\
\frac{1}{\delta-\sigma}\,\left(-x+\frac {1}{2q}+\delta\right)
& ~ & \frac{1}{2q} + \sigma & \leq~x~\leq &  \frac{1}{2q} + \delta
\\
0 & ~ & \frac{1}{2q} + \delta & \leq~x~\leq &  1 \\
\end{array}
\right. \ , \nn&&{~~~~}^{~~}_{~~}\nn
g(x)&=&\left\{\new
\begin{array}{ccc}
\sqrt{f(x)-f(x)^2}
& ~ & \frac{1}{2q} - \delta~\leq~x~\leq~\frac{1}{2q} - \sigma \\
0 & ~ & \mbox{\rm otherwise}
\end{array}\right. \ ,
\label{bumpfns}
\end{eqnarray}
where $\sigma<\delta<\frac{1}{2q}$ are positive quantities which
are fixed by two conditions. The first one is simply that the trace of
$f$ be $\beta=p'-q'\,\theta$, i.e. $\int_0^1\dd x~f(x)=\beta$.
{}From the explicit form in \eqn{bumpfns} it is easy to see that
the integral is just $\delta+\sigma$. Thus the first condition is
\be
\delta+\sigma=\beta \ .
\label{reqdelta}
\end{equation}
The second condition comes from the usage of the projections $\P_n$ in
the approximation scheme that we mentioned earlier and it ensures the
best possible transformation properties for $\P_n$ with respect to the
generators $U$ and $V$~\cite{ee}. The condition consists in choosing
$f$ to have the least possible slope in the two intervals
where it is not constant. The minimal slope is the larger of the two numbers
$\beta^{-1}$ and  $(1/q-\beta)^{-1}$ according to whether $\beta$ is
smaller or larger than $1/2q$. Again, from the explicit expression in
\eqn{bumpfns} the slope is just $(\delta-\sigma)^{-1}$. Thus the
second condition is
\be\label{slope}
\delta-\sigma=\left\{\new
\begin{array}{cclcr}
\beta &~& \beta & \leq & \frac 1{2q}\\
\frac 1q -\beta\ &~& \beta & \geq & \frac 1{2q}
\end{array}\right. ~.
\end{equation}
By putting together the conditions \eqn{reqdelta} and \eqn{slope} we get
\be
\sigma=\left\{\new
\begin{array}{c}
0 \\ \beta - \frac1{2q}
\end{array}\right. ~, ~~~~~
\delta=\left\{\new
\begin{array}{cclcr}
\beta &~& \beta & \leq & \frac 1{2q}\\
\frac{1}{2q} &~& \beta & \geq & \frac 1{2q}
\end{array}\right. ~.
\end{equation}
Examples of the functions $f$ and $g$ are plotted in Fig.~\ref{zawinul12}.

One also defines a number $\beta'$ by the relations
$(1/q-\beta)^{-1} := q/q' \beta'$. This is equivalent
to\footnote{\baselineskip=12pt
The sequence
    $\{\beta_{2n-1}\}$ for the $\beta'$'s is defined in
    appendix~\ref{appa}, eq.~(\ref{defbetaodd}).}
\be
\beta' = q\,\theta -p ~,
\end{equation}
from which we have the relation\footnote{See appendix~\ref{appa},
    eq.~(\ref{qbetaid1}).}
\be\label{diph}
q\,\beta + q'\,\beta' = 1 ~.
\end{equation}
The number $\beta'$ plays the same
role for the projection $\P'_n$ as $\beta$ does for $\P_n$.

By construction, the rank of $\P_n$ is $\beta$,
\be
\ncint \P_n = f_0 = \int\limits_0^1\dd x~f(x)= \beta = p'-q'\,\theta ~,
\end{equation}
while its monopole charge is $-q'$,
\be
c_1(\P_n)=- 6\,q'\,\int\limits_0^1\dd x~g(x)^2\,f'(x)= -q'~,
\end{equation}
where the last equality follows from the explicit choice \eqn{bumpfns}
for the bump functions. In a completely analogous manner one finds
\be
\ncint \P'_n = \beta' = -p + q\,\theta ~, ~~~ c_1(\P'_n) = q ~.
\end{equation}
Thus the projection $\P_n$ in (\ref{projdef}) represents a soliton
configuration carrying brane charges $(p_{2n-1},-q_{2n-1})$, and the integers
$p_{2n-1}$ and $q_{2n-1}$ thereby parametrize the vacua of the open
string field theory. The rank $\beta$ of $\P_n$ is the D-brane charge
after tachyon condensation. Analogously, the projection $\P'_n$ has brane
charges $(-p_{2n},q_{2n})$.

\begin{figure}[htb]
\epsfxsize=3.5 in
\bigskip
\centerline{\epsffile{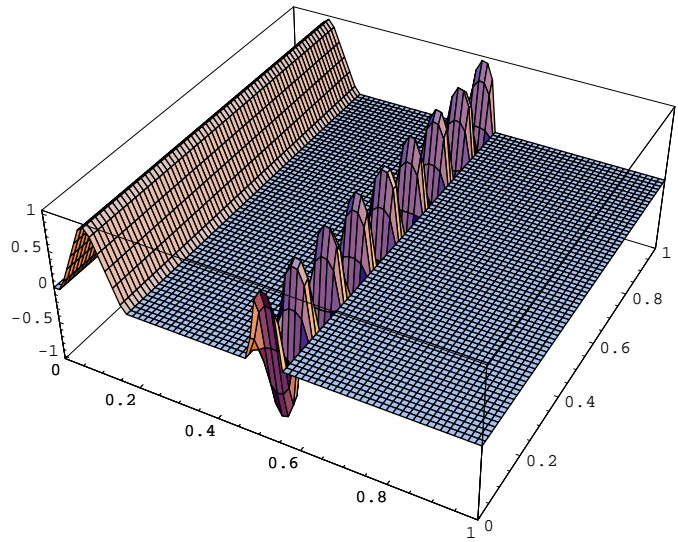}}
\caption{\baselineskip=12pt {\it The soliton field configuration
      corresponding to the projection operator $\P_n$ on the
      noncommutative torus. The noncommutativity parameter is as in
      Fig.~\ref{zawinul12}. The vertical axis is the Wigner function
      $\Omega^{-1}(\P_n)(x,y)$ and the horizontal plane is the
      $(x,y)$-plane.}}
\bigskip
\label{zawinul3}\end{figure}

Because these solitons will converge to generic fields on the
noncommutative torus, it is instructive to examine their spacetime
dependence as elements of $C^\infty(\torus^2)$. {}From
(\ref{nct}), (\ref{wigmap}) and (\ref{projdefprime}) we
can easily compute the Wigner function on $\torus^2$ corresponding to the
projection $\P_n$ in terms of the bump functions (\ref{bumpfns}) as
\be
\Omega^{-1}(\P_n)(x,y)=f(x)+
2\cos\left(2\pi\,q'\,y\right)\,g\left(x-\half \, q'\,\theta\right)\ .
\label{OmegaP}
\end{equation}
The soliton field (\ref{OmegaP}) represents a typical unstable
D7-brane projection configuration and its shape is plotted in
Fig.~\ref{zawinul3}. Note that each physical field configuration
(\ref{OmegaP}) is concentrated in two regions, each of which is
localized along the $x$-cycle of the torus but extended along the
$y$-direction. It therefore defines tachyonic lumps that have
strip-like configurations, unlike the standard point-like
configurations of GMS solitons on the noncommutative plane. The first
lump has a smooth locus of points and strip area $2\sigma$ depending
on both the D-brane charge and the monopole charge. The
second lump contains a periodically spiked locus of support points,
with period $q'$ and area $\delta-\sigma$. The spiking exemplifies
the UV/IR mixing property that generic noncommutative fields possess,
in that the size of the configuration decreases as its oscillation
period (the monopole charge) grows. Similar considerations can be made
for the Wigner function $\Omega^{-1}(\P'_n)(x,y)$.

\section{Soliton Regularization of Noncommutative Fields}\label{se:masu}
\setcounter{equation}{0}

We will now give the construction of the subalgebras
$\athetan$ and describe in precisely what sense these
subalgebras approximate the full algebra $\atheta$ of the
noncommutative torus~\cite{ee}. We will also describe how to appropriately
truncate fields to $\athetan$ in such a manner that they are 
recovered in the limit
$n\to\infty$. Throughout we shall keep in mind the physical
interpretations of these objects within the noncommutative D-brane
soliton picture. In this section we shall describe in some detail how
the pertinent matrix algebras emerge.

\subsection{From Solitons to Matrix Subalgebras}

For a fixed integer $n$, the subalgebra $\athetan$ is constructed
starting from the projections $\P_n$ and $\P'_n$ given
in~\eqn{projdef} and~\eqn{projdefprime}. These two projections
will give rise to two towers in $\atheta$ in which the two unitary
generators $U$ and $V$ are treated symmetrically: one of them is
modelled in one tower and the second in the other tower. A tower
in $\atheta$ of height $n$ is a family of $n$ orthogonal
projections in $\atheta$ all obtained from a single one by the
canonical action of a cyclic subgroup of $\IS^1=\IT^1$ of order
$n$. In the present case the first tower will be of height $q$,
with $q$ projections of trace $\beta = p'-q'\,\theta$, while the
second tower will be of height $q'$, with $q'$ projections of
trace $\beta' = q\,\theta -p$. The two towers will be orthogonal,
i.e. the sum of the projections making up the first tower is the
orthogonal complement of the sum of the projections making up the
second tower. In order to achieve this it is necessary to adjust
the second tower using the fact that any two projections in
$\atheta$ with the same $\K_0$-class are unitarily
equivalent~\cite{ri83}. From the orthogonality property we must then
have that
\beq
q\,(p'-q'\,\theta) + q'\,(q\,\theta-p) = q\,p'-q'\,p = 1~
\end{equation}
which is just the relation \eqn{diph} (see also \eqn{qbetaid1}).

For the rest of this subsection we shall simply write $\P=\P_n$ and
$\P'=\P'_n$. Given the projection $\P$, we first ``translate'' it by
the (outer) automorphism $\alpha:\atheta\to\atheta$ defined by
\be
\alpha(U)=\e^{2\pi\ii p/q}~U \ , ~~ \alpha(V)=V \ .
\label{rhoautodef}
\end{equation}
The corresponding Wigner function (\ref{OmegaP}) is translated
accordingly along the $x$-cycle of the torus~$\torus^2$,
\be
\Omega^{-1}\big(\alpha(\P)\big)(x,y)=
\Omega^{-1}(\P)(x+p/q,y) \ .
\label{Ptransl}
\end{equation}
By repeatedly applying $\alpha$ we can define new projections
\bea
\P^{ii} &:=& \alpha^{i-1}(\P) \nn
&\:=&V^{-q'}\,\rho\Big(g\big(x+(i-1)\,p/q\big)\Big)\nn &&
+\,\rho\Big(f\big(x+(i-1)\,p/q\big)\Big)+
\rho\Big(g\big(x+(i-1)\,p/q\big)\Big)\,V^{q'}
\label{transe}
\end{eqnarray}
for $i=1,\dots,q$. Since $\alpha^q ={\rm id}$, it follows that
$\P=\P^{11}=\P^{q+1,q+1}$. Moreover, using the explicit form of
(\ref{bumpfns}) it is straightforward to check that the elements
\eqn{transe} form a system of mutually orthogonal projection
operators, i.e. $\P^{ii}\,\P^{jj}=\delta_{ij}\,\P^{jj}$. As the
notation suggests, these projections are the diagonal elements of
a basis for a certain matrix subalgebra of $\atheta$ which we are
now going to describe.

Let $\ch_i\subset \ch =
L^2(\atheta\,,\,\ncint\,)$ be the range of the projection
$\P^{ii}$. Physically, if $\P^{ii}$ describes a collection of
noncommutative D-brane solitons, then $\ch_i$ is the corresponding
Chan-Paton space of the brane configuration, and
$\P^{ii}\,\atheta\,\P^{ii}$ is the algebra of endomorphims of this
Chan-Paton space. Of course, this space (and its endomorphism
algebra) need not be finite-dimensional, in which case the induced
D-brane worldvolume carries a $U(\infty)$ gauge symmetry after
tachyon condensation owing to the infinite collection of image
branes on the torus. This infinite-dimensional symmetry
corresponds to invariance of the noncommutative field theory under
symplectomorphisms of the D-brane worldvolume~\cite{LSZ2}. On each of the
$\ch_i$ the corresponding projection $\P^{ii}$ acts as the
identity $\id$, while for $j\neq i$ one has
$\ch_i\subset\ker(\P^{jj})$. In the D-brane picture, this means
that the dynamical degrees of freedom on any pair of distinct
non-BPS solitons acts on each other's massless open string states.

We will also need another set of operators which map one Chan-Paton
subspace into another, as they will be the off-diagonal elements of
the matrix algebra basis. For this, we consider the operator
\be
\Pi^{21}:=\P^{22}\,V\,\P^{11} \ .
\label{Pi21}
\end{equation}
This operator is a mapping from $\ch_1$ to $ \ch_2$, but is not
an isometry, i.e. $(\Pi^{21})^\dag\,\Pi^{21}\neq\id$. This fact may be
remedied somewhat by introducing a related {\it partial} isometry
$\P^{21}$, i.e.\ an operator for which $(\P^{21})^\dag\,\P^{21}$
and $\P^{21}\,(\P^{21})^\dag$ are projection operators, or
equivalently $\P^{21}\,(\P^{21})^\dag\,\P^{21}=\P^{21}$. Such an
operator is given by the partial isometry appearing in the polar
decomposition
\be
\Pi^{21}:=\P^{21}\,\left|\Pi^{21}\right| ~, ~~~~~\, \,
\left|\Pi^{21}\right|=\sqrt{\,\left(\Pi^{21}\right)^\dag\, \Pi^{21} }~ ,
\label{Pi21polar}
\end{equation}
which is well-defined since the operator (\ref{Pi21}) is
bounded. The decomposition (\ref{Pi21polar}) is understood as an equation in
the representation of the algebra $\atheta$ on the Hilbert space $\cal H$, so
that $\P^{21}\in\atheta$. The physical significance
of such an operator is that it is unitary in the orthogonal complement
to a kernel and a cokernel, and hence will produce localized solitons
(in the Wigner representation). The operator $\Pi_n^{21}$ and the
partial isometry $\P_n^{21}$ come arbitrarily close to each other in
the large $n$ limit~\cite{ee}, in the sense that
\be
\lim_{n\to\infty}\,\left\|\Pi_n^{21}-\P_n^{21}\right\|^{~}_0=0 \ .
\label{PiPlim}
\end{equation}

\begin{figure}
\epsfxsize=3.5 in
\bigskip
\centerline{\epsffile{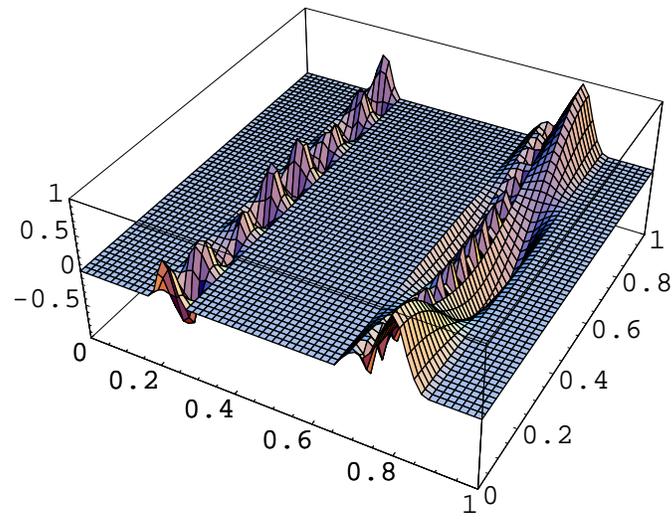}}
\epsfxsize=3.5 in
\medskip
\centerline{\epsffile{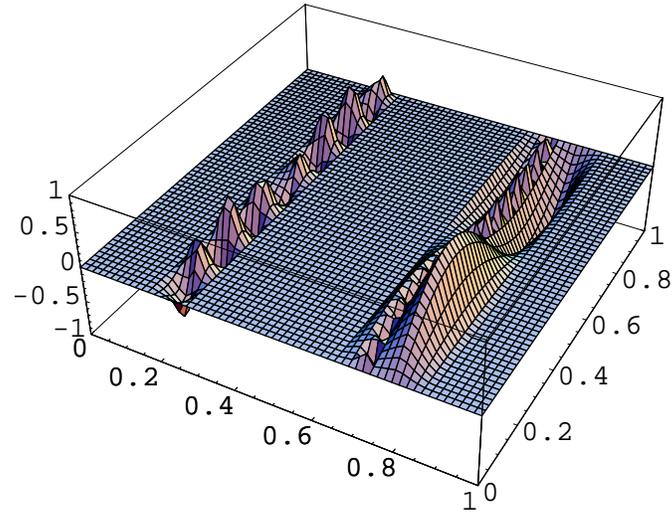}}
\epsfxsize=3.5 in
\medskip
\centerline{\epsffile{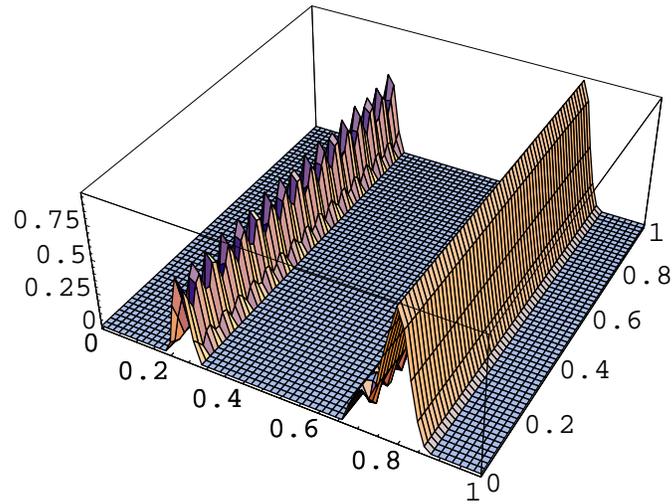}}
\caption{\baselineskip=12pt {\it The soliton field configuration
      corresponding to the operator $\Pi_n^{21}$ on the
      noncommutative torus. Displayed are its real part (top), imaginary part
      (middle), and modulus (bottom). Parameter values and axes are as
      in Fig.~\ref{zawinul3}.}}
\bigskip
\label{zawinul4}
\end{figure}

By using (\ref{nct}), (\ref{wigmap}) and (\ref{transe}), a
straightforward calculation gives the Wigner function on $\torus^2$
corresponding to the operator (\ref{Pi21}) in terms of the periodic
bump functions (\ref{bumpfns}) as
\bea
\e^{-2\pi\ii y}~\Omega^{-1}\left(\Pi^{21}\right)(x,y)&=&
f\left(x+\mbox{$\frac pq-\frac\theta2$}\right)\,f\left(x+
\mbox{$\frac\theta2$}\right)+g\left(x+\mbox{$\frac pq-\frac\theta2$}
\right)\,g\left(x+\mbox{$\frac\theta2$}\right)\nn&&
+\,g\left(x+\mbox{$\frac pq-\frac{(2q'+1)
\,\theta}2$}\right)\,g\left(x+\mbox{$\frac{(2q'-1)\,\theta}2$}
\right)\nn&&+\,\e^{4\pi\ii q'\,y}\,g\left(x+\mbox{$\frac pq-\frac{(2q'+1)
\,\theta}2$}\right)\,g\left(x+\mbox{$\frac\theta2$}\right)\nn&&+\,
\e^{-4\pi\ii q'\,y}\,g\left(x+\mbox{$\frac pq-\frac\theta2$}\right)\,
g\left(x-\mbox{$\frac{(q'-1)\,\theta}2$}\right)\nn&&
+\,\e^{2\pi\ii q'\,y}\,\left[f\left(x+\mbox{$\frac pq-\frac{(q'+1)\,
\theta}2$}\right)\,g\left(x-\mbox{$\frac{(q'-1)\,\theta}2$}\right)
\right.\nn&&
{}~~~~~~~~~~~~
+\left.f\left(x+\mbox{$\frac{(q'+1)\,\theta}2$}\right)\,g\left(x+
\mbox{$\frac pq-\frac{(q'+1)\,\theta}2$}\right)\right]\nn&&
+\,\e^{-2\pi\ii q'\,y}\,\left[f\left(x+\mbox{$\frac pq+
\frac{(q'-1)\,\theta}2$}\right)\,g\left(x-\mbox{$\frac{(q'-1)
\,\theta}2$}\right)\right.\nn&&
{}~~~~~~~~~~~~
+\left.f\left(x-\mbox{$\frac{(q'-1)\,\theta}2$}\right)
\,g\left(x+\mbox{$\frac pq-\frac{(q'+1)\,\theta}2$}\right)\right] \ .
\label{OmegaPi21}
\end{eqnarray}
According to (\ref{PiPlim}), the function (\ref{OmegaPi21}) represents
the typical stable D7-brane soliton partial isometry (at least for
sufficiently large approximation level $n$). Its shape is plotted in
Fig.~\ref{zawinul4}. Again, the multi-soliton image is
apparent, with smooth and periodically spiked support loci. Note that
while the modulus of the function expectedly displays
the
characteristic strips of projection solitons, the lumps of its real and
imaginary parts are point-like configurations.

Using (\ref{Pi21}) and
(\ref{Pi21polar}), we may now define translated partial isometries
analogously to what we did in \eqn{transe} as
\be
\P^{i+2,i+1}:=\alpha^i\left(\P^{21}\right) \ , ~~ i=1,\ldots, q-2 \ ,
\label{ejkdef}
\end{equation}
where $\alpha$ is the automorphism defined in \eqn{rhoautodef}.
Finally, we also define
\be
\P^{ji}:=\left(\P^{ij}\right)^\dag \ .
\label{Pjidagdef}
\end{equation}
The important fact, proven in appendix~\ref{appb}, is that for the
operators (\ref{transe}) and (\ref{ejkdef}) which we have defined,
there is a set of relations
\be
\P^{ij}\,\P^{kl}=\delta_{jk}\,\P^{il} \ . \label{matrixmult}
\end{equation}
These relations suggest the definition of $q^2$ operators $\P^{ij}$,
$1\leq i,j\leq q$. The remaining cases ($j\neq i$ and $j\neq i\pm1$) are
{\it defined} by (\ref{matrixmult}). For example,
$\P^{13}:=\P^{12}\,\P^{23}$, and so on.
Recall that $\ch_i\subset \ch = L^2(\atheta\,,\,\ncint\,)$ is the range
of the projection $\P^{ii}$. Then,
the newly defined operators
$\P^{ij}$ obtained in this way are partial isometries which are mappings
from $\ch_j$ to $\ch_i$, i.e. elements of
$\P^{ii}\,\atheta\,\P^{jj}$. For the collection of all of  them
$\{\P^{ij}\}_{1\leq i,j\leq q}$, the relation~\eqn{matrixmult}
     holds. In this way we can complete the sets of operators
     (\ref{transe}) and (\ref{ejkdef}) into a system of
matrix units which generate a $q\times q$ matrix algebra. Because of
(\ref{matrixmult}), a generic element of this algebra is a complex
linear combination $\sum_{i,j}\,a_{ij}~\P^{ij}$ and the product is the
usual matrix multiplication.

There is, however, a caveat. The operators $\P^{i+2,i+1}$
in~\eqn{ejkdef} are only defined for $i\leq q-2$, and this is in fact
sufficient to define all of the $\P^{ij}$ using (\ref{matrixmult}), including
\be
\P^{1q}:=\P^{12}\,\P^{23}\cdots\P^{q-1,q} \ . \label{defe1q}
\end{equation}
On the other hand, we can also define
\be
\tilde\P^{1q}:=\alpha^{q-1}\left(\P^{21}\right) \ . \label{defetilde1q}
\end{equation}
For the $q\times q$ matrix algebra to close, it would be necessary
that the two operators defined by~\eqn{defe1q} and~\eqn{defetilde1q}
coincide. This is \emph{not} the case. However, although they are not
identical, both of these operators are isometries from $\ch_q$ to $\ch_1$. As
a
consequence, they are related by an operator $z$ which is unitary on
$\ch_1$, i.e. a
unitary element of
$\P^{11}\,\atheta\,\P^{11}$, and which is therefore a partial isometry
on the full Hilbert space $\ch$. We therefore have
\be
\tilde\P^{1q}:=z~\P^{1q} \ .
\end{equation}

This means that the matrix units $\P^{ij}$, {\it along with} the
partial isometry $z$, close a subalgebra of $\atheta$, in which,
using (\ref{matrixmult}), a generic element is a complex linear
combination of the form
\be \label{Az}
\A(z)=\sum_{k\in\IZ}~\sum_{i,j=1}^qa_{ij;k}~z^k
\,\P^{ij} \ .
\end{equation}
By regarding $z$ as the unitary generator of a circle $\IS^1$, this subalgebra
is (naturally isomorphic to) the algebra $\mat_q(C^\infty(\IS^1))$ of
$q\times q$ matrix-valued functions on the circle. Since we are interested in
continuous and differentiable functions, we will always assume that
the complex expansion coefficients $a_{ij;k}$ in (\ref{Az}) are of
sufficiently rapid descent as $k\to\infty$,
i.e. $\{a_{ij;k}\}\in\IM_q(\IC)\otimes S(\IZ)$. The identity element
of this subalgebra is
\be
\id_q=\sum_{i=1}^q\P^{ii} \ .
\label{idqPii}
\end{equation}
   From the above definitions it follows that the trace of the
matrix units is given by
\be
\ncint\P^{ij}=\beta\,\delta_{ij} \ .
\label{Pijtrace}
\end{equation}
In particular, the identity element (\ref{idqPii}) has trace
$\ncint\id_q=q\,\beta$.

In the same way, by starting from the projection $\P'=\P'_n$ in
(\ref{projdefprime}), a second set of dual projections
$\{\hat\P^{\prime\,i'i'}\}_{1\leq i'\leq q'}$ can be built. This
is tantamount to using the $\zed_4$ Fourier transformation $U\mapsto
V,V\mapsto U^{-1}$ and
$(p,q)\leftrightarrow(-p',-q')$ in the above construction. The
dual set of projections is {\it not} orthogonal in $\atheta$ to the
first set above. However, because of (\ref{Pijtrace}) and the
Diophantine property of appendix~\ref{appa}, eq.~(\ref{qbetaid1}), the
second set is complementary to the first in the sense that the $\K_0$-class of
$\sum_i\P^{ii}+\sum_{i'}\hat\P^{\prime\,i'i'}$ is equal to the class
$(1,0)$ of the unit element of $\atheta$. It follows that
$\sum_{i'}\hat\P^{\prime\,i'i'}$ is unitarily equivalent to
$\id-\sum_i\P^{ii}$ (as we have mentioned, any two projections
    in $\atheta$ with the same $\K_0$-class are unitarily
equivalent \cite{ri83}).

We can therefore ``rotate'' the dual set of projections by conjugating
it with the corresponding unitary operator $w$, and thereby obtain a
gauge equivalent set of projections which is orthogonal to the first
set. This unitary operator can be chosen in such a manner
that
$\lim_{n\to\infty}\|[U,w_n]\|^{~}_0=\lim_{n\to\infty}\|[V,w_n]\|^{~}_0=0$.
This is essential to ensure that the orthogonal direct sum of the two algebras
built from
each set of projections contains elements approximating the unitary
generators $U$ and $V$ of the noncommutative torus $\atheta$, as
will be analysed in more detail in the next subsection. With the gauge
transformed dual projections
$\P^{\prime\,i'i'}:=w\,\hat\P^{\prime\,i'i'}\,w^\dag$, we can now
build another set of matrix units $\P^{\prime\,i'j'}$, $1\leq i',j'\leq q'$
which again close a $q'\times q'$ matrix algebra up to
a partial isometry $z'$.

By proceeding as before, for each integer $n$, one generates
an algebra which is isomorphic to a matrix algebra
\be
\athetan~\cong~\mat_{q_{2n}}\left(C^\infty(\circles^1)\right)\oplus
\mat_{q_{2n-1}}\left(C^\infty(\circles^1)\right) \ . \label{An}
\end{equation}
The direct sum arises from the orthogonality of the two towers based on
the projections $\P_n$ and $\P'_n$, respectively.
As we will discuss further later on, it is essential to have {\it two}
copies of such matrix algebras as in (\ref{An}), for K-theoretic
reasons. In what follows we will often use a matrix notation for the
elements of $\athetan$. The matrix elements are always understood to be
multiplied by the operators $\P^{ij}$ and $\P^{\prime\,i'j'}$ when
regarding them as elements of $\atheta$.

\subsection{The Approximation\label{Approx}}

We are now ready to describe in which precise sense the algebra $\athetan$ in
(\ref{An}) approximates the full noncommutative torus $\atheta$ \cite{ee}.
A derivation of this approximation by means of Morita-Rieffel equivalence
bimodules is presented in \cite{el} (see also \cite{bo}). We stress that
$\athetan$, being constructed out of elements of the noncommutative
torus, is a {\it subalgebra} of $\atheta$. The fact that this
subalgebra approximates the noncommutative torus resides in the
fact that for each element $a\in\atheta$, it is possible to construct
a corresponding element $\a_n\in\athetan$ which approximates it
in norm. The key ingredients in the construction are two unitary
elements $\U_n,\V_n\in\athetan$ which closely approximate  the
generators $U$ and $V$ of $\atheta$. The approximation improves as
$n\to\infty$, whereby the distance, in norm, between $\U_n$ and $U$
and between $\V_n$ and $V$ becomes arbitrarily small. We will give the
matrix expressions for $\U_n$ and $\V_n$ and the estimate of their
difference from $U$ and $V$ without proof, referring to~\cite{ee} for
details.  In this subsection we shall reintroduce the subscripts
$n$ on all quantities in order to be able to take limits.

As we recall in appendix~\ref{appa}, one can approximate the
noncommutativity parameter $\theta$ by
sequences of even and odd order approximants
$\theta_{2n}<\theta<\theta_{2n-1}$ with $\theta_k= {p_k}/{q_k}$. For each
level $n$ we introduce roots of unity
\be
\omega_n=\e^{2\pi\ii\theta_{2n}} \ , ~~
\omega_n'=\e^{2\pi\ii\theta_{2n-1}} ~,
\label{omegandef}
\end{equation}
with $(\omega_n)^{q_{2n}}=1=(\omega_n')^{q_{2n-1}}$. Define
\bea
\U_n&=&\left(\,\sum_{i=1}^{q_{2n}}(\omega_n)^{i-1}~\P_n^{ii}\right)\oplus
\left(\,\sum_{i'=1}^{q_{2n-1}-1}\P_n^{\prime\,i',i'+1}+z'\,
\P_n^{\prime\,q_{2n-1},1}\right) \nn
&=&\begin{pmatrix}\,{\cal C}_{q_{2n}}& (0)_{q_{2n}\times q_{2n-1}} \cr
(0)_{q_{2n-1}\times q_{2n}} &{\cal S}_{q_{2n-1}}(z'\,)\,\cr\end{pmatrix} \ ,
\nn
&& {~~~~~}^{~~}_{~~} \nn
\V_n&=&\left(\,\sum_{i=1}^{q_{2n}-1}
\P_n^{i,i+1}+z\,\P_n^{q_{2n},1}\right)\oplus\left(\,
\sum_{i'=1}^{q_{2n-1}}(\omega_n')^{i'-1}~\P_n^{\prime\,i'i'}\right) \nn
&=& \begin{pmatrix}\,{\cal S}_{q_{2n}}(z)& (0)_{q_{2n}\times q_{2n-1}}
\cr (0)_{q_{2n-1}\times q_{2n}} &{\cal C}_{q_{2n-1}}\,\cr
\end{pmatrix} \ ,
\label{defUVn}
\end{eqnarray}
where generally $(a)_{q\times r}$ denotes the $q\times r$ matrix whose
entries are all equal to $a\in\IC$.
In these expressions, for any pair of relatively prime integers $p, q$
with $q>0$, ${\cal C}_q$ is the $q\times q$ unitary clock matrix as in
\eqn{torfuz}, while for any $z\in\circles^1$, ${\cal S}_q(z)$ is the
generalized $q\times q$ unitary shift matrix
\be
{\cal S}_q(z)=\begin{pmatrix}\,0&1& & &0\,\cr &0&1& & \cr
    &  &\ddots&\ddots& \cr & & &\ddots&1\,\cr\,z& & & &0\,\cr
\end{pmatrix}
\label{zshift}
\end{equation}
with
\be
\big({\cal S}_q(z)\big)^q=z\,\id_q \ .
\label{zshiftid}\end{equation}
The generalization in the shift matrix (\ref{zshift}) resides in the
presence of the generic circular coordinate $z$. It becomes the usual
shift matrix in \eqn{torfuz} when $z$ is taken to be equal to~$1$,
${\cal S}_q={\cal S}_q(1)$.

As mentioned in section~\ref{FTNT}, the clock and shift matrices
${\cal C}_q$ and ${\cal S}_q(1)$ form a basis for the
finite-dimensional algebra  $\IM_q(\IC)$ of $q\times q$
complex-valued matrices. By considering both $z$ and $z'$ to be the
unitary generators of two distinct copies of the algebra
$C^\infty(\circles^1)$, the matrices (\ref{defUVn}) generate the
infinite-dimensional algebra
$\athetan\cong\mat_{q_{2n}}\left(C^\infty(\circles^1)\right)
\oplus\mat_{q_{2n-1}}\left(C^\infty(\circles^1)\right)$
of matrix-valued functions on two circles. From their definition in
\eqn{defUVn}, one finds that
\be
(\U_n)^{q_{2n}\,q_{2n-1}}=\begin{pmatrix}\id_{q_{2n}}& (0)_{q_{2n}
\times q_{2n-1}} \cr (0)_{q_{2n-1}\times q_{2n}} &
z^{\prime\,q_{2n}}\,\id_{q_{2n-1}}\,\cr\end{pmatrix} \ , ~~
(\V_n)^{q_{2n}\,q_{2n-1}}=\begin{pmatrix}z^{q_{2n-1}}
\,\id_{q_{2n}}& (0)_{q_{2n}\times q_{2n-1}} \cr (0)_{q_{2n-1}
\times q_{2n}} &\id_{q_{2n-1}}\,\cr\end{pmatrix} \, ,
\label{UnVnq}
\end{equation}
and these matrices generate the center $C^\infty(\circles^1)\oplus
C^\infty(\circles^1)$ of the algebra $\athetan$.
Moreover, $\U_n$ and $\V_n$ have a commutation relation which
approximates the one (\ref{nct}) of $U$ and $V$,
\be
\V_n\,\U_n={\mbf\omega}_n~\U_n\,\V_n
\label{UnVnomegan}
\end{equation}
with
\be
{\mbf\omega}_n=\omega_n\,\sum_{i=1}^{q_{2n}}\P_n^{ii}
{}~\oplus~\omega'_n\,\sum_{i'=1}^{q_{2n-1}}\P_n^{\prime\,i'i'}
=\begin{pmatrix}\,\omega_n\,\id_{q_{2n}}& (0)_{q_{2n}\times q_{2n-1}}
\cr (0)_{q_{2n-1}\times q_{2n}}
     &\omega_n'\,\id_{q_{2n-1}}\,\cr\end{pmatrix} \ .
\end{equation}
In all of these expressions we have stressed the important double
interpretations of these generators. The first equality emphasizes
that they are still elements of the algebra $\atheta$ (i.e. they are
expandable in a basis of solitons on the noncommutative torus), while
the second equality reminds us that they are elements of a matrix
algebra (i.e. they are matrix-valued fields on two circles).

Following \cite{ee}, we will now argue that the matrix algebra $\athetan$
``approximates'' the noncommutative torus $\atheta$. Some of the
technical details are given in appendix~\ref{appc}.
As recalled in appendix~\ref{appa}, in the limit $n\to\infty$ both
sequences $\theta_{2n}$ and $\theta_{2n-1}$ converge to $\theta$ and
both sequences $q_{2n}$ and $q_{2n-1}$ diverge. Then, the generators
$\U_n$ and $\V_n$ of $\athetan$ converge in norm to the generators $U$
and $V$ of $\atheta$ as
\be
\left\|U-\U_n\right\|^{~}_0\leq\varepsilon_n \ , ~~
\left\|V-\V_n\right\|^{~}_0\leq\varepsilon_n \ ,
\label{UVconvn}
\end{equation}
where
\be
\varepsilon_n=\max\left(\frac1{q_{2n}}\,,\,\frac1{q_{2n-1}}\right)~
C\left(\frac{q_{2n-1}\,\beta_{2n-1}}{q_{2n}\,\beta_{2n}}\right)
\label{varepsilonn}
\end{equation}
and $C$ is a suitable bounded function.
For each $n$ one now constructs a projection
\bea
\Gamma_n\,:\,\atheta&\longrightarrow&\athetan \ , \nn
a=\sum_{(m,r) \in \IZ^2} a_{m,r}~U^m\,V^r&\longmapsto&
\Gamma_n(a)  =\sum_{(m,r) \in \IZ^2} a_{m,r}~(\U_n)^m\, (\V_n)^r \
,\label{Gammanhomo}
\end{eqnarray}
which using \eqn{UnVnq} can also be written as
\be
\Gamma_n(a) = \sum_{i,j=1}^{q_{2n}q_{2n-1}}
    A_{ij}^{(n)} ~(\U_n)^i\,(\V_n)^j
\label{GammanaA}\end{equation}
where
\be
A_{ij}^{(n)}= \sum_{(m,r) \in \IZ^2} a_{i+m\,q_{2n}\,,\,j+r\,q_{2n-1}}~
\begin{pmatrix}\,z^{r\, q_{2n-1}}\,\id_{q_{2n}}& (0)_{q_{2n}\times
     q_{2n-1}}
   \cr (0)_{q_{2n-1}\times q_{2n}} &
z^{\prime\, m\, q_{2n}}\,\id_{q_{2n-1}}\,\cr\end{pmatrix}~.
\label{Aexp}
\end{equation}
In particular, $\Gamma_n(U)=\U_n$ and $\Gamma_n(V)=\V_n$, which along with
\eqn{UnVnomegan} shows that $\Gamma_n$ is not an algebra
homomorphism. It becomes one, however, in the limit $n\to\infty$.
The crucial fact is that for any element $a\in\atheta$, its projection
$\Gamma_n(a)$ is very close to it in norm, in the sense that from
\eqn{UVconvn} it follows that their difference goes to zero in the limit,
\be
\lim_{n\to\infty} ~\big\|a-\Gamma_n(a)\big\|^{~}_0 ~ =~0 ~.
\end{equation}

Therefore, to each element of $\atheta$ there always corresponds an
element of the subalgebra $\athetan$ to within an arbitrarily small
radius. A generic element $a\in\atheta$ can be approximated to
arbitrarily good precision by a matrix-valued function on two circles
of the form
\be
\Gamma_n(a) = \a_n(z,z'\,)=\begin{pmatrix}\,\a_n(z)& (0)_{q_{2n}
\times q_{2n-1}} \cr (0)_{q_{2n-1}\times q_{2n}}
&\a_n'(z'\,)\,\cr\end{pmatrix} \ ,
\label{anzzprime}
\end{equation}
with $\a_n(z)\in\mat_{q_{2n}}(C^\infty(\circles^1))$ and
$\a_n'(z'\,)\in\mat_{q_{2n-1}}(C^\infty(\circles^1))$. Only the
information about the higher momentum modes $a_{m,r}$ of the expansion
of $a$ is lost (i.e. for $m,r>q_{2n}\,q_{2n-1}$), and these
coefficients are small for Schwartz sequences. Hence the approximation
for large $n$ is good.

It is possible to prove an even stronger result~\cite{ee} which gives
a concrete realization of the noncommutative torus as the
\emph{inductive limit} $\atheta=\bigcup_{n=0}^\infty\bthetan$
of an appropriate inductive system of algebras
$\{\bthetan,\iota_n\}_{n\geq0}$, together with injective unital $*$-morphisms
$ \iota_n : \bthetan \hookrightarrow \cb_{{n+1}}$. It turns out that,
for K-theoretical reasons, the finite level algebras $\bthetan$ must
be taken as $\bthetan=\ca_{2n+1}$, with the latter algebra of the form
\eqn{An}. The crucial issue here is that the embeddings from one
algebra to the next one must be taken in such a way that, in the
limit, the K-theory groups (\ref{K0Atheta}) and (\ref{K1Atheta}) of
the noncommutative torus are obtained. That a judicious choice here is indeed
possible follows from the K-theoretic properties
$\K^0(\circles^1)=\K^1(\circles^1)=\zed$ of the circle, so that by
Morita equivalence the K-theory groups of the matrix algebras
$\athetan$ are given by
\be
\K_0(\athetan)=\K_1(\athetan)=\zed\oplus\zed \ ,
\label{K01Athetan}
\end{equation}
with the canonical ordering $\K_0^+(\athetan)=\IN_0\oplus\IN_0$
for the dimension group. The details are described in
appendix~\ref{appd}. A very heuristic explanation for the
necessity of using two towers in the matrix regularization will be
given in the next section.

The physical interpretation of the projection~(\ref{GammanaA})
should be clear. On the original noncommutative torus, there is an
infinite number of image D-branes parametrized by the momentum lattice
$\IZ^2$ of the quotient space $\IT^2=\IR^2/\IZ^2$ used to
construct the brane configurations from the universal cover of the
torus. The mapping~(\ref{GammanaA}) thereby corresponds to a
truncation of fields on the noncommutative torus in such a way
that there are only a finite number $q_{2n}\,q_{2n-1}$ of image
D-branes on $\torus^2$ at each level $n$, corresponding to the
collection of physical open string modes which are invariant under
the action of the cyclic group
$\zed_{q_{2n}\,q_{2n-1}}\times\zed_{q_{2n}\,q_{2n-1}}$. The Wigner
map can also be used to determine the finite two-cocycle that
appears in the twisted convolution of the image of the product in
the finite algebra, giving the analog of (\ref{twco}) for the
noncommutative torus, although we shall not investigate this
matter here.

Instead, in what follows it will be more useful to
encode the noncommutativity of the algebra $\athetan$ by using the
usual matrix multiplication of functions on a circle $\circles^1$.
We then obtain an expansion of noncommutative fields in terms of
both stable and unstable D-brane solitons on the torus $\torus^2$.
The remarkable fact about this soliton expansion is that it leads
to a description of the dynamics of a noncommutative field in a
very precise way in terms of a one-dimensional matrix model, whose
(inductive) limit reproduces {\it exactly} the original continuum
dynamics. This is quite unlike the situation with the
zero-dimensional matrix model regularizations of noncommutative
field theory~\cite{AMNS1,lls,AMNS2}, whereby the finite-dimensional
matrix algebras can never realize the noncommutative torus as an
inductive limit~\cite{pvb}. In the present case the regularization in
fact mimicks most properties of the continuum field theory already at the
finite level, owing to this much stronger limiting behaviour. In the
following we shall explore the implications of the soliton
regularization within this context.

\section{Noncommutative Field Theory as Matrix Quantum
    Mechanics\label{NCFTMQM}}
\setcounter{equation}{0}

In this section we shall go back to the setting of
section~\ref{DBraneNCSoliton} and consider open superstring field
theory on the background ${\cal M}\times\torus^2$. As
discussed there, in the presence of a constant $B$-field the tachyon
fields $T$ are functions on ${\cal M}\to\atheta$. The generic
situation we will therefore consider is that there is a set of fields,
which we denote collectively by $\Phi$, with Lagrangian density $\cal
L$, all of which are functions on ${\cal M}\to\atheta$.
By remembering that on the algebra $\atheta$
the integration  is given by the trace (\ref{ncintadefa00}),
the action for noncommutative string field theory compactified on a
two-torus can be written schematically as
\be
S=\frac{g_s\,\mu_9}{G_s}\,\int\limits_{\cal M}\sqrt{\det G}~
\ncint{\cal L}[\Phi,\partial_\mu\Phi] \ ,
\label{NCstrfield}
\end{equation}
where $G_s$ and $G_{\mu\nu}$ are the effective coupling and metric
felt by the open strings in the presence of the constant $B$-field
along $\torus^2$, $g_s$ is the closed string coupling constant, and
$\mu_9$ is the spacetime-filling D-brane tension in the absence of the
$B$-field. The derivatives $\partial_\mu$, $\mu=1,2$ are the canonical linear
derivations on $\atheta$ defined in (\ref{t2act}).

We will now use the mapping (\ref{Gammanhomo}) onto the approximating
subalgebra $\athetan$ in (\ref{An}) to build a matrix field theory which
regulates the field theory (\ref{NCstrfield}) on $\atheta$. Using
(\ref{anzzprime}) we replace the fields $\Phi$ on $\atheta$ by the
fields $\mPhi_n(z,z'\,)$ on $\athetan$ which are direct sums of
$q_{2n}\times q_{2n}$ matrix fields $\mPhi_n(z)$ and $q_{2n-1}\times
q_{2n-1}$ matrix fields $\mPhi_n'(z'\,)$ on $\circles^1$.
We need to examine the actions of the trace (\ref{ncintadefa00}) and
derivative
(\ref{t2act}) on the algebra $\athetan$. We will reinterpret them
as operations on the matrix algebras $\athetan$, without reference to
their embeddings as subalgebras of the noncommutative torus $\atheta$,
which in the $n\to\infty$ limit converge to the trace and derivative
on $\atheta$. The resulting matrix quantum mechanics can be regarded as a
non-perturbative regularization of the original continuum field theory on the
noncommutative torus, which is obtained as the limit $n\to\infty$.

\subsection{Spacetime Averages}

Let us start with the canonical trace. On elements of $\atheta$
the trace (\ref{ncintadefa00}) is determined through the definition $\ncint
U^m\,V^r:=\delta_{m0}\,\delta_{r0}$. To
determine the trace of corresponding elements of
$\athetan$, we note that because of (\ref{Pijtrace}), traces of powers
$(\U_n)^m\,(\V_n)^r$ of the generators (\ref{defUVn}) vanish unless
the corresponding powers of both the clock and shift operators are
proportional to the identity elements $\id_{q_{2n}}$ or
$\id_{q_{2n-1}}$, which happens whenever $m$ and $r$ are arbitrary integer
multiples of $q_{2n}$ or $q_{2n-1}$. From the definitions of the
unitaries $z$ and $z'$, we further have
\be
\ncint z^m\,\id_{q_{2n}}=q_{2n}\,\beta_{2n}\,\delta_{m0} \ , ~~
\ncint z^{\prime\,m}\,\id_{q_{2n-1}}=q_{2n-1}\,\beta_{2n-1}\,
\delta_{m0} \ ,
\label{ncintzzprime}
\end{equation}
from which it follows that
\be
\ncint(\U_n)^m\,(\V_n)^r=\sum_{k\in\IZ}\left(q_{2n}\,
\beta_{2n}\,\delta_{m\,,\,q_{2n}\,k}\,\delta_{r0}+q_{2n-1}\,
\beta_{2n-1}\,\delta_{m0}\,\delta_{r\,,\,q_{2n-1}\,k}\right) \ .
\label{ncintUnVnpowers}
\end{equation}

In the large $n$ limit, by using (\ref{qbetaid1}) we see that the trace
$\ncint\Gamma_n(a)$ is therefore well approximated by $a_{0,0}$, since the
correction terms $a_{q_{2n}\,k\,,\,0}$ and $a_{0\,,\,q_{2n-1}\,k}$ for
$k\neq0$ are then small for Schwartz sequences. It is now clear how to rewrite
$\ncint\Gamma_n(a)$ in terms of operations which are intrinsic to the matrix
algebras (\ref{An}).
The trace (\ref{ncintzzprime}) can be reproduced on functions on $\circles^1$
by integration over the circle, while the trace of the matrix degrees of
freedom are ordinary $q_{2n}\times q_{2n}$ and $q_{2n-1}\times q_{2n-1}$
matrix
traces $\Tr$ and $\Tr'$, respectively, accompanied by the appropriate
normalizations
$q_{2n}\,\beta_{2n}$ and $q_{2n-1}\,\beta_{2n-1}$. In $\athetan$, we regard
$z$
and $z'$ as ordinary circular coordinates and set $z:=\e^{2\pi\ii\tau}$,
with $\tau\in[0,1)$, and $z':=\e^{2\pi\ii\tau'}$, with
$\tau'\in[0,1)$. It then follows that the trace of a generic element
(\ref{anzzprime}) can be written solely in terms of matrix quantities
as
\be
\ncint\a_n=\beta_{2n}\,\int\limits_0^1
\dd\tau~\Tr\,\a_n(\tau)+\beta_{2n-1}
\,\int\limits_0^1\dd\tau'~\Tr'\,\a_n'(\tau'\,) \ .
\label{ncintazzprime}
\end{equation}

\subsection{Kinetic Energies\label{Kinetic}}

The definition of the equivalent of the derivations~\eqn{t2act} is
somewhat more involved. We will define two \emph{approximate}
derivations $\mnabla_\mu$, $\mu=1,2$ on $\athetan$, which in the limit
$n\to\infty$ approach the $\del_\mu$'s. They are approximate derivations
in the sense that the Leibniz rule holds only in the limit. They are,
however, sufficient for the definition of the regulated action.
For this, let us look more closely at the map
$a\mapsto\a_n=\Gamma_n(a)$ defined in~(\ref{GammanaA},\ref{Aexp}), and
express it as a power series expansion
\bea
\a_n(z,z'\,)&=&\a_n(z)~\oplus~\a_n'(z'\,)\nn&=&
\sum_{i,j=1}^{q_{2n}}~\sum_{k\in\IZ}\,
\alpha^{(n)}_{i+\big[\frac{q_{2n}}2\big]_0,j;k}~z^k\,\left({\cal
C}_{q_{2n}}\right)^i\, \big({\cal
S}_{q_{2n}}(z)\big)^j\nn&&\oplus~\sum_{i',j'=1}^{q_{2n-1}}
{}~\sum_{k'\in\IZ}\,\alpha^{\prime\,(n)}_{i',j'+\big[\frac{q_{2n-1}}2
\big]_0;k'}~z^{\prime\,k'}
\,\left({\cal S}_{q_{2n-1}}(z'\,)\right)^{i'}\,\big({\cal
C}_{q_{2n-1}}\big)^{j'} \ , \label{Aexp2}
\end{eqnarray}
where $[\,\cdot\,]^{~}_0$ denotes the integer part. Notice that the
roles of the clock and generalized shift matrices are interchanged
between the two towers. The shift in the first
index of the expansion coefficients $\alpha^{(n)}$ in the first tower
effectively sets the range of the powers of the clock operators to lie
symmetrically about~$0$ in the range
$-\big[\frac{q_{2n}}2\big]_0,\dots,\big[\frac{q_{2n}}2\big]_0$. It is
made for technical reasons that will become clearer below. An
analogous argument holds for the second index of
${\alpha^{\prime\,(n)}}$ in the second tower. The remaining momentum
modes lie in the range $j=1,\dots,q_{2n}$ and
$i'=1,\dots,q_{2n-1}$. While differences between the various index range
conventions
vanish in the limit $n\to\infty$, they do affect the convergence
properties of the finite level approximations.

The expansion coefficients of~(\ref{Aexp2}) may be computed from
(\ref{GammanaA},\ref{Aexp}) to get
\bea
\alpha_{i,j;k}^{(n)}&=&\sum_{l\in\IZ}
a_{i+q_{2n}l-\big[\frac{q_{2n}}2\big]_0\,,\,j+q_{2n}k} \ , \nn
\alpha^{\prime\,(n)}_{i',j';k'}&=&\sum_{l'\in\IZ}
a_{i'+q_{2n-1}k'\,,\,j'+q_{2n-1}l'-\big[\frac{q_{2n-1}}2\big]_0}
  \ . \label{Aexp2Gamma}
\end{eqnarray}
In the first tower the coefficients of the high momentum modes of $U$ are
summed to low momentum ones. However, for Schwartz sequences this
``correction'' is small and vanishes as $q_{2n}\to\infty$. The choice
of the range of powers of the clock matrices made in~\eqn{Aexp2} was
in fact motivated by the necessity to be able to ignore these small
coefficients. In the second tower it is the high momentum modes of $V$
which are lost. The interplay between the
two towers is such that they neglect different high momentum
modes, so that in a certain sense the two errors ``compensate'' each
other. The representation~(\ref{Aexp2},\ref{Aexp2Gamma}) thereby provides a
nice heuristic insight into the role of the two towers in the
matrix regularization.

Let us now look at the projection of $\del_1
a=2\pi\ii\,\sum_{m,r}m\,a_{m,r}~U^m\,V^r$ in the two towers. By using
(\ref{Aexp2Gamma}) the corresponding expansion coefficients may be
written as
\bea
\big(\Gamma_n(\del_1
a)\big)_{i,j;k}&=&2\pi\ii\,\sum_{l\in\IZ}
\left(i+q_{2n}l-\left[\mbox{$\frac{q_{2n}}2$}\right]_0\,\right)\,a_{i+q_{2n}l-
\big[\frac{q_{2n}}2\big]_0\,,\,j+q_{2n}k}\nn&=&
2\pi\ii\,\left(i-\left[\mbox{$\frac
{q_{2n}}2$}\right]_0\,\right)\,\alpha^{(n)}_{i,j;k}+O
\left(q_{2n}\,a_{i-\big[\frac{q_{2n}}2\big]_0\,,\,q_{2n}k}\right) \ ,
  \nn&&{~~~~}^{~~}_{~~}\nn\big(\Gamma_n(\del_1
a)\big)_{i',j';k'}'&=&2\pi\ii\,\sum_{l'\in\IZ}
\left(i'+q_{2n-1}k'\,\right)\,
a_{i'+q_{2n-1}k'\,,\,j'+q_{2n-1}l'-\big[\frac{q_{2n-1}}
2\big]_0}\nn&=& 2\pi\ii\,\left(i'+q_{2n-1}k'\,\right)\,\alpha^{(n)}_{i',j';k'}
\ .
\label{derivexpcoeffs}
\end{eqnarray}
In the first set of equalities in (\ref{derivexpcoeffs}) the neglected terms in
the
second equality vanish for Schwartz sequences as $n\to\infty$,
while in the second set no approximation is necessary. The same
reasoning can be repeated for the projection of $\del_2a$, and together these
results
suggest the definitions
\bea
&&\mnabla_1\a_n(z,z'\,)~=~2\pi\ii\left[\,
\sum_{i,j=1}^{q_{2n}}~\sum_{k\in\IZ}\,
   i\,\alpha_{i+\big[\frac{q_{2n}}2\big]_0,j;k}^{(n)}~z^k\,
   \left({\cal C}_{q_{2n}}\right)^i\,
\big({\cal S}_{q_{2n}}(z)\big)^j\right.\nn&&~~~~~~
\oplus\left.~\sum_{i',j'=1}^{q_{2n-1}}~\sum_{k'\in\IZ}\,
\left(i'+q_{2n-1}k'\,\right)\,\alpha_{i',j'+\big[\frac{q_{2n-1}}2
\big]_0;k'}^{\prime\,(n)}~z^{\prime\,k'}
\,\big({\cal S}_{q_{2n-1}}(z'\,)\big)^{i'}\,
\left( {\cal C}_{q_{2n-1}}\right)^{j'}\right] ~, \nn&&{~~~~}^{~~}_{~~}\nn
&&\mnabla_2\a_n(z,z'\,)~=~2\pi\ii\left[\,
\sum_{i,j=1}^{q_{2n}}~\sum_{k\in\IZ}\,
(j+q_{2n-1}k)\,\alpha_{i+\big[\frac{q_{2n}}2\big]_0,j;k}^{(n)}~
z^k\,\left({\cal C}_{q_{2n}} \right)^i\,\big({\cal
S}_{q_{2n}}(z)\big)^j\right.\nn&&~~~~~~\oplus\left.~
\sum_{i',j'=1}^{q_{2n-1}}~\sum_{k'\in\IZ}\,j'\,
{\alpha}_{i',j'+\big[\frac{q_{2n-1}}2\big]_0;k'}^{\prime\,(n)}~
z^{\prime\,k'}\big( {\cal S}_{q_{2n-1}}(z'\,)\big)^{i'}\,
\left({\cal C}_{q_{2n-1}}\right)^{j'}\right] ~. \label{derin}
\end{eqnarray}
These two operations converge to the canonical linear derivations
on the algebra $\atheta$ and satisfy an approximate Leibniz rule
which is proven in appendix~\ref{appf}.

We now wish to express the ``derivatives'' (\ref{derin}) as operations
acting on $\a_n(z,z'\,)$ expressed as a pair of matrix-valued
functions on circles,
\be
\a_n(z,z'\,)=\sum_{k\in\IZ}~\sum_{i,j=1}^{q_{2n}}a_{ij;k}^{(n)}~
z^k\,\P_n^{ij}~\oplus~\sum_{k'\in\IZ}~
\sum_{i',j'=1}^{q_{2n-1}}a_{i'j';k'}^{\prime\,(n)}~z^{\prime\,k'}
\,\P_n^{\prime\,i'j'} \ .
\label{anmatrixbasis}\end{equation}
For this, we need to find the appropriate change of matrix basis
between the two expansions (\ref{Aexp2}) and (\ref{anmatrixbasis}). We
will do this explicitly below only for the first tower, the analogous
formul{\ae} for the second tower being the obvious modifications.

The key formula which enables this change of basis is provided by the identity
\be
\left({\cal C}_{q_{2n}}\right)^i\,\big({\cal S}_{q_{2n}}(z)\big)^j
=\sum_{s=1}^{q_{2n}-j}(\omega_n)^{i(s-1)}~\P_n^{s,s+j}+z\,
\sum_{s=q_{2n}-j+1}^{q_{2n}}
(\omega_n)^{i(s-1)}~\P_n^{s,s+j-q_{2n}} \ ,
\label{keyidentity}\end{equation}
which is readily derived from the orthonormality relation
(\ref{matrixmult}). A straightforward consequence of
(\ref{keyidentity}), the trace formula (\ref{Pijtrace}), and the identity
\be
\frac1{q_{2n}}\,\sum_{t=1}^{q_{2n}}(\omega_n)^{t(s-s'\,)}=\delta_{ss'}
{}~~~~{\rm for}~~s,s'\in\IZ_{q_{2n}}
\label{rootdelta}\end{equation}
is that the elements of the matrix basis of the
expansion (\ref{Aexp2}) are orthogonal,
\be
\Tr\left[\big({\cal S}_{q_{2n}}(z)^\dag\big)^j\,\left( {\cal
C}_{q_{2n}}^\dag\right)^i\,\right]\,\left[\left( {\cal
C}_{q_{2n}}\right)^s\,\big({\cal S}_{q_{2n}}(z)
\big)^t\,\right]=q_{2n}\,\beta_{2n}~\delta_{is}\,\delta_{jt} \ .
\label{clockshiftortho}\end{equation} From~(\ref{clockshiftortho})
it follows that the expansion coefficients of (\ref{Aexp2}) may
thereby be computed as
\be
\alpha_{i+\big[\frac{q_{2n}}2\big]_0,j;k}^{(n)}
=\frac1{q_{2n}\,\beta_{2n}}\,\oint
\frac{\dd z}{2\pi\ii z^{k+1}}~\Tr\,\a_n(z)\,\big({\cal
S}_{q_{2n}}(z)^\dag \big)^j\,\left({\cal C}_{q_{2n}}^\dag\right)^i
\ . \label{Aexp2coeffs}
\end{equation}
By substituting~(\ref{anmatrixbasis}) into~(\ref{Aexp2coeffs}),
and using~(\ref{keyidentity}) along with~(\ref{Pijtrace}), after
some algebra we arrive at the change of basis
$a_{ij;k}^{(n)}\mapsto \alpha_{i,j;k}^{(n)}$ in the form
\be
\alpha_{i+\big[\frac{q_{2n}}2\big]_0,j;k}^{(n)}=
\frac1{q_{2n}}\,\left[\,\sum_{s=1}^{q_{2n}-j}a_{s,s+j;k}^{(n)}~
(\omega_n)^{-i(s-1)}+\sum_{s=q_{2n}-j+1}^{q_{2n}}a^{(n)}_{s,s+j-q_{2n};k+1}~
(\omega_n)^{-i(s-1)}\right] \ . \label{aAchange}\end{equation}
On the other hand, from~(\ref{Pijtrace}) it follows that the
expansion coefficients of (\ref{anmatrixbasis}) may be computed
from
\be
a_{ij;k}^{(n)}=\frac1{\beta_{2n}}\,\oint\frac{\dd z} {2\pi\ii
z^{k+1}}~\Tr\,\a_n(z)\,\P_n^{ji} \ .
\label{anmatrixcoeffs}\end{equation} By substituting (\ref{Aexp2})
into (\ref{anmatrixcoeffs}), and again applying
(\ref{keyidentity}) and (\ref{Pijtrace}), we arrive at the change
of basis $\alpha_{i,j;k}^{(n)}\mapsto a_{ij;k}^{(n)}$ in the form
\be
a_{ij;k}^{(n)}~=~\sum_{s=1}^{q_{2n}}\,(\omega_n)^{s(i-1)}~\times~
\left\{\begin{matrix}\alpha^{(n)}_{s+\big[\frac{q_{2n}}2\big]_0,j-i;k}&i<j\\
\,\alpha^{(n)}_{s+\big[\frac{q_{2n}}2\big]_0,q_{2n}+j-i;k-1}&i\geq
j\end{matrix}\right. \ . \label{Aachange}
\end{equation}

Let us now deal with the derivative $\mnabla_1$ acting on the
first tower in (\ref{derin}), and substitute
$i\,\alpha_{i+\big[\frac{q_{2n}}2\big]_0,j;k}^{(n)}$ in place of
$\alpha_{i+\big[\frac{q_{2n}}2\big]_0,j;k}^{(n)}$
in~(\ref{Aachange}) to obtain the canonical matrix elements
$b_{i,j;k}^{(n)}$ of the expansion of $\mnabla_1\a_n(z,z'\,)$
analogous to (\ref{anmatrixbasis}). For $i<j$, it follows
from~(\ref{aAchange}) that they are given by
\bea
b_{i,j;k}^{(n)}&=&\frac1{q_{2n}}\,\sum_{s=1}^{q_{2n}}s\,(\omega_n)^{si}
\,\left[\,\sum_{s'=1}^{q_{2n}-j+i}a_{s',s'+j-i;k}^{(n)}~
(\omega_n)^{-ss'}\right.\nn && ~~~~~~~~~~~~~~~~~~~~ +\left.
\sum_{s'=q_{2n}+i-j+1}^{q_{2n}}a^{(n)}_{s',s'+j-i-q_{2n};k+1}~
(\omega_n)^{-ss'}\right] \ , \label{bijkn}
\end{eqnarray}
with a similar expression in the case $i\geq j$.
To understand the geometrical meaning of the expression
(\ref{bijkn}), we recall that the translation generators on the ordinary
torus $\IT^2$ are given by
\be
\e^{-x_0\,\del_1}~f(x,y)~\e^{x_0\,\del_1}=f(x+x_0,y)
\end{equation}
plus the analogous expression for the shift in $y$. The canonical
derivations on the fuzzy torus, i.e. the discrete versions of these
operators, are realized in a unitary fashion (rather than Hermitian) and are
given by clock and shift matrices as~\cite{AMNS1,AMNS2}
\bea
\left({\cal C}_{q_{2n}}^\dag\,\a_n\,{\cal
C}^{~}_{q_{2n}}\right)_{i,j;k} &=&
(\omega_n)^{j-i}~a_{ij;k}^{(n)} \ , \nn\left({\cal
S}_{q_{2n}}^\dag \,\a_n\,{\cal
S}^{~}_{q_{2n}}\right)_{ij;k}&=&a^{(n)}_{[i-1]_{q_{2n}}\,,\,
[j+1]_{q_{2n}}\,;\,k} \label{clockshiftderiv}
\end{eqnarray}
with
$[\,\cdot\,]_q^{~}$ denoting the integer part modulo $q$. Given
that the periodic delta-function on the cyclic group
$\IZ_{q_{2n}}$ is represented by the finite Fourier transform
(\ref{rootdelta}) in terms of the $q_{2n}$-th roots of unity
$\omega_n$, the sum $\sum_t\,t\,(\omega_n)^{t(s-s')}$ may be
formally identified as being proportional to a discrete derivative
of the delta-function\footnote{\baselineskip=12pt To understand
this identification better, it is instructive to recall the
Fourier integral representation of the Dirac delta-function
$\delta(x)=\int_\IR\dd k~\e^{\ii k\,x}$ on the real line $\IR$.
{}From this formula it follows that $\delta'(x)=\ii\int_\IR\dd
k~k~\e^{\ii k\,x}$ is the Fourier expansion of the derivative of
the delta-function.} on $\IZ_{q_{2n}}$.

The relation~(\ref{bijkn}) thereby identifies a finite shift operator
$\Sigma$ acting on functions $f:\IZ_{q_{2n}}\to\IC$. It may be
regarded as the ``infinitesimal'' version of the shift operation in
(\ref{clockshiftderiv}), according to the definition
\be
-\frac{2\pi\ii}{q_{2n}}\,\sum_{s,t=1}^{q_{2n}}t\,(\omega_n)^{t(s-s')}~f(s)
:=\Sigma f(s'\,) \ . \label{finitedeltaderiv}
\end{equation}
In components this operator can be written as $\Sigma
f(s'\,)=\sum_s\Sigma_{ss'}\,f(s)$ with
\be
\Sigma_{ss'}=-\frac{2\pi\ii}{q_{2n}}\,\sum_{t=1}^{q_{2n}}
t\,(\omega_n)^{t(s-s')} \ .
\label{Sigmassprime}\end{equation}
For the action on matrices we define
\be
(\SSigma\a_n)_{ij;k}:=b_{ij;k}^{(n)} \ . \label{bijknSigma}
\end{equation}
In components its action on matrix-valued functions on a circle
is given by the expression
$\SSigma\a_n(\tau)_{ij}=\sum_{s,t}\SSigma(\tau)_{ij,st}\,\a_n(\tau)_{st}$,
with
\be
\SSigma(\tau)_{ij,st}~=~\Sigma_{is}~\times~\left\{\begin{matrix}
 
\delta_{t,s+j-i}&i<j&1\leq s\leq q_{2n}+i-j\\\,
 
\delta_{t,s+j-i-q_{2n}}~\e^{2\pi\ii\tau}
     &i<j&q_{2n}+i-j+1\leq 
s\leq q_{2n}\\\delta_{t,s+i-j}&j\leq
     i&1\leq s\leq 
q_{2n}+j-i\\\delta_{t,s+i-j-q_{2n}}~
 
\e^{-2\pi\ii\tau}&j<i&q_{2n}+j-i+1\leq s\leq q_{2n}
 
\end{matrix}\right. \ .
\label{SSigmaijkl}\end{equation}
The 
skew-adjoint shift operator $\SSigma$ defines the finite
analog of 
the derivative $\partial_1$ acting on the matrix part of
the 
expansion (\ref{anmatrixbasis}).

Proceeding to the derivative 
$\mnabla_2$ acting on the first tower in
(\ref{derin}), by defining 
the ``infinitesimal'' clock operator
\be
\Xi_{ij}:=2\pi\ii 
j~\delta_{ij} \label{Xidef}
\end{equation}
we may write the canonical 
matrix expansion coefficients $c_{ij;k}^{(n)}$ 
of
$j\,\alpha_{i+\big[\frac{q_{2n}}2\big]_0,j;k}^{(n)}$ using 
(\ref{aAchange}) and
(\ref{Aachange})
as
\be
c_{ij;k}^{(n)}=(j-i)\,a_{ij;k}^{(n)}=\frac1{2\pi\ii}\,
\big[\,\Xi\,,\,\a_n\,\big]_{ij;k} \ . \label{cijkn}\end{equation}
The operator $\Xi$ defines the ``infinitesimal'' version of the
clock operation in (\ref{clockshiftderiv}) and yields the finite
analog of the derivative $\partial_2$ acting on the matrix part of
the expansion (\ref{anmatrixbasis}). In this sense, the derivative
terms in this matrix model are more akin to the derivatives
obtained by expanding functions on the noncommutative \emph{plane} in
a soliton basis~\cite{MRW,LS1,BGI,LVZ,GW,LSZ}.

Finally, the components of the
derivations in (\ref{derin}) which are proportional to the
circular Fourier integers $k$ are evidently proportional to the
derivative operators $z\,\dd/\dd z$ of $\circles^1$ acting on
$\a_n(z)$. Completely analogous formul{\ae} hold also for the
second tower in~(\ref{derin}). In this way we may represent the
derivatives (\ref{derin}) acting on matrix-valued functions
(\ref{anmatrixbasis}) on $\circles^1$ as
\bea
\mnabla_1\a_n(\tau,\tau'\,)&=&\SSigma\a_n(\tau)~\oplus~\left(q_{2n-1}~
\dot{\a}_n'(\tau'\,)+ \big[\Xi'\,,\, \a_n'(\tau'\,)\big]\right) \ , \nn
\mnabla_2\a_n(\tau,\tau'\,)&=&\left(q_{2n}~\dot{\a}_n(\tau)+
\big[\Xi\,,\,\a_n(\tau)\big]\right)~\oplus~\SSigma'\a_n'(\tau'\,)
\ , \label{nablafinite}
\end{eqnarray}
where
$\dot{\a}_n(\tau):=\dd\a_n(\tau)/\dd\tau$ and
$\dot{\a}_n'(\tau'\,):=\dd\a_n'(\tau'\,)/\dd\tau'$.

\subsection{Approximate Actions}

We can now write down an action defined on elements of $\athetan$ which
approximates well the action functional (\ref{NCstrfield}) as
\bea
S_n&=&\frac{g_s\,\mu_9}{G_s}\,\int\limits_{\cal M}\sqrt{\det G}~
\left\{\beta_{2n}\,\int\limits_0^1\dd\tau~
\Tr\,{\cal L}\big[\mPhi_n(\tau)\,,\,\mnabla_\mu\mPhi_n(\tau)\big]
\right.\nn&& ~~~~~~~~~~~~~~~~~~~~~~~+\left.\beta_{2n-1}\,\int\limits_0^1
\dd\tau'~\Tr'\,{\cal L}\big[\mPhi_n'(\tau'\,)\,,\,\mnabla_\mu
\mPhi_n'(\tau'\,)\big]\right\} \ ,
\label{NCstrfieldn}
\end{eqnarray}
with $\mnabla_\mu$, $\mu=1,2$ given by (\ref{nablafinite}). The
noncommutativity of the torus has now been transformed into matrix
noncommutativity. Note, however, that this is \emph{not} the Morita
equivalence of noncommutative field theories, which would connect a field
theory on the noncommutative torus to a matrix theory on the regular torus
$\torus^2$. Here the matrix model is defined on a \emph{sum} of two circles,
and the procedure is exact in the limit, in the sense that the algebras
$\athetan$ converge to $\atheta$ in the manner explained in the previous
section. The fact that (\ref{NCstrfieldn}) already involves continuum 
fields
is also the reason that the derivations (\ref{nablafinite}) 
are infinitesimal
versions of the usual lattice ones 
(\ref{clockshiftderiv}), and in the present
case the removal of the 
matrix regularization does not require a complicated
double scaling 
limit involving a small lattice spacing parameter. In the 
next
section we shall study some explicit examples of this 
approximation to field
theories on the noncommutative torus and, in 
particular, describe some
aspects of their quantization in the matrix 
representation.

\section{Applications 
\label{se:applications}}
\setcounter{equation}{0}

In this section we 
will briefly describe three simple applications
of the matrix quantum 
mechanics formalism. First, we shall analyse
how the perturbative 
expansion of noncommutative field theories is
described by the matrix 
model. We show that at any finite level
$n$, there is no UV/IR mixing 
present in quantum amplitudes, but
that the standard divergences are 
recovered in the limit
$n\to\infty$. This suggests that the present 
matrix formalism
could be a good arena to explore the renormalization 
of
noncommutative field theories. Second, we examine a simple 
model
for the energy density in string field theory. We show that 
the
correct tension of a D-brane in processes involving 
tachyon
condensation is already reproduced at a finite level in the 
matrix
model. This feature fits well with the recent proposals on 
the
description of tachyon dynamics in open string field theory, 
using
one-dimensional matrix models for strings in 
two-dimensional
target 
spaces~\cite{McGV1,Martinec1,KMS1,McGTV1,AKK1}. Finally, we
briefly 
initiate a nonperturbative analysis of gauge dynamics on
the 
noncommutative torus by exploiting a Hamiltonian formulation
of the 
matrix quantum mechanics, and indicate how the results
compare with 
the known exact solution of noncommutative Yang-Mills
theory in two 
dimensions~\cite{PS1}. More complicated exactly
solvable models are 
also readily analysed in principle, in
particular by exploiting the 
fact that the ``time'' direction of
the matrix quantum mechanics is 
compactified on a circle so that
the regulated theory is really a 
finite temperature field theory.
For example, if one considers a 
$2+1$~dimensional field theory
with space taken to be the 
noncommutative torus, then our
regularization technique provides a 
dimensional reduction of the
model to a $1+1$~dimensional matrix 
field theory with spacetime a
cylinder 
$\IR\times\IS^1$.

\subsection{Perturbative Dynamics}

In this 
subsection we will demonstrate that perturbation theory within
the 
framework of the matrix quantum mechanics is easily tractable, 
in
contrast to some other matrix regularizations of noncommutative 
field
theory, and show how various novel perturbative aspects arise 
within
the matrix approximation scheme. For definiteness, we will concentrate
on the real scalar $\phi^4$-theory which on the
noncommutative torus is defined by the action
\be
S[\phi]=\ncint\left[\frac12\,\phi\,\left(\Box+\mu^2\right)\phi+
    \frac g{4!}\,\phi^4\right] \ ,
\label{phi4actioncont}\end{equation}
where $\phi$ is a Hermitian element of the algebra $\atheta$ and
$\Box=(\partial_1)^2+(\partial_2)^2$ is the Laplacian, while $\mu$ and
$g$ are respectively dimensionless mass and coupling
parameters. Following the general prescription of the previous section, we
approximate this field theory by the Hermitian matrix quantum mechanics with
Euclidean action
\bea
&&S_n[\mphi_n,\mphi_n']~=~\beta_{2n}\,\int\limits_0^1\dd\tau~
\Tr\left[\frac12\,
    \mphi_n(\tau)\,\big((\mnabla_1)^2+(\mnabla_2)^2+\mu^2\big)
    \mphi_n(\tau)+\frac g{4!}
    \,\mphi_n(\tau)^4\right]\nn&&~~~~~+\,
\beta_{2n-1}\,\int\limits_0^1\dd\tau'~
    \Tr'\left[\frac12\,\mphi_n'(\tau'\,)\,\big((\mnabla_1)^2+(\mnabla_2)^2+
    \mu^2\big)
    \mphi'_n(\tau'\,)+\frac g{4!}\,\mphi_n'(\tau'\,)^4\right]
\ . \nn&&
\label{phi4actionmatrix}
\end{eqnarray}
Everything we say in this subsection will hold independently and
symmetrically in both towers of the finite level algebra $\athetan$,
and so for brevity we will only analyse the first tower explicitly.

To deal with this quantum mechanics in perturbation theory, it is most
convenient to use a power series expansion of the form (\ref{Aexp2})
and expand the Hermitian scalar fields $\mphi_n\in u(q_{2n})\otimes
C^\infty(\IS^1)$ in the first tower as
\be
\mphi_n(z)=\sum_{i,j=0}^{q_{2n}-1}~\sum_{k\in\IZ}\,
\varphi_{ij;k}^{(n)}~z^k\,\left({\cal C}_{q_{2n}}\right)^i\,
\big({\cal S}_{q_{2n}}(z)\big)^j \ .
\label{mphinzexp}\end{equation}
For the quantum theory, we will use path integral quantization, defined
by treating the complex expansion coefficients $\varphi_{ij;k}^{(n)}$
as the dynamical variables and integrating over the corresponding
configuration space ${\cal C}_n:=u(q_{2n})\otimes S(\IZ)$. Quantum
correlation functions are then defined as
\be
\left\langle\varphi_{i_1j_1;k_1}^{(n)}\cdots\varphi_{i_Lj_L;k_L}^{(n)}
    \right\rangle~:=~\frac{~\int\limits_{{\cal C}_n}
    \DD\varphi^{(n)}~\e^{-S_n[\varphi^{(n)}]}\,
    \varphi_{i_1j_1;k_1}^{(n)}\cdots\varphi_{i_Lj_L;k_L}^{(n)}}
{\int\limits_{{\cal C}_n}\DD\varphi^{(n)}~\e^{-S_n[\varphi^{(n)}]}} \ 
,
\label{matrixcorrfns}\end{equation}
where the integration measure 
is given 
by
\be
\DD\varphi^{(n)}:=\prod_{i,j=0}^{q_{2n}-1}~\prod_{k\in\IZ}\,
\dd\varphi_{ij;k}^{(n)}
\label{pathintmeas}\end{equation}
with 
$\dd\varphi_{ij;k}^{(n)}$ ordinary Lebesgue measure on $\IC$. Note 
that
the expression (\ref{pathintmeas}) is still formal because of 
the
infinitely many Fourier modes on $\circles^1$. Nevertheless,
as 
we show in the following, the finiteness of the range of the 
matrix
indices in $\IZ_{q_{2n}}$ is sufficient to
regularize the 
original noncommutative field theory.

We now substitute the 
expansion (\ref{mphinzexp}) into the action
(\ref{phi4actionmatrix}), 
compute derivatives using the definitions
(\ref{derin}) and the 
$(\mphi_n)^4$ interaction term using the
commutation relations 
(\ref{UnVnomegan}), and then apply the
orthogonality relations 
(\ref{clockshiftortho}). The quadratic form in
the free part of the 
action (\ref{phi4actionmatrix}) is then diagonal,
and from its 
inverse we arrive at the free 
propagator
\be
\triangle^{(n)\,kl}_{ij;st}:=\left\langle\varphi_{ij;k}^{(n)\,\dag}\,
\varphi_{st;l}^{(n)}
\right\rangle_{g=0}=\frac1{(2\pi)^2\,q_{2n}\,(\beta_{2n})^2}\,
\frac1{i^2+(j+q_{2n}\,k)^2+\mu^2}~\delta_{is}~\delta_{jt}~\delta_{kl} 
\ .
\label{freepropmatrix}\end{equation}
Furthermore, the vertices 
for the $\phi^4$ field theory in the matrix
representation are given 
by
\be
S_n\left[\varphi^{(n)}\right]_{\rm 
int}=\prod_{a=1}^4~
\sum_{i_a,j_a=0}^{q_{2n}-1}~\sum_{k_a\in\IZ}\,
\varphi_{i_1j_1;k_1}^{(n)\,\dag}\,\varphi_{i_2j_2;k_2}^{(n)\,\dag}\,
\varphi_{i_3j_3;k_3}^{(n)}\,\varphi_{i_4j_4;k_4}^{(n)}~
V_{i_1j_1;i_2j_2;i_3j_3;i_4j_4}^{(n)\,k_1,k_2,k_3,k_4} 
\ 
,
\label{phi4verts}\end{equation}
where
\be
V_{i_1j_1;i_2j_2;i_3j_3;i_4j_4}^{(n)\,k_1,k_2,k_3,k_4}=\frac 
g{4!}\,
q_{2n}\,(\beta_{2n})^2\,(\omega_n)^{i_3j_2-i_1j_4}~
\delta_{i_1+i_2,i_3+i_4}~\delta_{j_1+j_2,j_3+j_4}~
\delta_{k_1+k_2,k_3+k_4} 
\ .
\label{vertexfn}\end{equation}
The vertex function 
(\ref{vertexfn}) is invariant under cyclic
permutations of its 
arguments $(i_a,j_a,k_a)$, $a=1,\dots,4$.

The propagator 
(\ref{freepropmatrix}) in this representation
is rather simple in 
form, in contrast to the usual matrix regularizations 
of
noncommutative field theory wherein the kinetic terms of the 
action
generically have a very complicated form~\cite{AMNS2,MRW,LSZ}. 
This is what
makes the
matrix quantum mechanics approach much more 
fruitful. The matrix
quantum mechanics provides an exact,
finite 
regulated theory which precisely mimicks the properties of 
the
original continuum model. This is a physical manifestation of the 
fact
that the finite level algebras $\athetan$ converge to $\atheta$. 
In
particular, from (\ref{freepropmatrix}) and (\ref{vertexfn}) we 
see
that the Feynman graphs have a natural ribbon structure with 
an
additional label by the circular momentum modes of the fields, and 
the
notion of planarity in the matrix model is the same as that in 
the
noncommutative field theory. Again, all of these features are 
in
contradistinction to the usual matrix model formulations.

As an 
explicit example of how perturbation theory works within 
this
setting, let us compute the quadratic part of the
effective 
quantum action in the matrix representation. The one-loop dynamics
is 
obtained by contracting two legs in (\ref{vertexfn}) using 
the
propagator (\ref{freepropmatrix}). All eight possible 
contractions of
two neighbouring legs are identical and sum up to 
give the total
planar 
contribution
\be
8\,\sum_{i',j'=0}^{q_{2n}-1}~\sum_{s',t'=0}^{q_{2n}-1}~
\sum_{k',l'\in\IZ}\,\triangle_{i'j';s't'}^{(n)\,k'l'}~
V_{ij;i'j';s't';st}^{(n)\,k,k',l',l}~=~\frac 
g3~
\delta_{is}~\delta_{jt}~\delta_{kl}~I_{\rm 
p}^{(n)}\left(\mu^2\right) \ 
,
\label{planarcontr}\end{equation}
where
\be
I^{(n)}_{\rm 
p}\left(\mu^2\right)=\frac1{(2\pi)^2}\,
\sum_{i'=0}^{q_{2n}-1}~\sum_{r\in\IZ}\,
\frac1{i^{\prime\,2}+r^2+\mu^2} 
\ .
\label{Inplanar}\end{equation}
In (\ref{Inplanar}) we have 
transformed the sums over
$j'\in\IZ_{q_{2n}}$ and $k'\in\IZ$ into a 
single sum over
$r:=j'+q_{2n}\,k'\in\IZ$. The infinite sum in 
(\ref{Inplanar}) can be
evaluated explicitly to 
give
\beq
I^{(n)}_{\rm
p}\left(\mu^2\right)=\frac1{4\pi}\,\sum_{i'=0}^{q_{2n}-1}\,\frac{\coth
\left(\pi\,\sqrt{i^{\prime\,2}+\mu^2}\,\right)}{\sqrt{i^{
\prime\,2}+\mu^2}} \ .
\label{Inplanarexpl}\end{equation}

For any 
finite $q_{2n}$ the function (\ref{Inplanarexpl}) is finite,
for all 
$\mu^2$, and thus the matrix quantum mechanics
naturally regulates 
the ultraviolet divergence of the one-loop scalar
tadpole diagram. In 
the limit $n\to\infty$, whereby $q_{2n}\to\infty$, the 
sum
(\ref{Inplanarexpl}) diverges, and the leading divergent 
behaviour can
be straightforwardly worked out to be given 
by
\beq
\lim_{n\to\infty}\,I_{\rm 
p}^{(n)}\left(\mu^2\right)\simeq
\frac1{4\pi}\,\ln(q_{2n}) \ 
,
\label{Inplanarlim}\end{equation}
reproducing the standard 
logarithmic ultraviolet divergence of scalar
field theory in two 
dimensions. In particular, we see that the matrix
rank $q_{2n}$ plays 
the role of an ultraviolet regulator in the matrix
quantum mechanics. 
This is the characteristic feature of a fuzzy
approximation to a 
field theory. In (\ref{Inplanarexpl}), the limit
$q_{2n}\to\infty$ 
requires an infinite mass renormalization
$\mu=q_{2n}\,\tilde\mu$, 
keeping $\tilde\mu$ fixed, in order to obtain a
massive scalar field 
theory in the limit.

There are also four possible contractions of 
opposite legs in
(\ref{vertexfn}), which all agree and sum to give 
the total
non-planar 
contribution
\be
4\,\sum_{i',j'=0}^{q_{2n}-1}~\sum_{s',t'=0}^{q_{2n}-1}~
\sum_{k',l'\in\IZ}\,\triangle_{i'j';s't'}^{(n)\,k'l'}~
V_{ij;i'j';st;s't'}^{(n)\,k,k',l,l'}~=~\frac 
g6~
\delta_{is}~\delta_{jt}~\delta_{kl}~I_{\rm 
np}^{(n)}\left(\mu^2\right)_{ij} 
\
,
\label{nonplanarcontr}\end{equation}
where
\be
I^{(n)}_{\rm
np}\left(\mu^2\right)_{ij}=\frac1{(2\pi)^2}\,\sum_{i'=0}^{q_{2n}-1}~
\sum_{r\in\IZ}\,
\frac{(\omega_n)^{i'j-ir}}{i^{\prime\,2}+r^2+\mu^2}
\label{Innonplanar}\end{equation}
and 
we have used the fact that $(\omega_n)^{q_{2n}}=1$. The infinite sum 
in
(\ref{Innonplanar}) can be evaluated explicitly in
terms of the 
generalized hypergeometric function
\be
{}^{~}_3F^{~}_2\left.\left\{\begin{matrix}\lambda_1~~\lambda_2~~\lambda_3\cr
{}~~~\eta_1~~\eta_2~~~ \end{matrix}\right|w\right\}:=
\sum_{p\in\IN_0}\frac{(\lambda_1)_p\,(\lambda_2)_p\,(\lambda_3)_p}
{(\eta_1)_p\,(\eta_2)_p}~\frac{w^p}{p!}
\label{3F2def}\end{equation}
with $w,\lambda_a,\eta_b\in\IC$, $a=1,2,3$, $b=1,2$, and
$(\lambda)_p:=\lambda(\lambda+1)\cdots(\lambda+p)$. One finds
\bea
&&I^{(n)}_{\rm
    np}\left(\mu^2\right)_{ij}~=~\frac1{(2\pi)^2}\,
\frac1{\mu^2}~{}^{~}_3F^{~}_2\left.\left\{
    \begin{matrix}1~~\ii\mu~~-\ii\mu\cr
    1+\ii\mu~~1-\ii\mu\,\end{matrix}\right|(\omega_n)^j\right\}\nn&&~~~~~-\,
\frac1{(2\pi)^2}\,\frac1{(q_{2n})^2+\mu^2}~{}^{~}_3F^{~}_2\left.\left\{
\begin{matrix}1~~q_{2n}+\ii\mu~~q_{2n}-\ii\mu\cr
    1+q_{2n}+\ii\mu~~1+q_{2n}-\ii\mu\,\end{matrix}\right|(\omega_n)^j
\right\}\nn&&~~~~~+\,\frac1{2\pi^2}\,
\sum_{i'=0}^{q_{2n}-1}\,\frac{(\omega_n)^{2j-i}}{1+i^{\prime\,2}+\mu^2}
    ~{}^{~}_3F^{~}_2\left.\left\{\begin{matrix}
    1~~1+\ii\sqrt{i^{\prime\,2}+\mu^2}~~1-\ii
    \sqrt{i^{\prime\,2}+\mu^2}\,\cr2+\ii\sqrt{i^{\prime\,2}+\mu^2}~~2-\ii
    \sqrt{i^{\prime\,2}+\mu^2}\end{matrix}\right|(\omega_n)^{-i}\right\}
\ .\nn&&
\label{Innonplanarexpl}
\end{eqnarray}

One can show from (\ref{3F2def}) that the leading large $n$ behaviour
of (\ref{Innonplanarexpl}) is given by
\be
\lim_{n\to\infty}\,I^{(n)}_{\rm
    np}\left(\mu^2\right)_{ij}\simeq\frac1{(2\pi)^2\,(q_{2n})^2}\,\left(
    \frac{2\,(\omega_n)^{2j}}{1-(\omega_n)^i}-\frac1{1-(
    \omega_n)^j}\right)
\label{Innonplanarlim}\end{equation}
with $\omega_n\to\e^{2\pi\ii\theta}$ in the limit. This quantity
thereby vanishes, {\it except} when either $i=0$, $j=0$ or $\theta=0$
in which case it diverges. This is simply the UV/IR mixing property of
the noncommutative field theory~\cite{MVRS1}. Integrating out
infinitely many degrees of freedom in the non-planar loop diagram
generates an infrared singularity, making the amplitude singular at vanishing
external
momentum and giving it a pole in the noncommutativity parameter
at $\theta=0$.

On the other hand, at any finite level $n<\infty$, the non-planar
matrix contribution is {\it always} finite. Generically, the
generalized hypergeometric function (\ref{3F2def}) has a branch
point at $w=1$, and for ${\rm
Re}(\lambda_1+\lambda_2+\lambda_3-\eta_1-\eta_2)<0$ the series 
is
absolutely convergent everywhere on the unit disc $|w|\leq1$. 
Thus
for finite matrix rank $q_{2n}$, the 
function
(\ref{Innonplanarexpl}) is an analytic function of 
the
noncommutativity parameter, even for vanishing external 
momenta
$i$ or $j$, i.e. there is {\it no} UV/IR mixing in the 
matrix
regulated theory, at least at one-loop order. Of course, 
there
must be a transition regime in $q_{2n}$ wherein a 
non-analyticity
develops, as it appears in the limit 
(\ref{Innonplanarlim}). But
integrating out all degrees of freedom in 
the loop does not
generate an infrared singularity in the regulated 
model.

The absence of UV/IR mixing at finite level $n$, along with 
the
simplicity of the propagator (\ref{freepropmatrix}), implies that 
the
matrix quantum mechanics is a good arena to explore 
the
renormalizability of noncommutative field theories, as the usual 
mixing of
high and low momentum scales would typically appear to make 
standard
Wilsonian renormalization to all orders of perturbation 
theory
hopeless. In particular, it confirms the expectations that an 
appropriate
non-perturbative regularization could wash away these 
effects. On the
other hand, it also seems to suggest that exotic 
non-perturbative
phenomena, such as the existence of vacuum phases 
with broken
translational symmetry~\cite{GS1}, are unobservable at 
finite level. Indeed,
for
$n<\infty$ and generic $\mu^2<0$, there 
does not appear to be any
qualitative difference in the infrared 
behaviour of the noncommutative
propagator from the case $\mu^2>0$. 
It is most likely that there is
again some transition regime for 
$n\gg1$ wherein the exotic broken
symmetry phases dominate the vacuum 
structure of the theory, and it
would be interesting to find an 
analytic approach to detect these
phases in the matrix quantum 
mechanics. Heuristically, their existence
can be deduced by looking 
at the soliton expansions (\ref{anmatrixbasis}) of
$\mphi_n(z)$ in 
the finite level algebra $\athetan$ directly in terms
of the 
projections and partial isometries of the noncommutative
torus. 
Recall that these solitons displayed themselves 
momentum
non-conserving stripe patterns (see 
figs.~\ref{zawinul3}
and~\ref{zawinul4}). A striped phase in the 
scalar field theory would
then occur when, for $n$ sufficiently 
large, the mode numbers of the vacuum
expectation value 
$\langle\varphi^{(n)}_{ij;k}\rangle$ freeze about a
particular value 
corresponding to a single projection or partial
isometry $\P_n^{ij}$, 
and thereby yielding the characteristic stripe
patterns. From this 
argument it is tempting to speculate that they may
be due to a 
Kosterlitz-Thouless type phase transition in the matrix
quantum 
mechanics which occurs in the large $n$ limit, whereby a
condensation 
of vortices in the vacuum is responsible for the breaking
of the 
translational symmetry.

Let us further remark that UV/IR mixing is 
also absent in the matrix
model regularizations of noncommutative 
field theory that
are derived by soliton expansion on the 
noncommutative plane~\cite{LS1,LSZ} (to
be
discussed in 
section~\ref{GMSExp}), but {\it not} in those which are 
derived
through lattice regularization. In these latter cases, UV/IR 
mixing is
already present non-perturbatively as a generic kinematical 
property of the
lattice regularization of noncommutative field 
theory~\cite{AMNS2}. Indeed, in
the
reduced models, striped phases of 
the theory are observable for
relatively small values of the matrix 
dimension~\cite{AC1,BHN1}. The relation between rational
noncommutative theories and matrix-valued commutative theories on the
torus is applied to UV/IR mixing in~\cite{GHLL1}.

\subsection{Tachyon Dynamics}

We will 
now examine how the matrix quantum mechanics can be used
to describe 
D-branes as the decay products in tachyon condensation
on unstable D-branes in string field theory. We begin with the
Type~IIA case (equivalently bosonic strings) described in
section~\ref{DBraneNCSoliton}. We are interested in the
noncommutative field theory of the open string tachyon and gauge
field on a system of unstable D9-branes. This depends on the
specification of a projective module over $\atheta$ (see
section~\ref{SolDyn}) in order to define the anti-Hermitian
connection gauge field $A_\mu$, but for simplicity we consider
here only the free module $(\atheta)^{\oplus N}$ provided by $N$
copies of the noncommutative torus algebra itself, which
corresponds to a topologically trivial connection on the
worldvolume field theory of $N$ noncommutative D9-branes. The
components of the curvature of the gauge connection are denoted
$F_{\mu\nu}$. The tachyon field is Hermitian and lives in the
adjoint representation of the gauge group, and its covariant
derivatives are denoted $D_\mu T$.

The action is given explicitly by~\cite{HKLM1,HKL1}
\bea
S^{\rm IIA}&=&\frac{g_s\,\mu_9}{G_s}\,\int\limits_{\cal M}\sqrt{\det G}~
\ncint\left[\frac12\,f\left(T^2-\id\right)\,G^{\mu\nu}\,D_\mu T\,D_\nu
    T-V\left(T^2-\id\right)\right.\nn&&
{}~~~~~~~~~~~~~~~~~~~~~~~~~~~~~~-\left.\frac14\,h\left(T^2-\id\right)\,
    F_{\mu\nu}F^{\mu\nu}+\dots\right] \ ,
\label{SIIAdef}
\end{eqnarray}
where here and in the following repeated indices are always understood
to be summed over, and indices are raised by the inverse open string
metric $G^{\mu\nu}$. The dots in (\ref{SIIAdef}) denote possible
higher-derivative contributions to the effective action, but will not be
required in the ensuing low-energy analysis. The functions $f$ and
$h$, and the tachyon potential $V$, are not known explicitly, but they
are constrained to satisfy certain conditions in accord with the
conjectures surrounding open string tachyon condensation~\cite{Sen1}. In
particular, the tachyon potential has a local maximum at $T=0$
representing the unstable D9-branes, with $\mu_9=V(-1)$ giving their
tension. It also has local minima $V(0)=0$ at $T=\pm\,\id$
corresponding to the closed string vacuum. The functions $f$ and $h$
vanish in the closed string vacuum, $f(0)=h(0)=0$, while
$f(-1)=h(-1)=1$.

We will consider a simplified version of this model to make
the results as transparent as possible. We study a
$2+1$-dimensional noncommutative field theory, i.e. take ${\cal
    M}=\IR^1$, and consider the action near $T=0$. The corresponding
energy functional
\be
E^{\rm IIA}[A,T]=\ncint\left[\frac12\,D_\mu T\,D_\mu T+
    V\left(T^2-\id\right)+\frac12\,F^2\right]
\label{EIIAAT}\end{equation}
is then that of a D2-brane wrapped around $\torus^2$ in the presence
of a constant $B$-field, with $D_\mu T=\partial_\mu T-[A_\mu,T]$,
$\mu=1,2$ and
\be
F=\partial_1A_2-\partial_2A_1+[A_1,A_2] \ .
\label{Fdef}\end{equation}
The simplest classical extrema of (\ref{EIIAAT}) are given by spatially uniform
tachyon fields $\Omega^{-1}(T)$ on $\torus^2$ ($\Omega^{-1}$ being the
Wigner map \eqn{wigmap}) which are
critical points of the potential $V(T^2-1)$, and vanishing gauge fields
$F=A_\mu=0$, $\mu=1,2$. We will now proceed to analyse these vacua
within the matrix approximation. In the Hermitian matrix quantum mechanics, we
replace the energy functional (\ref{EIIAAT}) by the Euclidean action
\bea
E_n^{\rm IIA}[\T_n,\T_n']&=&\beta_{2n}\,\int\limits_0^1\dd\tau~
\Tr\left[\frac12\,\mnabla_\mu\T_n(\tau)\,\mnabla_\mu\T_n(\tau)+V
    \big(\T_n(\tau)^2-\id_{q_{2n}}\big)\right]  \nonumber \\ &&+\,\beta_{2n-1}\,
\int\limits_0^1\dd\tau'~\Tr'\left[\frac12\,\mnabla_\mu\T_n'(\tau'\,)
    \,\mnabla_\mu\T'_n(\tau'\,)+V
    \big(\T'_n(\tau'\,)^2-\id_{q_{2n-1}}\big)\right] \ . \nonumber\\&&
\label{EnIIATn}
\end{eqnarray}

Focusing for the time being on the first tower, we shall seek
time-independent extrema of the energy functional (\ref{EnIIATn}),
$\dot\T_n=0$. From (\ref{Xidef}) and (\ref{nablafinite}) it then
follows that the $q_{2n}(q_{2n}+1)/2$ equations for the critical
points are given by
\be
-(2\pi)^2\,(i-j)^2\,(\T_n)_{ij}-
(\SSigma\T_n\SSigma)_{ij}=\Big(\T_n\,V'\big((\T_n)^2-
\id_{q_{2n}}\big)\Big)_{ij} \ , ~~ 1\leq i\leq j\leq q_{2n} \ .
\label{critptsTn}\end{equation}
We have used the fact both $\mnabla_1$ and $\mnabla_2$ satisfy an
``integration by parts'' rule
\beq
\Tr\big(\a_n^\dag\,(\mnabla_\mu\b_n)\big)=-\Tr\big((\mnabla_\mu
\a_n)^\dag\,\b_n\big) \ , ~~ \mu=1,2 \ .
\label{mnablaparts}\end{equation}
It is straightforward to see from these equations that the off-diagonal
elements of the matrix $\T_n$ vanish. This follows from the
explicit form (\ref{SSigmaijkl}) of the shift operator $\SSigma$,
which for $i<j$ would produce a $\tau$-dependence in the
second term of (\ref{critptsTn}), while the other two terms are
time-independent. Thus $\T_n$ must be a diagonal matrix, which we
write explicitly in terms of projections on $\atheta$ as
\be
\T_n=\sum_{i=1}^{q_{2n}}\eta_i~\P_n^{ii} 
\ .
\label{Tndiag}\end{equation} The moduli $\eta_i\in\IR$ of 
this
solution have constraints which may be found by 
substituting~(\ref{Tndiag})
into~(\ref{critptsTn}), 
using~(\ref{SSigmaijkl}) along with the fact
that 
$\SSigma_{ij,ss}\neq 0$ only if $i=j$, and by 
using
$\Tr\SSigma=\pi\ii(q_{2n}+1)$. We can thus write the equation 
for the
$\eta_i$'s 
as
\be
\eta_i\,V'\big((\eta_i)^2-1\big)=\pi^2\,(q_{2n}+1)^2\,\sum_{j=1}^{q_{2n}}
\eta_j
\label{etaieqns}\end{equation}
which 
must be satisfied for each $i=1,\dots,q_{2n}$.

This result is fairly 
generic. It states that time-independent field
configurations on the 
noncommutative torus correspond to diagonal
matrices in the matrix 
quantum mechanics. In particular, all classical
ground states commute 
with each other. This is reminescent of what
happens in the BFSS 
matrix model of M-theory~\cite{BFSS1}, whereby the
vacuum corresponds 
to static, commuting spacetime matrix
coordinates for D0-branes. 
Moreover, this solution shows that 
the
time-independent
configurations of the matrix quantum mechanics 
are naturally
projection-type solitons on the noncommutative torus. 
We will see in
the next section how projections on $\atheta$ also 
arise by a
somewhat different dynamical mechanism.

A class of 
solutions to the equations (\ref{etaieqns}) can be
constructed by 
demanding that $\T_n$ be a critical point of the
tachyon potential 
$V(T^2-\id)$,
i.e. $\T_n\,V'((\T_n)^2-\id_{q_{2n}})=0$. For this,
we 
assume that $n$ is sufficiently large, that the matrix 
dimension
$q_{2n}$ is even, and that $V(\lambda)$ is a 
polynomial
potential. Let $\{\lambda_I\}_{I\geq1}$ be a set of 
distinct, real critical
points of $V(\lambda)$ which are each bounded 
from below as
$\lambda_I\geq-1$. We then arrange the collection of 
real numbers
$\{\eta_{j}\}_{j=1}^{q_{2n}}$ pairwise according to the 
rule
\be
\eta_I=\sqrt{1+\lambda_I} \ , ~~ 
\eta_{q_{2n}-I+1}=-\sqrt{1+
   \lambda_I} \ , ~~ 
I=1,2,\dots
\label{etaIpairs}\end{equation}
with the remaining 
$\eta_i$'s set equal to $\pm\,1$ in pairs. The
solution 
(\ref{Tndiag}) then 
obeys
\be
\left(\T_n\right)^2-\id_{q_{2n}}=\sum_{I\geq1}\lambda_I\,
\left(\P_n^{II}+\P_n^{q_{2n}-I+1,q_{2n}-I+1}\right)
\label{Tnid}\end{equation}
and 
thereby satisfies the required extremization condition. In this
case, 
both sides of (\ref{etaieqns}) vanish.

The energy of these classical 
solutions in the matrix quantum
mechanics may now be found by 
substituting (\ref{Tnid}) into
(\ref{EnIIATn}). An identical analysis 
proceeds in the second tower,
producing a solution parametrized by 
another set
$\{\lambda_{I'}'\}_{I'\geq1}$ of critical points. After 
recalling the
definition of the sequence $\beta_k$ from 
appendix~\ref{appa}, one
finds
\be
E_n^{\rm 
IIA}\{\lambda_I,\lambda_{I'}'\}=2\,(p_{2n-1}-q_{2n-1}\,\theta)\,
\sum_IV(\lambda_I)-2\,(p_{2n}-q_{2n}\,\theta)\,\sum_{I'}V(\lambda'_{I'}) 
\ .
\label{EnIIAlambdaI}\end{equation}
The lightest excitation 
corresponds to the configuration whereby all
$\lambda_I$'s and 
$\lambda_{I'}'$'s vanish except $\lambda_1=-1$. Then this
formula 
gives the standard contribution to the mass-shell relation
from the tension $\mu_2=V(-1)$ of the D2-branes, as can be found from
the appropriate Born-Infeld action for the D-brane dynamics. What is
remarkable about this term is that it has the correct
$\theta$-dependence already at a {\it finite} level in the matrix
model. Since the induced mass of a D0-brane bound to the D2-branes due
to the background $B$-field is given by $\mu_0=\theta\,\mu_2$, the
term $(p_{2n-1}-q_{2n-1}\,\theta)\,V(-1)$ arising in this way from
(\ref{EnIIAlambdaI}) represents the energy of $p_{2n-1}$
D2-branes carrying $-q_{2n-1}$ units of D0-brane monopole charge. Thus at any
finite level $n$, it gives the appropriate mass-shell relation on the
noncommutative torus with energy bounded between $0$ and
$\mu_2$.

Let us now turn to the Type~IIB case. Following the prescription of
section~2.2, the appropriate version of the string field theory action
(\ref{SIIAdef}) can be written down~\cite{HKL1}. By
using the same steps as above, the analog of the regulated energy
functional (\ref{EnIIATn}) in the first tower reads
\bea
E_n^{\rm IIB}[\T_n]&=&\beta_{2n}\,\int\limits_0^1\dd\tau~\Tr\biggl[
\big(\mnabla_\mu\T_n(\tau)\big)^\dag\,\big(\mnabla_\mu\T_n(\tau)\big)
\biggr.\nn&&~~~~~~+\biggl.U\big(\T_n(\tau)^\dag\,
\T_n(\tau)-\id_{r_{2n}}\big)+U\big(\T_n(\tau)\,\T_n(\tau)^\dag-
\id_{q_{2n}}\big)\biggr] \ ,
\label{EnIIBTn}
\end{eqnarray}
where now the regulated tachyon field $\T_n(\tau)$ is a $q_{2n}\times
r_{2n}$ complex-valued matrix, with $\T_n(\tau)^\dag$ its Hermitian
conjugate. This functional describes an approximation to the
noncommutative field theory of a D2 brane-antibrane system wrapping
$\IT^2$. We will take the arbitrary integers $r_{2n}\leq q_{2n}$ for
definiteness, with $r_{2n}\to\infty$ in the limit $n\to\infty$.
Varying (\ref{EnIIBTn}) on time-independent configurations
$\dot\T_n=0$ yields the critical point equations
\be
-(2\pi)^2\,(i-j)^2\,(\T_n)_{ij}+(\SSigma^2\T_n)_{ij}=\big(\T_n\,U'
(\T_n^\dag\,\T_n-\id_{r_{2n}})+U'(\T_n\,\T_n^\dag-\id_{q_{2n}})\,\T_n
\big)_{ij}
\label{critptsTnB}\end{equation}
with $1\leq i\leq q_{2n}$ and $1\leq j\leq r_{2n}$, plus the analogous
equations for the conjugate matrix elements $(\T_n^\dag)_{ij}$.

As before, it is straightforward to show from (\ref{critptsTnB}) that all
$i\neq j$ matrix elements of $\T_n$ vanish. We can thereby write down
solutions as the $q_{2n}\times r_{2n}$ complex matrices
\be
\T_n=\begin{pmatrix}\sum\limits_{i=1}^{r_{2n}}\sigma_i~\P_n^{ii}
\\\,(0)_{(q_{2n}-r_{2n})\times r_{2n}}\,\end{pmatrix}
\label{Tnl2n}\end{equation}
of generic rank $r_{2n}$ 
with
\be
\T_n^\dag\,\T_n=\sum_{i=1}^{r_{2n}}|\sigma_i|^2~\P_n^{ii} \ 
, 
~~
\T_n\,\T_n^\dag=\sum_{i=1}^{r_{2n}}|\sigma_i|^2~\P_n^{ii}~\oplus~
(0)_{q_{2n}\times 
q_{2n}} \ ,
\label{dimkerTs}\end{equation}
where the moduli 
$\sigma_i\in\IC$ satisfy an equation completely
analogous to 
(\ref{etaieqns}). Generically, these solutions are
evidently 
determined by finite-dimensional partial isometries 
on
$\IC^{\,r_{2n}}\to\IC^{\,q_{2n}}$. By taking $|\sigma_i|=\eta_i$, 
with
$\eta_i$ as in (\ref{etaIpairs}), and 
substituting
(\ref{Tnl2n},\ref{dimkerTs}) into (\ref{EnIIBTn}), we 
find that the
energy of this solution is given by
\be
E_n^{\rm 
IIB}[\T_n]=\beta_{2n}\,\left(4\,\sum_IU(\lambda_I)
+(q_{2n}-r_{2n})\,U(-1)\right) 
\ .
\label{EnIIBTnl2n}\end{equation}

When $r_{2n}=q_{2n}$, the 
energy (\ref{EnIIBTnl2n}) is precisely
{\it twice} that of the 
Type~IIA case (\ref{EnIIAlambdaI}). By
adjusting parameters as 
before, this is the energy appropriate to
$q_{2n-1}$ D0-branes and 
$q_{2n-1}$ D0-antibranes inside the original
D2-$\overline{\rm D2}$ 
system. For $r_{2n}<q_{2n}$, the second
term of (\ref{EnIIBTnl2n}) 
dominates in the limit $n\to\infty$. From
appendix~\ref{appa}, 
eqs.~(\ref{convbound}) and~(\ref{defbetaeven}),
we have 
$\beta_{2n}\simeq1/q_{2n-1}$ in the large $n$ limit, so
that by 
taking $r_{2n}\simeq q_{2n}$ in this limit, we may adjust the
second 
term so that it yields the appropriate continuum mass-shell
relation 
for the D0 brane-antibrane system in the D2-$\overline{\rm
  D2}$ 
system. Furthermore, from (\ref{dimkerTs}) it follows that, 
for
generic moduli $\sigma_i\in\IC$, the index of the regulated 
tachyon field
configuration is given by
\be
{\rm 
index}(\T_n)=\Tr\left(\id_{r_{2n}}-\T_n^\dag\,\T_n\right)-
\Tr\left(\id_{q_{2n}}-\T_n\,\T_n^\dag\right)=r_{2n}-q_{2n} 
\ ,
\label{indexTn}\end{equation}
and thus it carries the correct 
monopole charge of
$q_{2n}$ D0-branes and $r_{2n}$ D0-antibranes.

We 
conclude that the finite-level matrix quantum mechanics 
captures
quantitative properties of D-brane projection solitons in 
open string
field theory, through the standard mechanism of tachyon 
condensation
on unstable D-branes. Heavier excitations correspond to 
more complicated
configurations of D0-branes in the matrix model. It 
would be
interesting to characterize also time-dependent solutions of 
the
matrix quantum mechanics. In this context, the classical ground 
states
may mimick those of the BMN matrix model for M-theory in a 
plane wave
background~\cite{BMN1}, which admits a multitude of 
supersymmetric
time-dependent classical configurations. A 
particularly interesting
class of finite-dimensional $\frac12$-BPS 
configurations describes
rotating non-spherical giant gravitons with 
the noncommutative
geometry of a fuzzy torus~\cite{Mik1,Park1}. The 
solution depends on
two moduli $\mu,\zeta\in\IR$ and is given 
explicitly by
\be
{\sf 
Z}_n(\tau)=\e^{\mu\,\tau/3}\,\left(\,
\sum_{i=1}^{q_{2n}-1}\alpha_i~\P_n^{i,i+1}+\alpha_{q_{2n}}~
\P_n^{q_{2n},1}\right) 
\ ,
\label{ZnBMN}\end{equation}
where the parameters 
$\alpha_i(\mu,\zeta)\in\IC$ are constrained by
the pertinent BPS 
equations. Time-dependent solutions are thereby
expected to 
dynamically generate off-diagonal elements of the soliton
basis.

\subsection{Yang-Mills Matrix Quantum Mechanics}

Let $p,q>0$ be a pair of relatively prime integers, and ${\cal
   E}_{p,q}$ a Heisenberg module over a ``dual'' noncommutative
   torus ${\cal A}_\alpha$ to $\atheta$\footnote{\baselineskip=12pt
   These projective modules and dual algebras will be described explicitly
   in section~\ref{SolDyn}.}. Choose a connection on ${\cal E}_{p,q}$
   with corresponding anti-Hermitian gauge fields $A_\mu\in\atheta$, $\mu=1,2$
and
   curvature given by (\ref{Fdef}). Yang-Mills gauge theory on
   ${\cal A}_\alpha$ is then defined by the classical action
\be
S_{\rm YM}[A_1,A_2]=\frac1{2g^2}\,\ncint F^2 \ ,
\label{NCYMaction}\end{equation}
with $g$ the dimensionless Yang-Mills coupling constant. The
corresponding matrix quantum mechanics is the one-dimensional field
theory of four anti-Hermitian matrix fields with action
\bea
&&S_n[\X_n,\Y_n;\X_n',\Y_n']=\frac{\beta_{2n}}{2g^2}\,\int\limits_0^1
\dd\tau~\Tr\left(\mnabla_1\X_n(\tau)-\mnabla_2\Y_n(\tau)+
\big[\X_n(\tau)\,,\,\Y_n(\tau)\big]\right)^2\nn&&~~~~~~+\,
\frac{\beta_{2n-1}}{2g^2}\,\int\limits_0^1
\dd\tau'~\Tr'\left(\mnabla_1\X_n'(\tau'\,)-\mnabla_2\Y_n'(\tau'\,)+
\big[\X_n'(\tau'\,)\,,\,\Y_n'(\tau'\,)\big]\right)^2 \ .
\label{SnNCYM}
\end{eqnarray}

In this subsection we will exploit the fact that the
time direction of the matrix quantum mechanics (\ref{SnNCYM}) is
Euclidean and compactified on the unit circle $\IS^1$. This implies that the
corresponding path integrals compute quantum averages of the system in
a thermal ensemble. The vacuum energy, for example, is given by the
usual statistical mechanical partition function
\be
{\cal Z}_n=\Tr^{~}_{\cal G}\left(\e^{-\hat H_n}\right)=
\sum_{\lambda_n\in{\rm spec}(\hat H_n)}\e^{-\lambda_n} \ ,
\label{calZndef}\end{equation}
where the trace is over the Hilbert space $\cal G$ of physical states of the
matrix
quantum mechanics and $\hat H_n$ is the quantum Hamiltonian operator,
represented on $\cal G$, corresponding to the action
(\ref{SnNCYM}). The partition function may thereby be readily obtained
by computing the eigenvalues of $\hat H_n$ in canonical quantization of
the model (\ref{SnNCYM}) on $\IR$. Quantum gauge theory on the
noncommutative torus is known to be an exactly solvable model which is
given exactly by its semi-classical approximation~\cite{PS1}. In the
following we will study the manner in which its matrix approximation
captures this feature. While we will not completely solve the problem
at the level of the matrix quantum mechanics, our analysis will
illustrate in a straightforward manner what properties to seek
in the 
search for exactly solvable noncommutative field
theories. Analogous 
computations for the lattice regularizations of
noncommutative gauge 
theory in two dimensions can be found
in~\cite{PS2,GS2}.

As always, 
we focus on the first tower in (\ref{SnNCYM}), and 
use
(\ref{nablafinite}) to write the action explicitly 
as
\be
S_n[\X_n,\Y_n]=\frac{\beta_{2n}}{2g^2}\,\int\limits_0^1\dd\tau~
\Tr\left(\SSigma\X_n(\tau)-q_{2n}\,\dot\Y_n(\tau)+\big[\X_n(\tau)-
\Xi\,,\,\Y_n(\tau)\big]\right)^2 
\ .
\label{SnNCYM1st}\end{equation}
The canonical momentum conjugate 
to $\Y_n$ is given by
\be
(\mPi_n)_{ij}:=\frac{\delta 
S_n}{\delta(\dot\Y_n)_{ij}}=
-\frac{q_{2n}\,\beta_{2n}}{g^2}\,\big(\SSigma\X_n-q_{2n}\,
\dot\Y_n+[\X_n-\Xi,\Y_n]\big)_{ji} 
\ ,
\label{Pindef}\end{equation}
while the momentum of the $\X_n$ 
field vanishes since
(\ref{SnNCYM1st}) involves no time derivatives 
of $\X_n$. The matrix
field $\X_n$ is thus non-dynamical and serves 
simply as a Lagrange
multiplier imposing the constraints 
$\frac{\delta
  S_n}{\delta(\X_n)_{ij}}=0$ for $1\leq i,j\leq 
q_{2n}$. By using
(\ref{Pindef}) these constraint equations may be 
written in the simple
matrix 
form
\be
\G_n:=\SSigma\mPi_n+[\Y_n,\mPi_n]=0 \ 
.
\label{constreqs}\end{equation}
The Hamiltonian corresponding to 
the action (\ref{SnNCYM1st}) is given
using (\ref{Pindef}) 
as
\be
H_n=\Tr\left(\frac{g^2}{2(q_{2n})^2\,\beta_{2n}}\,(\mPi_n)^2+
\frac1{q_{2n}}\,\big(\SSigma\X_n+[\X_n-\Xi,\Y_n]\big)\,\mPi_n
\right) 
\ ,
\label{Hndef}\end{equation}
which after imposing the constraints 
(\ref{constreqs}) and using
cyclicity of the trace can be written 
as
\be
H_n=\Tr\left(\frac{g^2}{2(q_{2n})^2\,\beta_{2n}}\,(\mPi_n)^2-
\frac1{q_{2n}}\,[\Xi,\Y_n]\,\mPi_n\right) 
\ .
\label{Hnconstr}\end{equation}
Note that the constraints 
(\ref{constreqs}) are explicitly
time-dependent, while the 
Hamiltonian (\ref{Hnconstr}) on the
constraint surface is independent 
of $\tau$.

In canonical quantization, we promote the matrix fields 
$\Y_n,\mPi_n$ to
operators obeying the commutation 
relations
\be
\big[(\mPi_n)_{ij}\,,\,(\Y_n)_{kl}\big]=-\ii\delta_{ik}\,
\delta_{jl} 
\ .
\label{cancommrels}\end{equation}
We will represent 
(\ref{cancommrels}) in the Schr\"odinger
polarization wherein the 
physical states are the wavefunctions
$\Psi(\Y_n)\in L^2\big(\ii 
u(q_{2n})\big)=L^2\big(\IR^{(q_{2n})^2}\big):={\cal
G}$
and the 
canonical momenta are represented as the derivative 
operators
\be
(\mPi_n)_{ij}=-\ii\,\frac\partial{\partial(\Y_n)_{ij}}
\label{mPinderiv}\end{equation}
acting 
on $\cal G$. By choosing a normal ordering prescription, the
quantum 
Hamiltonian operator on $\cal G$ is then given from
(\ref{Hnconstr}) 
as
\be
\hat
H_n=\sum_{i,j=1}^{q_{2n}}\left(-\frac{g^2}{2(q_{2n})^2\,\beta_{2n}}
\,\frac{\partial^2}{\partial(\Y_n)_{ij}\,\partial(\Y_n)_{ji}}+
\frac{2\pi}{q_{2n}}\,(i-j)\,(\Y_n)_{ij}\,\frac\partial
{\partial(\Y_n)_{ji}}\right) 
\ .
\label{hatHn}\end{equation}

The constraints (\ref{constreqs}) 
should now be implemented on the
Hilbert space $\cal G$, which 
truncates it to the subspace of physical
wavefunctions $\Psi$ obeying
\be
\G_n\Psi=0 \ .
\label{GnPsi}\end{equation}
{}From (\ref{mPinderiv}) and the explicit form (\ref{SSigmaijkl}) of the
shift operator, we thereby arrive at the set of $(q_{2n})^2$
differential equations
\bea
\left[\,\sum_{s=1}^{q_{2n}+i-j}\Sigma_{is}\,\frac\partial{
\partial(\Y_n)_{s,s+j-i}}+
\e^{2\pi\ii\tau}\,\sum_{s=q_{2n}+i-j+1}^{q_{2n}}\Sigma_{is}\,
\frac\partial{\partial(\Y_n)_{s,s+j-i-q_{2n}}}\right.&&\nn+
\left.\sum_{s=1}^{q_{2n}}\left((\Y_n)_{is}
\,\frac\partial{\partial(\Y_n)_{sj}}-(\Y_n)_{sj}\,
\frac\partial{\partial(\Y_n)_{is}}\right)\right]\Psi(\Y_n)&=&0~~~~{\rm
for}~~i<j
\ , \nn&&{~~~~}^{~~}_{~~}\nn
\left[\,\sum_{s=1}^{q_{2n}+j-i}\Sigma_{is}\,\frac\partial{
\partial(\Y_n)_{s,s+i-j}}+
\e^{-2\pi\ii\tau}\,\sum_{s=q_{2n}+j-i+1}^{q_{2n}}\Sigma_{is}\,
\frac\partial{\partial(\Y_n)_{s,s+i-j-q_{2n}}}\right.&&\nn+
\left.\sum_{s=1}^{q_{2n}}\left((\Y_n)_{is}
\,\frac\partial{\partial(\Y_n)_{sj}}-(\Y_n)_{sj}\,
\frac\partial{\partial(\Y_n)_{is}}\right)\right]\Psi(\Y_n)&=&0~~~~{\rm
for}~~i\geq j \ . \nn&&
\label{constreqsexpl}
\end{eqnarray}
These constraints have to be imposed at all times $\tau$. With the
understanding that the $i=j$ part is set trivially to $0$, the
$\tau$-dependent parts of (\ref{constreqsexpl}) combine into the
constraints
\be
\sum_{s=q_{2n}-|i-j|+1}^{q_{2n}}\Sigma_{is}\,
\frac{\partial\Psi(\Y_n)}{\partial(\Y_n)_{s,s+|i-j|-q_{2n}}}=0
\label{constrtau}\end{equation}
which hold {\it for all} $i,j=1,\dots,q_{2n}$. One can show from
(\ref{Sigmassprime}) that the shift matrix $(\Sigma_{ss'})_{1\leq
   s,s'\leq q_{2n}}$ is invertible. It follows then from
(\ref{constrtau}) that the physical wavefunctions $\Psi(\Y_n)$ are
independent of the off-diagonal elements $(\Y_n)_{ij}$ for $i<j$. By
anti-Hermiticity, they are also independent of
$(\Y_n)_{ij}=-\overline{(\Y_n)_{ji}}$ for $i>j$. By setting
$i=j$ in (\ref{constreqsexpl}) the same argument shows that
$\Psi(\Y_n)$ are independent of all diagonal matrix elements of
$\Y_n$, and thus must vanish in $L^2(\ii u(q_{2n}))$.

It follows that there are no physical propagating modes left in the
quantum theory, and the quantum Hamiltonian (\ref{hatHn}) vanishes on
the physical state space, $\hat H_n=0$. This feature is the earmark of a
topological quantum field theory in which only global, topological degrees of
freedom play a role. It is exactly what is anticipated 
in
two-dimensional noncommutative Yang-Mills theory~\cite{PS1}, 
whereby
the gauge invariance of the theory under 
area-preserving
diffeomorphisms of $\IT^2$ kills all local degrees of 
freedom in the model. We
may take the present analysis in the matrix 
model to be a direct proof of the
topological nature of 
noncommutative gauge theory in two dimensions.

While this feature 
would appear to make the statistical sum
(\ref{calZndef}) trivial, 
this is not the case, as there is a large
moduli space of field 
configurations obeying the constraints
(\ref{constreqs}). The 
continuum version of the quantum theory is
given exactly by the 
semi-classical expansion, and we would expect the
matrix 
regularization to capture this property in some way. For this, we 
write
the
thermal partition function (\ref{calZndef}) explicitly as the formal
path integral
\be
{\cal Z}_n=\int\limits_{({\cal C}_n)^2}\DD\X_n~\DD\Y_n~
\e^{-S_n[\X_n,\Y_n]} \ ,
\label{calZnpathint}\end{equation}
where the configuration space is ${\cal C}_n:=\ii u(q_{2n})\otimes
C^\infty(\IS^1)$ and the integration measure is the formal Feynman
measure
\be
\DD\X_n:=\prod_{i,j=1}^{q_{2n}}~\prod_{\tau\in[0,1)}\dd\X_n(\tau)_{ij}
\ .
\label{DDXndef}\end{equation}
After a simple shift of the $\X_n$ field in (\ref{SnNCYM1st}), one is
left with a functional Gaussian integration in (\ref{calZnpathint})
which may be formally carried out to yield
\bea
{\cal Z}_n&=&\int\limits_{{\cal C}_n}\DD\Y_n~\frac1{\det'\left(
\frac{\beta_{2n}}{2g^2}\,(\SSigma-
{\rm ad}^{~}_{\Y_n})\right)^2}\nn&&+\,
\int\limits_{{\cal C}_n}\DD\Y_n~\int\limits_{\ker(\SSigma-{\rm
     ad}^{~}_{\Y_n})}\DD\X_n~\exp\left[-\frac{\beta_{2n}}{2g^2}\,
\int\limits_0^1\dd\tau~\Tr\left(q_{2n}\,\dot\Y_n(\tau)-\big[
\Xi\,,\,\Y_n(\tau)\big]\right)^2\right] \ . \nn&&
\label{calZnYndet}
\end{eqnarray}
The prime on the determinant in (\ref{calZnYndet}) indicates that zero
modes are excluded in its evaluation, while the second contributions
come from the flat directions $\X_n(\tau)$ of the operators $\SSigma-{\rm
   ad}^{~}_{\Y_n}$ for each field configuration $\Y_n(\tau)$. We will
always ignore irrelevant (infinite) constants arising from the
functional integrations.

The large $n$ limit of (\ref{calZnYndet}) yields the fluctuation
determinant that is intractable directly in the continuum
theory~\cite{PS1}, and our matrix model provides a systematic means of
evaluating such complicated objects. Moreover, the second term can be
expected to lead in the limit to the exact sum over instantons of
two-dimensional noncommutative Yang-Mills theory~\cite{PS1}. Let us
indicate how this may arise. For this, we need to study the structure of the
space $\ker(\SSigma-{\rm ad}^{~}_{\Y_n})$, or equivalently the moduli
space of solutions to the equations
\be
\SSigma\X_n=[\Y_n,\X_n] \ .
\label{SSigmaXnYn}\end{equation}

To get a feeling for the type of solution spaces that occur, we
first seek time-independent solutions of (\ref{SSigmaXnYn}).
Repeating the argument of the previous subsection, this implies
that the configurations $\X_n$ are 
diagonal,
\be
\X_n=\sum_{i=1}^{q_{2n}}\ii x_i~\P_n^{ii} \ 
,
\label{Xndiag}\end{equation}
with moduli $x_i\in\IR$. Then 
(\ref{SSigmaXnYn}) 
becomes
\be
\pi\,(q_{2n}+1)\,\delta_{ij}\,\sum_{k=1}^{q_{2n}}x_k=
\ii(x_i-x_j)\,(\Y_n)_{ij} 
\ .
\label{SSigmaXnYncomps}\end{equation}
Setting $i=j$ in 
(\ref{SSigmaXnYncomps}) yields the 
constraint
\be
\sum_{i=1}^{q_{2n}}x_i=0 \ . 
\label{sumxi0}\end{equation} For each
$i\neq j$, the equations 
(\ref{SSigmaXnYncomps}) imply that either
$x_i=x_j$ or 
$(\Y_n)_{ij}=0$. It is straightforward to
characterize the number of 
independent moduli in terms of the
integer $r$ with $0\leq 
r\leq(q_{2n}-1)/2$ which specifies how
many non-vanishing components 
$(\Y_n)_{ij}$, $i\neq j$ there are
in the given solution. If there 
are $r<q_{2n}-1$ such matrix
elements, then the corresponding pairs 
$x_i=x_j$ are equal and
each eliminate one degree of freedom. From 
the constraint
(\ref{sumxi0}) it is straightforward to see that there 
are in all
$q_{2n}-r-1$ real moduli $x_i$. If $r\geq q_{2n}-1$, then 
the
constraint (\ref{sumxi0}) eliminates all the $x_i$'s. In 
both
cases there are $q_{2n}$ real diagonal elements and $r$ 
complex
off-diagonal elements of the $\Y_n$ matrices. It follows that 
the
kernel of the operator $\SSigma-{\rm
  ad}^{~}_{\Y_n}$ admits an 
orthogonal decomposition into subspaces
corresponding to constant 
$(\X_n,\Y_n)$ configurations as
\be
\ker\left(\SSigma-{\rm 
ad}^{~}_{\Y_n}\right)=
\bigoplus_{r=0}^{q_{2n}(q_{2n}-1)/2}{\cal K}_r 
\ ,
\label{kerdecomp}\end{equation}
where
\be
{\cal 
K}_r=\left\{\begin{matrix}\,\IR^{q_{2n}-r-1}\oplus\IR^{q_{2n}}
\oplus\IC^r&0\leq 
r<q_{2n}-1\\\IR^{q_{2n}}\oplus\IC^r&
q_{2n}-1\leq r\leq 
q_{2n}(q_{2n}-1)/2\end{matrix}\right. \ 
.
\label{calKrdef}\end{equation}

In the time-dependent case, we can exploit the invariance of the path
integration in (\ref{calZnYndet}) under arbitrary unitary
transformations of the matrices $\X_n(\tau)$ to diagonalize them. Then
our analysis of the contributions to the second integral in (\ref{calZnYndet})
carries
through in exactly the same manner as described above. For each of them, the
path integral over the matrix trajectories $\Y_n(\tau)$ is Gaussian
and yields the determinant of the operator
$\frac{\beta_{2n}}{2g^2}\,(q_{2n}\,\frac\dd{\dd\tau}+{\rm
   ad}^{~}_\Xi)^2$ on the unit circle and restricted to the subspaces
of $\ii u(q_{2n})$ in which $\Y_n$ has $r$ non-zero off-diagonal
matrix elements. We may evaluate (\ref{calZnYndet}) in this way to the
formal expression
\bea
{\cal Z}_n&=&\int\limits_{{\cal C}_n}\DD\Y_n~
\frac1{\det'\left(\frac{\beta_{2n}}{2g^2}\,(
\SSigma-{\rm ad}^{~}_{\Y_n})\right)^2}\nn&&+\,
\sum_{r=0}^{q_{2n}(q_{2n}-1)/2}{\rm V}_r\,\sum_{i_1,\dots,i_r=1}^{q_{2n}}~
\sum_{j_1,\dots,j_r=1}^{q_{2n}}~\prod_{\stackrel{\scriptstyle k,l=1}
{\scriptstyle i_k\neq j_l}}^r~\prod_{m\in\IN_0}\,
\Biggl(\,\frac{2g^2/\beta_{2n}\,(q_{2n})^2}
{m^2-\left(\frac{i_k-j_l}{q_{2n}}\right)^2}
\Biggr)^2 \ ,
\label{calZnfinal}
\end{eqnarray}
where ${\rm V}_r$ is the suitably regulated volume factor ${\rm
   V}_r={\rm vol}(\IR^{q_{2n}-r-1})$ for $0\leq r<q_{2n}-1$, while
   ${\rm V}_r=1$ otherwise. It would be interesting to examine now how
   the appropriately regulated form of the expression
   (\ref{calZnfinal}) reproduces the partition sum of the corresponding
   continuum theory in the limit $n\to\infty$. Although we have not
   completely solved the problem here, the expression
   (\ref{calZnfinal}) once again illustrates the exact solvability of
   the gauge theory.

\section{Moduli Spaces and Soliton Regularization on the
    Noncommutative Plane \label{se:moduli}}
\setcounter{equation}{0}

In this final section we shall describe the relationship between our soliton
approximation of field theory on the noncommutative torus and the
standard solitons on the noncommutative plane. We shall deal only with
the GMS solitons which are obtained in the limit of large Moyal
noncommutativity~\cite{GMS}. We will see that the matrix regularization of
noncommutative field theory in this context provides illuminating
results concerning the moduli spaces of these solitons. To help
motivate the analysis, we will begin by showing how a special class of
projections on the noncommutative torus naturally arise as the
classical solutions of a model for the dynamics of solitons on 
the
noncommutative torus. We will then use these projections to 
obtain the
matrix analogs of GMS solitons, which among other things 
provides the
starting point for the construction of one-dimensional 
matrix model
regularizations of field theories on the noncommutative 
plane.

\subsection{Soliton Dynamics on the Noncommutative 
Torus\label{SolDyn}}

We will begin by describing how to model the 
dynamics of projection
solitons on the noncommutative torus in an 
adiabatic
approximation. Usually, one would proceed by introducing a 
K\"ahler
metric on the moduli space of fixed rank projection 
operators, which
is typically an infinite-dimensional Grassmannian 
manifold. The
K\"ahler form may be obtained as the curvature of a 
determinant line
bundle over the Grassmannian, and with it one may 
construct a
non-linear $\sigma$-model describing the moduli space 
dynamics of
solitons~\cite{LRvU1,GHS1}. The motion of the solitons 
may thereby be
studied by calculating geodesics on the moduli space 
in the obtained
K\"ahler metric. A non-trivial, curved geometry then 
corresponds to
velocity dependent forces between the solitons. Here 
we shall instead
follow the approach of~\cite{dkl} where non-linear 
$\sigma$-models in
the context of noncommutative geometry were 
proposed. This approach
exploits the inherent non-linearity of the 
space of
projections of $\atheta$ directly, without explicit 
reference to any
K\"ahler geometry.

We define a noncommutative field 
theory whose configuration space is
the collection $\cp_\theta$ of 
all projections in the algebra
$\atheta$. The $\sigma$-model dynamics 
is governed by the action
functional $S:\cp_\theta\to\IR^+$ defined 
by
\be
S(\P) = \frac{1}{2 \pi}\,\ncint 
\partial_{\mu}\P\,\partial_{\mu}\P=
\frac{1}{\pi}\,
\ncint\P\,\partial_{\mu}\P\,\partial_{\mu}\P 
~, \label{actfun}
\end{equation}
where $\partial_\mu$, $\mu=1,2$ are 
the two linear derivations defined in
(\ref{t2act}). This is just the 
standard action that
one would write down which captures the dynamics 
of multiple solitons
(within a certain energy range), except that we 
utilize a flat metric
in its definition. The second equality follows 
from the constraint
$\P^2=\P$ and the Leibniz rule. The positivity of 
the trace $\ncint$
guarantees that (\ref{actfun}) is always a 
positive real number.

We will seek critical points of the action 
functional (\ref{actfun})
in a given connected component of 
$\cp_\theta$, corresponding to an
equivalence class of projections of 
fixed rank and fixed topological
charge. For this, we need 
to
carefully take into account the non-linear structure of the 
space
$\cp_\theta$. An element $\delta\P\in T_\P(\cp_\theta)$ in the 
tangent
space to $\cp_\theta$ at a given point $\P$ is not arbitrary 
but must
fulfill two requirements. First of all, it must be 
Hermitian,
$(\delta\P)^\dag=\delta\P$. Secondly, it must 
obey
$(\P+\delta\P)^2=\P+\delta\P+O((\delta\P)^2)$, which implies 
that
$(\id-\P)\,\delta\P=\delta\P\,\P$. It follows that the most 
general
tangent vector in $T_\P(\cp_\theta)$ is of the 
form
\be
\delta\P=(\id-\P)\,c\,\P+\P\,c^\dag\,(\id-\P)
\label{gentanvec}\end{equation}
with 
$c$ arbitrary elements of the algebra $\atheta$.

The equations of 
motion now follow as usual from the 
variational
principle
\be
0=\delta 
S(\P)=-\frac1{2\pi}\,\ncint\Box(\P)\,\delta\P \ 
,
\label{deltaSP0}\end{equation}
where 
$\Box=\partial_\mu\,\partial_\mu$ is the Laplacian. We have used
the 
Leibniz rule, along with invariance and cyclicity of the trace. 
By
substituting in (\ref{gentanvec}) and using the fact 
that
$c\in\atheta$ is arbitrary, we arrive at the field 
equations
\be
\P\,\Box(\P)\,(\id-\P)=0 \ , ~~ 
(\id-\P)\,\Box(\P)\,\P=0 \ ,
\label{fieldeqns}\end{equation} which 
together are equivalent to
\be
\P\,\Box(\P)-\Box(\P)\,\P=0 \ . 
\label{eom}
\end{equation}
These are non-linear equations of second 
order which are rather
difficult to solve explicitly. However, as we 
shall show
presently, the absolute minima of the action 
functional
(\ref{actfun}) in a given connected component of 
$\cp_{\theta}$
actually satisfy first order equations which are 
easier to solve.

For this, we recall from section~\ref{ProjSeq} that 
for any projection
$\P\in\cp_\theta$, there is a topological charge 
(the first Chern
number) defined by \eqn{topcha} with 
$c_1(\P)\in\IZ$. Then, just as in
four dimensional Yang-Mills theory, 
this topological quantity yields a
bound on the action functional. 
Due to positivity of the trace
$\ncint$ and its cyclic property, we 
have
\be\label{pro} \ncint
\big(\partial_\mu(\P)\,\P\pm\ii 
\epsilon_{\mu\nu}\,
\partial_\nu(\P)\,\P 
\big)^\dag\,\big(
\partial_\mu(\P)\,\P\pm\ii\epsilon_{\mu\beta}\,\partial_\beta(\P)\,\P
\big)\geq 
0~.
\end{equation}
By expanding out the left-hand side of (\ref{pro}) 
and comparing it with
(\ref{topcha}) and (\ref{actfun}), we then arrive at the inequality
\be\label{bpbou} S (\P) \geq \pm\,2\,c_1(\P) ~.
\end{equation}
The inequality \eqn{bpbou}, which gives a lower bound on the action,
is the analog of the one for ordinary two-dimensional $\sigma$-models
\cite{bp}. {}From \eqn{pro} it is clear that equality in \eqn{bpbou} occurs
exactly when the projection $\P$ satisfies the {\it self-duality} or
{\it anti-self-duality} equations
\be
\label{sd0} \big(\partial_\mu\P\pm\ii
\epsilon_{\mu\nu}\,\partial_\nu\P \big)\,\P= 0 ~.
\end{equation}
The two equations \eqn{sd0} can be reduced to
\be
\label{sd}
\bar{\partial} (\P)\,\P= 0~, ~~ \partial (\P)\,\P= 0~,
\end{equation}
respectively; here  $\partial = \half\,(\partial_1 - \ii
\partial_2)$ and $\bar{\partial} = \half\,(\partial_1 + \ii
\partial_2) $. Simple manipulations show directly that each of the
equations \eqn{sd} implies the field equations \eqn{eom}, as they
should. Solutions of \eqn{sd} are called $\sigma$-model instantons.

The non-linear equations (\ref{sd}) can be reduced to linear ones by
   introducing gauge degrees of freedom and by lifting them to a 
bundle
  (a module)~\cite{dkl,dkl03,dkl03b}. The particular module is 
dictated by the
  given homotopy class that we are working in, which 
is in turn
  determined by the rank $\ncint\P=p+q\,\theta$, 
$p,q\in\IZ$ and
  topological charge $c_1(\P)=q$ of the projection 
solutions to
  (\ref{sd}). We will identify the algebra $\atheta$ as 
the
  endomorphism algebra of a suitable bundle and regard any 
projection
  $\P$ as an operator on this bundle. For this, we need 
to
consider the representation theory of another copy $\aalpha$ of 
the
noncommutative torus with unitary generators $Y$ and $Z$ obeying 
the relation
\be
Z\,Y=\e^{2\pi\ii\alpha}~Y\,Z \ . 
\label{nctbis}\end{equation} When
$\alpha$ is an irrational number, every finitely generated
projective module over the algebra $\aalpha$ which is not free is
isomorphic to a Heisenberg module. As these modules will also be
of central importance in the following, we shall review their
basic properties here~\cite{cr}.

As already mentioned, any
such module ${\cal E}_{p,q}$ is characterized by two integers $p,q$
satisfying $p+q\,\alpha>0$, which can be taken to be relatively prime
with $q>0$, or $p=0$ and $q=1$. As a vector space, the module
\be
{\cal E}_{p,q}=S(\IR)\otimes\IC^q
\label{calEpqdef}\end{equation}
is the space of Schwartz functions of one continuous variable
$s\in\IR$ and one discrete variable $k\in\IZ_q$. By introducing the
notation
\be
\varepsilon=p/q-\alpha
\label{varepsilondef}\end{equation}
the space (\ref{calEpqdef}) is made into a {\it right} module over
$\aalpha$ by defining
\bea
(\xi Y)^{~}_k(s)&:=&\xi^{~}_{[k-p]_q}(s-\varepsilon) \ , \nn
(\xi Z)^{~}_k(s)&:=&\e^{2\pi\ii(s-k/q)}~\xi^{~}_k(s)
\label{rgtmod}
\end{eqnarray}
for $\xi\in{\cal E}_{p,q}$, with the relations (\ref{nctbis}) being
easily verified. On the module (\ref{calEpqdef}) one defines an
$\aalpha$-valued Hermitian structure
\be
\langle\,\cdot\,,\,\cdot\,\rangle^{~}_\alpha\,:\,{\cal E}_{p,q}
\times{\cal E}_{p,q}~\longrightarrow~\aalpha
\label{hermAalpha}\end{equation}
by the formula
\be
\langle\xi,\eta\rangle^{~}_\alpha:=
\sum_{(m,r)\in\IZ^2}~\sum_{k=0}^{q-1}~
\int\limits_\IR\dd s~\overline{\xi^{~}_{[k-m\,p]_q}(s-m\,
\varepsilon)}~\eta^{~}_k(s)~\e^{-2\pi\ii r\,(s-k/q)}~Y^m\,Z^r
\label{hsrgt}\end{equation}
for $\xi,\eta\in{\cal E}_{p,q}$. Note the antilinearity of the first
factor.

The endomorphism algebra ${\rm End}_{\aalpha}({\cal E}_{p,q})$, which
acts from the left on $\heisen$, can be identified with the original copy
$\atheta$ of the noncommutative torus where the noncommutativity
parameter $\theta$ is ``uniquely'' determined by $\alpha$ in the
following way. Since $p$ and $q$ are relatively prime, there exist
integers $a,b\in\IZ$ such that $b\,q-a\,p=1$. Then the
noncommutativity parameter is given by the discrete M\"obius
transformation
\be
\theta=\frac{a\,\alpha-b}{q\,\alpha-p} \ .
\label{flt}\end{equation}
Notice that given any other pair of integers $a',b'\in\IZ$ with
$b'\,q-a'\,p=1$, one has $\theta'-\theta\in\IZ$ so that ${\cal
    A}_{\theta'}\cong\atheta$. It follows that the algebra ${\rm
    End}_{\aalpha}({\cal E}_{p,q})$ is generated by the two unitary operators
$U$ and $V$ which act from the {\it left} on $\heisen$ by
\bea
(U\xi)^{~}_k(s)&:=&\xi^{~}_{[k-1]_q}(s-1/q) \ , \nn(V\xi)^{~}_k(s)&:=&
\e^{\frac{2\pi\ii}q\,(s/\varepsilon+a\,k)}~\xi^{~}_k(s) \ ,
\label{lftmod}
\end{eqnarray}
and one easily verifies the defining relations
(\ref{nct}) of the algebra $\atheta$.

On $\heisen$ there is also an $\atheta$-valued inner product
\be
\langle\,\cdot\,,\,\cdot\,\rangle^{~}_\theta\,:\,{\cal E}_{p,q}
\times{\cal E}_{p,q}~\longrightarrow~\atheta
\label{hermAtheta}\end{equation}
which is given explicitly by
\be
\langle\xi,\eta\rangle^{~}_\theta:=\frac1{|q\,\varepsilon|}\,
\sum_{(m,r)\in\IZ^2}~\sum_{k=0}^{q-1}~\int\limits_\IR
\dd s~\xi^{~}_k(s)~\overline{\eta^{~}_{[k-m]_q}(s-m/q)}~\e^{-\frac{2\pi\ii}{q
\,\varepsilon}\,(s-m/q+a\,k\,\varepsilon)}~U^m\,V^r
\label{hslft}\end{equation}
for $\xi,\eta\in\heisen$. Notice that now the antilinearity is in the
second factor. The key feature of this Hermitian structure is that it
is compatible with the $\aalpha$-valued one
(\ref{hermAalpha},\ref{hsrgt}),
\be
\langle\xi,\eta\rangle^{~}_\theta\,\zeta=\xi\,\langle\eta,\zeta
\rangle^{~}_\alpha
\label{morcom}\end{equation}
for all $\xi,\eta,\zeta\in\heisen$. This means that 
the
$\atheta$-$\aalpha$ bimodule $\heisen$ provides a Morita 
equivalence
between the two algebras $\atheta$ and 
$\aalpha$.
Physically, the compatibility condition 
(\ref{morcom})
corresponds to T-duality between the vertex operator 
algebras of
$(p'-2)$-$(p'-2)$ and $p'$-$p'$ strings acting on the 
Hilbert space of
$p'$-$(p'-2)$ open string states in the low-energy 
limit~\cite{LLSNC,SW}. These
open
string modes stretch between a 
single D$(p'-2)$-brane and a collection
of $p$ coincident 
D$p'$-branes carrying $q$ units of vortex
D$(p'-2)$-brane 
charge.

Using the previous construction one can now 
build
projections on the algebra $\atheta$ by picking suitable 
elements
$\xi'\in\heisen$ with
$\langle\xi',\xi'\rangle^{~}_\alpha 
=\id$. The bimodule property
(\ref{morcom}) then implies that $\P 
=
\langle\xi',\xi'\rangle^{~}_\theta$ is a non-trivial projection 
in
$\atheta$. Furthermore, by using the 
identification
$\atheta\cong{\rm End}_{{\cal A}_{\alpha}}({\cal 
E}_{p,q})$, any such a
projection may be equivalently written in the 
more suggestive form
\be
\P = 
\left\langle\xi\left(\langle\xi,\xi\rangle_{\alpha}^{~}
\right)^{-1/2}~,~\xi\left(\langle\xi,\xi\rangle^{~}_{\alpha}
\right)^{-1/2}\,\right\rangle_\theta^{~}
= 
|\xi\rangle~\frac\id{\langle\xi,\xi\rangle_{\alpha}^{~}}~
\langle\xi| 
\ ,
\label{psugg}\end{equation}
where for each element 
$|\xi\rangle\in{\cal E}_{p,q}$ the
corresponding dual element 
$\langle\xi|\in({\cal E}_{p,q})^*$ is
defined by means of the 
Hermitian structure 
as
$\langle\xi|(\eta)=\langle\xi,\eta\rangle^{~}_{\alpha}\in{\cal
A}_{\alpha}$, and we have only assumed now 
that
$\langle\xi,\xi\rangle_\alpha$ is an invertible element of the 
algebra
$\aalpha$.

In order to translate the self-duality equations 
(\ref{sd}) for
$\P$ to equations for $\xi$, we need to introduce a 
gauge
connection on the right $\aalpha$-module $\ce_{p,q}$. This is 
done
by means of two covariant derivatives\footnote{The 
covariant
derivative $\nabla$ introduced here should not be confused 
with
the approximate derivation $\mnabla$ introduced 
in
section~\ref{Kinetic}.} $\nabla_1, 
\nabla_2:\,
\ce_{p,q}\rightarrow\ce_{p,q}$ which are given explicitly 
by
\be
\label{con} (\nabla_1 \xi)^{~}_k(s) := \frac{2\pi\ii}\varepsilon
\,s\,\xi^{~}_k(s)~,
{}~~(\nabla_2 
\xi)^{~}_k(s) := \frac{\dd \xi_k^{~}(s)}{\dd 
s}~.
\end{equation}
Notice that the discrete index $k$ is not 
touched. This connection has
constant 
curvature
\be
[\nabla_1,\nabla_2]=-\frac{2\pi\ii}\varepsilon\,\id \ 
,
\label{constcurv}\end{equation}
and the two operators (\ref{con}) 
satisfy a Leibniz rule with respect
to the right 
action,
\be
\nabla_\mu (\xi a) = (\nabla_\mu \xi)a + \xi 
(\partial_\mu a) \ , ~~\mu=1,
2~,
\label{leirgt}\end{equation}
for 
any $\xi\in\ce_{p,q}$ and $a\in\ca_{\alpha}$. They are 
also
compatible with the $\aalpha$-valued Hermitian 
structure,
\be
\partial_\mu \hs{\xi}{\eta}_\alpha = \hs{\nabla_\mu 
\xi}{\eta}_\alpha +
\hs{\xi}{\nabla_\mu \eta}_\alpha~,
{}~~\mu=1, 
2~,
\label{comrgt}\end{equation}
for any $\xi, \eta \in \ce_{p,q}$. 
Furthermore, by using compatibility
(\ref{comrgt}) and the right 
Leibniz rule (\ref{leirgt}) one can show
that the induced derivations 
on the endomorphism algebra,
\be
\delta_1, \delta_2\,:\,{\rm 
End}_{\ca_{\alpha}}(\ce_{p,q})~\longrightarrow~
{\rm 
End}_{\ca_{\alpha}}(\ce_{p,q}) ~,
{}~~ \delta_\mu (T):=[\nabla_\mu, 
T] ~,
\end{equation}
are proportional to the generators of the 
infinitesimal action of the
commutative torus $\IT^2$ on $\atheta 
\cong{\rm
 End}_{\ca_{\alpha}}(\ce_{p,q})$,
\be
\delta_\mu=\frac1{q\,\varepsilon}\,\partial_\mu 
\ .
\label{deltaproptopartial}
\end{equation}
It is because of this 
property that we select the particular connection
\eqn{con}.

Then, 
by using these ingredients, it is straightforward to show that 
the
projection
$\P$ in (\ref{psugg}) satisfies the self-duality 
equations \eqn{sd} if
and only if there exists an element 
$\rho\in\ca_{\alpha}$ such 
that
\be
\label{nsd}
\overline{\nabla}\xi=\xi\rho \ 
,
\end{equation}
with $\overline{\nabla} = 
\half\,(\nabla_{1}+\ii\nabla_{2})$. This
equation follows from a 
simple computation with the 
element
$\rho=(\langle\xi,\xi\rangle_\alpha)^{-1}\,\langle\xi,\overline{\nabla}
\xi\rangle_\alpha$. 
When $\rho$ is   constant (i.e. it is
proportional to the unit of 
$\ca_{\alpha}$, $\rho=\lambda\,\id$ with
$\lambda\in\IC$), the 
equation (\ref{nsd}) reduces to the simple
differential 
equation
\be
\frac{\dd\xi}{\dd 
s}+\left(\frac{2\pi\,s}\varepsilon+2\ii\lambda
 \right)\,\xi=0
\end{equation}
whose solutions are the Gaussian 
fields
\be\label{gauss}
\xi_{\lambda}(s)=\mbf{A}~\e^{-\pi\,s^{2}/\varepsilon-2\ii\lambda\,s} 
\ .
\end{equation}
The vector $\mbf{A}=(A_1, \dots, A_q) \in\IC^{q}$ 
can be taken to lie in the
complex projective space $\IC \IP^{q-1}$ 
by removing an
inessential normalization. It is possible to prove 
that,
for all values of the deformation parameter $\theta$, the 
norms
$\hs{\xi_{\lambda}}{\xi_{\lambda}}_\alpha\in\aalpha$ 
are
invertible \cite{bo0,wa}. Accordingly, the Gaussian 
functions
\eqn{gauss} provide a complex one-parameter family of 
solutions $\P_\lambda =
\ket{\xi_{\lambda}} ( \hs{\xi_{\lambda}} 
{\xi_{\lambda}}_\alpha )^{-1}
\bra{\xi_{\lambda}}$ of the 
self-duality equations \eqn{sd}. The
projection 
$\P_\lambda\in\atheta$ has rank
$\ncint\P_\lambda=p+q\,\theta$ and 
topological charge
$c_1(\P_\lambda)=q$.

The physically relevant 
values of the complex parameter $\lambda$ can
be restricted by gauge 
symmetry. Any two elements $\xi$ and
$\xi'$ of $\ce_{p,q}$ provide 
different projections (\ref{psugg}) if
and only if they belong to 
different orbits of the action of the group
$GL(\ca_{\alpha})$ of 
invertible elements of $\ca_{\alpha}$ which
acts on the right on 
$\ce_{p,q}$,
\be
|\xi\rangle~\longmapsto~|\xi^g\rangle=|\xi\rangle g 
\ , ~~ g\in GL(\aalpha) \
{}.
\label{gaugeorb}\end{equation}
Note 
that we do not require $g$ to be unitary. The action
(\ref{gaugeorb}) 
preserves the invertibility of $\hs{\xi}{\xi}_\alpha$
and leaves the 
corresponding projection (\ref{psugg})
invariant. Furthermore, from 
the Leibniz rule for the gauge connection
it follows that if $\xi$ is 
a solution of the self-duality equation
(\ref{nsd}), then the 
transformed vector $\xi^g$ also solves an
equation of the form 
(\ref{nsd}),
$\overline{\nabla}\xi^g=\xi^g\rho^g$, withthe 
element
$\rho\in\aalpha$ modified to
\be \label{gauge} 
\rho~\longmapsto~\rho^g=g^{-1}\,\rho\,
g+g^{-1}\,\overline{\partial}g 
\ .
\end{equation}
Elements of the group $GL(\ca_{\alpha})$ thereby 
play the role of complex
gauge transformations.

In the case of the 
Gaussian fields (\ref{gauss}), it is
straightforward to show from 
(\ref{gauge}) with $\rho=\lambda\,\id$
and $\rho^g=\lambda'\,\id$ 
that $\xi_{\lambda}$ and $\xi_{\lambda'}$
are gauge equivalent if and 
only if
$\xi_{\lambda'}=\xi_{\lambda}\,U^{m}V^r$ for some pair of 
integers
$(m,r)\in\IZ^2$. The parameters of the Gaussian functions 
are then
related by
\be
\lambda'= \lambda + \pi \ii (m+ \ii r) \ 
.
\end{equation}
It follows that the gauge inequivalent parameters 
$\lambda$ make up an
ordinary torus $\IT^2$. As we will see later on, 
the moduli
$\lambda\in\IT^2$ correspond to the locations of the 
solitons on the
underlying torus. The moduli space of Gaussian 
fields
(\ref{gauss}) is thus $\IC \IP^{q-1} \times \IT^2$. For 
further
aspects of these and other constructions, 
see~\cite{dkl,dkl03,dkl03b}.

\subsection{The Boca 
Projection}\label{se:boca}

Let us now describe explicitly a 
particular instance of the globally
minimizing soliton projections of 
the previous subsection; it will
play an important role in the 
following. The Boca projection on the
noncommutative torus comes from 
choosing the simplest bimodule ${\cal
   E}_{0,1}=S(\IR)$, for which 
$\varepsilon=1/\theta$, $a=p=0$, $b=q=1$,
and $\alpha=-1/\theta$ in 
the above construction. With the Gaussian Schwartz
function
\be
\xi(s)=\xi_{\lambda=0}(s)=\e^{-\pi\,\theta\,s^2} \,
\label{GaussSchwartz} \ ,\end{equation} which is such that the
element $\langle\xi,\xi\rangle^{~}_{-1/\theta}\in{\cal
A}_{-1/\theta}$ is invertible \cite{bo0,wa}, it follows that
\be
\B_\theta:=\left\langle\xi\left(\langle\xi,\xi\rangle_{-1/\theta}^{~}
\right)^{-1/2}~,~\xi\left(\langle\xi,\xi\rangle^{~}_{-1/\theta}
\right)^{-1/2}\,\right\rangle_\theta^{~}
\label{Btheta}\end{equation}
is a projection on $\atheta$ which is homotopic to the Powers-Rieffel
projection (\ref{Ptheta}).

The general form of the Boca projection (\ref{Btheta}) may be deduced 
by using the
${\cal A}_{-1/\theta}$-action (\ref{rgtmod}) on (\ref{GaussSchwartz})
and the inner product (\ref{hsrgt}) to explicitly compute the element
$\langle\xi,\xi\rangle^{~}_{-1/\theta}\in{\cal A}_{-1/\theta}$. The
square root may be computed by using holomorphic functional calculus,
and one thereby finds~\cite{MM1}
\bea
\xi(s)\left(\langle\xi,\xi\rangle^{~}_{-1/\theta}\right)^{-1/2}
&=&\e^{-\pi\,s^2/\theta}\,\left[1-
\sum_{k\in\IN}\frac{(2k-3)!!}{k!}\,\left(\frac\theta8\right)^{k/2}
\right.\nn&&~~~~~\times\Biggl.
\sum_{\stackrel{\scriptstyle(\boldsymbol{m},\boldsymbol{r})\in
\IZ^k\times\IZ^k}{\scriptstyle(\boldsymbol{m},\boldsymbol{r})
\neq(\boldsymbol{0},\boldsymbol{0})}}
\e^{-Q_k(\boldsymbol{m},\boldsymbol{r})}\,\prod_{i=1}^k
\e^{-(2\pi\,s/\theta)\,(m_i+\ii r_i)}\Biggr] \ ,
\label{b12xisgen}
\end{eqnarray}
with the convention $(-1)!!:=1$ and $Q_k$ the quadratic form
on $\IZ^k\times\IZ^k$ defined by
\beq
Q_k(\boldsymbol{m},\boldsymbol{r})=\frac\pi{2\,\theta}\,\sum_{i=1}^k
\left[(m_i)^2+(r_i)^2\,\right]+\frac\pi\theta\,\left(\,\sum_{i=1}^km_i
\right)^2+\frac{\pi\ii}\theta\,
\left(\,\sum_{i=1}^km_i\,r_i+2\,\sum_{i<j}m_i\,r_j\right) \ .
\label{Qkform}\end{equation}
The corresponding projection is given by
(\ref{Btheta}) and (\ref{hslft}), and it will be used in the following to
relate the soliton basis of Section~\ref{se:masu} to matrix
noncommutative solitons on~$\IR^2$.

At the special rational values $\theta=1/k$, $k\in\IN$ (in which case
the algebra ${\cal A}_{-1/\theta}\cong C^\infty(\IT^2)$ is
commutative), the Boca projection can be expressed in terms of
theta-functions of the generators $U$ and $V$ of $\atheta$. For this,
we introduce the elliptic Jacobi-Erderlyi theta-functions
\be
\vartheta^{~}_{\frac
   aN\,,\,b}(\nu\,|\ii\sigma)=\sum_{m\in\IZ}\e^{-\pi\,\sigma\,
   (m+a/N)^2+2\pi\ii(m+a/N)\,b}~\e^{2\pi\ii(Nm+a)\,\nu}
\label{genus1Jacobi}\end{equation}
for $a\in\IZ$, $b\in\IR$ and $N\in\IN$, which are holomorphic in $\nu\in\IC$
for moduli
$\sigma\in\IC$ with ${\rm Re}(\sigma)>0$. We use the convention that
$N=1$ when $a=0$ in (\ref{genus1Jacobi}). Then by taking $k\in\IN$
even for definiteness, and using the Weyl maps $\rho$ and $\rho'$ of
(\ref{xsubalg}) and (\ref{ysubalg}), the Boca projection
(\ref{Btheta}) may be expressed succinctly as~\cite{bo0}
\bea
\B_{1/k}&=&\frac1{k\,\rho\Bigl(\vartheta^{~}_{0,0}\left(x^k\,\left|\,
\mbox{$\frac{\ii k}2$}\right.\right)\Bigr)\,
\rho'\Bigl(\vartheta^{~}_{0,0}\left(y^k\,\left|\,
\mbox{$\frac{\ii k}2$}\right.\right)\Bigr)}\nn&&\times\,
\sum_{l,m=0}^{k-1}\e^{-\pi\ii l\,m/k}~
\rho\Bigl(\vartheta_{\frac mk\,,\,\frac l2}\left(x\,\left|\,
\mbox{$\frac{\ii k}2$}\right.\right)\Bigr)\,
\rho'\Bigl(\vartheta_{\frac lk\,,\,\frac m2}\left(y\,\left|\,
\mbox{$\frac{\ii k}2$}\right.\right)\Bigr) \ .
\label{Bocaspecial}
\end{eqnarray}
The Wigner function on $\IT^2$ corresponding to the projection
(\ref{Bocaspecial}), which is real and exhibits localized bump
configurations, is displayed in~\cite{GHS1}. This form will be
used later on to give a physical interpretation to the relationship
between torus solitons and solitons on the noncommutative plane.

\subsection{GMS Soliton Expansions\label{GMSExp}}

Let us now consider the noncommutative plane $\IR^2_\Theta$, which is
defined heuristically by the Heisenberg commutation relation
$[y,x]=2\ii\Theta$. We will assume that $\Theta>0$ for
definiteness. The algebra $C^\infty(\IR_\Theta^2)$ may be
identified as the appropriate completion of the
polynomial algebra $F(\IR^2)/I_\Theta$, where
$F(\IR^2)=\IC\,\langle \id,x,y \rangle$ is the free unital algebra on 
two generators
$x,y$,   and $I_\Theta$ is the two-sided ideal of
$F(\IR^2)$ generated by the element $y\,x - x\,y-2\ii\Theta \,\id$. As a
Heisenberg algebra, it has a unique irreducible representation which 
is the usual Fock
space
\be
{\cal F}=\ell^2(\IN_0)=\overline{{\rm span}^{~}_{\IC}}\,
\big\{\,|m\rangle~\big|~m\in\IN_0\big\}
\label{Fockspace}\end{equation}
for the Schr\"odinger representation of quantum mechanics. In
(\ref{Fockspace}), the vectors $|m\rangle$ are the elements of the
usual orthonormal number basis of a one-dimensional harmonic
oscillator, $\langle m|n\rangle=\delta_{mn}$, and the 
appropriate
completion is indicated which will always be implicitly 
understood in
what follows. In particular, a basis for
the algebra of 
bounded linear operators on $\cal F$ is provided by the 
set
$\{\,|n\rangle\langle m|~|~m,n\in\IN_0\}$.

The Weyl map $\Omega$ 
and star-product on the noncommutative plane are
defined analogously 
to (\ref{weymap}), (\ref{wigmap}) and
(\ref{twco}) by using Fourier 
transformation of fields on $\IR^2$~\cite{Sz1}.
In
particular, the 
Wigner functions on $\IR^2$ corresponding to the rank one
Fock space 
operators $|n\rangle\langle m|$ are the 
Landau
wavefunctions
\be
\psi_{n,m}^{~}(w,\bw\,)=\frac1{\sqrt{4\pi\,\Theta}}~\Omega^{-1}
\big(|n\rangle\langle 
m|\big) \ ,
\label{Landauwavefn}\end{equation}
where $w,\bw$ are complex coordinates on the plane. They are the
orthonormal eigenfunctions in $L^2(\IR^2)$ of the Landau
Hamiltonian for a charged particle moving on $\IR^2$ under the
influence of a constant, perpendicularly applied magnetic field
$B=\Theta^{-1}$. The ground state wavefunction is the Gaussian field
\be
\psi^{~}_{0,0}(w,\bw\,)=\frac1{\sqrt{\pi\,\Theta}}\,\e^{-|w|^2/2\,\Theta} \ ,
\label{psi00}\end{equation}
while the higher Landau levels can be obtained from (\ref{psi00}) by
application of the differential creation operators
\be
a^\dag:=\frac12\,\left(-\sqrt\Theta~\frac\partial{\partial\bw}+
\frac w{\sqrt\Theta}\right) \ , ~~ b^\dag:=\frac12\,\left(-\sqrt\Theta~
\frac\partial{\partial w}+\frac\bw{\sqrt\Theta}\right)
\label{abdag}\end{equation}
as
\bea
\psi^{~}_{n,m}(w,\bw\,)&=&\frac{\left(a^\dag\right)^n}{\sqrt{n!}}\,
\frac{\left(b^\dag\right)^m}{\sqrt{m!}}\psi_{0,0}^{~}(w,\bw\,)
\nn&=&\frac{(-1)^{\min(n,m)}}{\max(n,m)!}\,\sqrt{\frac{n!\,m!}
{\pi\,\Theta}}\,\left(\frac{|w|^2}{\Theta}\right)^{|n-m|/2}~
\e^{\ii(n-m)\,{\rm arg}(w)}~\e^{-|w|^2/2\,\Theta}\nn&&\times\,
L_{\min(n,m)}^{|n-m|}\big(|w|^2/\Theta\big) \ ,
\label{psinm}\end{eqnarray}
where
\be
L_k^r(t)=\sum_{l=0}^k(-1)^l\,{k+r\choose k-l}\,\frac{t^l}{l!}
\label{Lkrt}\end{equation}
are the associated Laguerre polynomials.

{}From the Wigner representation (\ref{Landauwavefn}) it follows
immediately that these functions obey the star-product
projection relation
\be
\psi^{~}_{n,m}\star\psi^{~}_{n',m'}=\frac1{\sqrt{4\pi\,\Theta}}~
\delta^{~}_{mn'}~\psi^{~}_{n,m'} \ ,
\label{starprodprojrel}\end{equation}
and thereby determine solitonic configurations of noncommutative field
theory on $\IR^2$~\cite{GMS}. The basic Gaussian soliton (\ref{psi00}) can be
centered about any point on the plane by using the exact translational
symmetry of noncommutative field theory, and hence the one-soliton moduli
space as a complex manifold is isomorphic to $\IC$. The Wigner
function of the rank~$k$ projection
\be
\P_{(k)}=\sum_{m=0}^{k-1}|m\rangle\langle m|
\label{rankkproj}\end{equation}
describes $k$ solitons, and the corresponding moduli space is the
$k^{\rm th}$ symmetric product $\IC^k/S_k$ of the single soliton 
moduli
space, endowed with a {\it smooth} K\"ahler 
metric~\cite{GHS1}. The
basic Murray-von~Neumann partial
isometry is 
provided by the shift operator
\be
{\cal 
S}_\infty=\sum_{m\in\IN_0}|m+1\rangle\langle 
m|
\label{calSinfty}\end{equation}
with $({\cal 
S}_\infty^\dag)^k\,({\cal S}_\infty^{~})^k=\id$ and
$({\cal 
S}_\infty^{~})^k\,({\cal
   S}_\infty^\dag)^k=\id-\P_{(k)}$. Again 
the moduli space of partial
isometries $({\cal S}_\infty)^k$, and 
hence the moduli space of $k$
D-branes on $\IR^2$, is manifestly 
$\IC^k/S_k$.

The key feature of the Landau wavefunctions within the 
present context
is that they are complete in $L^2(\IR^2)$, so that 
any field $f\in
C^\infty(\IR^2)$ may be expanded 
as
\be
f(w,\bw\,)=\sum_{(n,m)\in\IN^2_0}f^{~}_{n,m}~\psi^{~}_{n,m}(w,\bw\,) 
\
, \label{Landaucomplete}
\end{equation}
where the expansion 
coefficients $\{f_{n,m}\}\in S(\IZ^2)$ are chosen
to yield finite 
Landau semi-norms
\beq
\|f\|^{~}_{{\rm 
L},k}:=\left(\,\sum_{(n,m)\in\IN_0^2}\Theta^{2k}\,(2n+1)^k\,
(2m+1)^k\,|f_{n,m}|^2\right)^{1/2}<\infty 
~~~~ \forall k\in\IN_0 \ .
\label{fkknorms}\end{equation}
This 
suggests a natural regularization of noncommutative fields on
$\IR^2$ 
whereby the Landau quantum numbers are truncated to a
finite range 
$0\leq n,m\leq N-1$, and the expansion coefficients
of 
(\ref{Landaucomplete}) are assembled into an $N\times N$ 
matrix
$(f_{n,m})\in\IM_N(\IC)$~\cite{LSZ}. Similar truncations have 
also
been used as approximations of a disc~\cite{LVZ}, a 
strip~\cite{BGK}
and a punctured plane~\cite{PZ}. Because of 
(\ref{starprodprojrel}), the
star-product
$f\star f'$ of two fields 
corresponds to the usual matrix product
of $(f_{n,m}^{~})$ and 
$(f'_{n,m})$ in $\IM_N(\IC)$, and the
noncommutativity of the plane 
$\IR_\Theta^2$ is thereby mapped
into the noncommutativity of matrix 
multiplication. In addition,
by orthonormality, spacetime integrals 
over $\IR^2$ of fields $f$
are given by traces of their matrices 
$(f_{n,m})$. Thus the
expansion of functions in the GMS soliton basis 
provides a very
natural way to map noncommutative field theory on 
$\IR^2$ onto a
zero-dimensional matrix model~\cite{LSZ}. The 
regularization provided by the
finite matrix dimension $N$ controls 
both ultraviolet and infrared
divergences at the same time~\cite{LS1} 
and avoids the renormalization
problems set in by UV/IR mixing. The 
limit $N\to\infty$ required
to map back onto the original continuum 
field theory corresponds
to the usual 't~Hooft planar 
limit~\cite{MRW,LSZ}. However, as mentioned
earlier,
this is a subtle 
point, as the algebra of the noncommutative plane
is not the 
inductive limit of finite-dimensional algebras, and so
one has to 
define this limit carefully as in~\cite{lls}. Moreover,
in general 
this limit will not commute with the scaling limits
used in ordinary 
field theoretic renormalization~\cite{LSZ}. In the following
we will 
relate the GMS soliton basis to that of the previous
sections and 
show how to regulate field theories on noncommutative
$\IR^2$ by 
means of one-dimensional matrix models.

\subsection{From Torus 
Solitons to GMS Solitons}\label{se:torustogms}

\begin{figure}
\bigskip
\centerline{\epsfxsize=3 in\epsffile{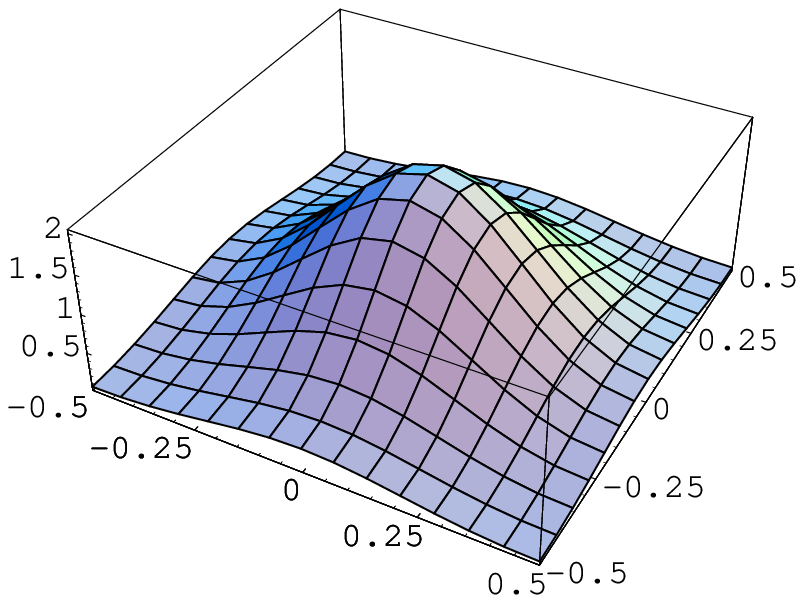}\hskip 0.3in
  \epsfxsize=3 in \epsffile{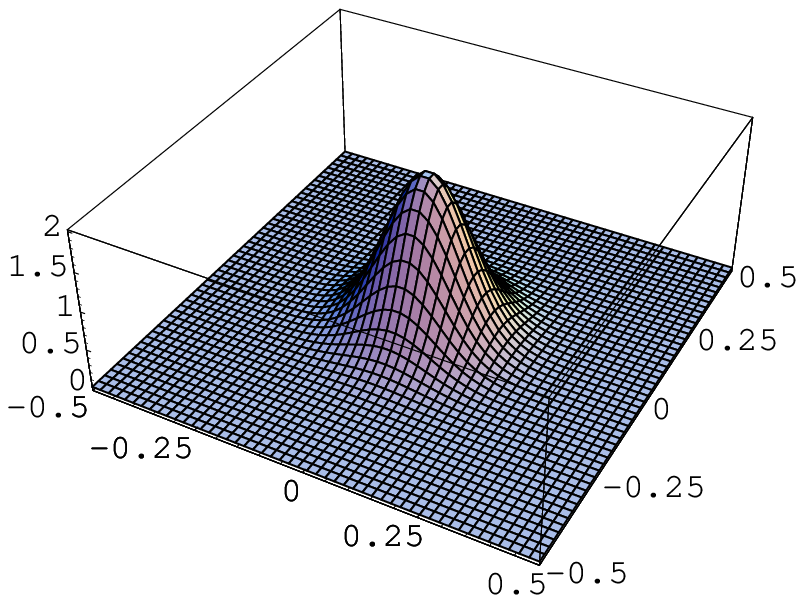}}
\caption{\baselineskip=12pt {\it The soliton field configuration
      corresponding to the Boca projection in the small $\theta$
      limit. Axes are as in Fig.~\ref{zawinul3}. Displayed are its
      shapes when the noncommutativity parameter is taken to be the
      inverse of the golden mean (left) and for $\theta=\frac1{10}$
(right).}}
\bigskip
\label{Bocafigs}\end{figure}

To relate the GMS 
soliton regularization above to that of the previous
sections, we 
shall first describe how to pass from solitons on the
noncommutative 
torus to the noncommutative solitons (\ref{psinm}) 
on
$\IR^2$~\cite{KS2}. For this, we consider the Boca 
projection
(\ref{Btheta}) in the limit $\theta\to0$. In that limit, 
the Schwartz
function (\ref{b12xisgen}) reduces 
to
\be
\xi(s)\big(\langle\xi,\xi\rangle^{~}_{-1/\theta}\big)^{-1/2}
=\left(\frac2\theta\right)^{1/4}~\e^{-\pi\,s^2/\theta}
+O\left(\e^{-2\pi\,s^2/\theta}\right) 
\ .
\label{b12xistheta0}\end{equation}
By substituting 
(\ref{b12xistheta0}) into (\ref{Btheta}) and
performing the resulting 
Gaussian integrals over $s$ in (\ref{hslft})
we arrive at the Boca 
projection in this limit 
as
\be
\mpsi^{~}_\theta:=\lim_{\theta\to0}\,\B_\theta=\theta\,\sum_{(m,r)\in\IZ^2}
\e^{-\frac{\pi\,\theta}2\,(m^2-2\ii 
m\,r+r^2)}~U^m\,V^r \ .
\label{Btheta0}\end{equation}
By using the 
map (\ref{wigmap}) it follows that the Wigner function
corresponding to the element (\ref{Btheta0}) of $\atheta$ decouples
the sums over $m$ and $r$. It thereby reduces to a product of elliptic
functions of the form
\be
\Omega^{-1}(\mpsi^{~}_\theta)(x,y)=\theta~\vartheta^{~}_{0,0}\left(x\,
\left|\,\mbox{$\frac{\ii\theta}2$}\right.\right)
\,\vartheta^{~}_{0,0}\left(y\,\left|\,\mbox{$\frac{\ii\theta}2$}\right.\right)
\ .
\label{wigmpsi}\end{equation}
The function (\ref{wigmpsi}) is plotted in Fig.~\ref{Bocafigs}. In
contrast to the projections used earlier, this soliton configuration
resembles the Gaussian GMS soliton (\ref{psi00}). In particular, for
any $\theta$ its height is always $2$, and as $\theta$ decreases its
width becomes smaller and a spike develops. In the limit $\theta=0$,
the function (\ref{wigmpsi}) vanishes everywhere except at the origin
of $\IR^2$ where it is finite. However, this limit is not smooth, and
for $\theta=0$ the soliton does not exist, as there are no non-trivial
projections in a commutative algebra.

By using the Jacobi 
inversion formula (equivalently Poisson 
resummation)
\be
\vartheta^{~}_{0,0}(\nu\,|\ii\sigma)=
\frac1{\sqrt\sigma}~\e^{-\pi\,\nu^2/\sigma}~
\vartheta^{~}_{0,0}\left(\left.\mbox{$\frac{\ii\nu}\sigma$}\,\right|
\mbox{$\frac\ii\sigma$}\right) 
\ ,
\label{Jacobiinv}\end{equation}
the $\theta\to0$ limit will pick 
out the $m=0$ mode of the
corresponding theta-functions in 
(\ref{wigmpsi}). This 
yields
\be
\Omega^{-1}(\mpsi^{~}_\theta)(x,y)=2~\e^{-2\pi\,(x^2+y^2)/\theta} 
\ .
\label{wiglargetheta}\end{equation}
We now map the local 
coordinates $(x,y)$ of the torus $\IT^2$ onto
those $(w,\bw\,)$ of 
its universal covering space $\IR^2$ by rescaling
the cycles of 
$\IT^2$ to give them a radius $R$ and then taking 
the
decompactification limit $R\to\infty$. This relates the two sets 
of coordinates
as
\be
(x+\ii y,x-\ii 
y)=\frac{(w,\bw\,)}{2\pi\,R}
\label{coordrescale}\end{equation}
which 
implies that the noncommutativity parameters are related 
through
\be
\theta=\frac\Theta{\pi\,R^2} \ 
,
\label{thetarescale}\end{equation}
consistently with the small 
$\theta$ limit used above.

It follows that in the large area limit 
of the toroidal theory, with the
noncommutativity parameter $\Theta$ 
of $\IR^2$ held fixed, the Wigner
function (\ref{wiglargetheta}) 
coincides with the basic Gaussian GMS
soliton 
(\ref{psi00}),
\be
\Omega^{-1}(\mpsi^{~}_\theta)(w,\bw\,)=\sqrt{4\pi\,\Theta}~
\psi_{0,0}^{~}(w,\bw\,) 
\ .
\label{wigmpsiGMS}\end{equation}
Equivalently, by using 
(\ref{Landauwavefn}) one may identify 
the
operators
\be
\mpsi^{~}_\theta=|0\rangle\langle0|
\label{mpsi00}\end{equation}
in 
the decompactification limit of the torus.
The higher Landau levels 
$\psi_{n,m}^{~}$ (or equivalently
$|n\rangle\langle m|$) can be 
similarly obtained by using
higher rank projections on the 
noncommutative torus, but we will
content ourselves here with the 
fact that they can obtained from
(\ref{wigmpsiGMS}) through the 
identity (\ref{psinm}).

A nice physical interpretation of this 
relationship between the
toroidal and planar noncommutative solitons 
may be given in the case
of rational $\theta=1/k$, with $k\in\IN$ 
even, as used in arriving at
(\ref{Bocaspecial}). For this, we note 
that the Landau wavefunction
(\ref{psi00}) can be multiplied by an 
arbitrary anti-holomorphic
function $\phi(\,\overline{w}\,)$ on 
$\IC$, which we may choose so as to give
the state
a momentum 
$p/2\,\Theta$ along the $x$ direction of $\IR^2$ and 
hence
write
\be
\psi^{~}_{0,0}(w,\overline{w}\,;p):=\phi^{~}_p(\,\overline{w}\,)\,
\psi^{~}_{0,0}(w,\overline{w}\,)=\frac1{\sqrt{\pi\,\Theta}}~
\e^{-(2\,p^2-4\ii 
p\,\overline{w}-\overline{w}^{\,2})/2\,\Theta}~
\e^{-|w|^2/2\,\Theta} \ .
\label{psi00p}\end{equation}
We will assume that the area of the torus $\IT^2$ is quantized such that
the quantity $\frac8{R^2}$ is an integer; for definiteness we
choose $R^2=8$.

We can map the wavefunction (\ref{psi00p}) to one on the torus by
regarding a soliton on the torus as an infinite lattice of
solitons on the plane. Taking the quotient of $\IR^2$ by the
momentum lattice $\IZ^2$ leads, using (\ref{thetarescale}), to a
quantization condition $p=4\pi m/k$, $m\in\IZ$ on the momentum in
the $x$ direction. On the other hand, $p$ is also the $y$
coordinate of the location of the soliton (\ref{psi00p}) on
$\IR^2$, and so for each quantum number $m$ it can assume $\frac
k2$ different values $p=\frac{4\pi r}k+2\pi m$, $r=1,\dots,\frac
k2$. By summing over all $m\in\IZ$ to take the quotient, we arrive
from (\ref{psi00p}) at $\frac k2$ basic Landau wavefunctions on
the torus. Using (\ref{coordrescale}) they may be written as
\bea
\psi^{~}_{0,0}(x,y;r)&=&\sum_{m\in\IZ}\psi^{~}_{0,0}(w,\overline{w}
\,;2\pi m+4\pi r/k)\nn&=&\frac1{\sqrt{8\pi^2\,\theta}}~
\e^{2\pi\,k\,(x-\ii y)^2}~\e^{-2\pi\,k\,(x^2+y^2)}~
\vartheta^{~}_{\frac{2r}k\,,\,0}\left(\left.\sqrt2\,k\,
(x-\ii y)\,\right|\,\mbox{$\frac{\ii k}2$}\right)
\label{psi00torus}
\end{eqnarray}
with $r=1,\dots,\frac k2$. One now takes an appropriate sum of the
$\frac k2$ functions (\ref{psi00torus}) in order to obtain a
star-product projection on $\IT^2$. By decoupling the theta-function
in (\ref{psi00torus}), it is possible to thereby show that the Weyl
image of the resulting wavefunction coincides with the Boca projection
in (\ref{Bocaspecial}). A similar derivation of this noncommutative
soliton on $\IT^2$ is employed in~\cite{GHS1}, based on the
construction of the noncommutative torus algebra $\atheta$ as the
commutant of $\IZ^2$ in the crossed product of the algebra
$C^\infty(\IR_\Theta^2)$ with the momentum lattice $\IZ^2$.

\subsection{Matrix GMS Projections}

We will now apply the approximation of $\atheta$ described in
section~\ref{Approx} to obtain a matrix regularization of the basic
GMS soliton projections. We start from the $\theta\to0$ limit of the
Boca projection in (\ref{Btheta0}), and apply the map
(\ref{Gammanhomo}) to get
\be
\Gamma_n(\mpsi^{~}_\theta)=\theta\,\sum_{(m,r)\in\IZ^2}
\e^{-\frac{\pi\,\theta}2\,(m^2-2\ii m\,r+r^2)}~(\U_n)^m\,
(\V_n)^r \ .
\label{Gammanmpsi}\end{equation}
We will evaluate (\ref{Gammanmpsi}) by regarding it as a matrix-valued
function on a pair of circles. Let us first examine the matrix
elements corresponding to the first tower. By using (\ref{Aexp2}),
(\ref{Aexp2Gamma}) and (\ref{keyidentity}), we may write the
$(ij)^{\rm th}$ matrix element of (\ref{Gammanmpsi}) with $1\leq i,j\leq
q_{2n}$ as
\bea
\big(\Gamma_n(\mpsi_\theta^{~})\big)_{ij}&=&\theta\,\sum_{(k,l)\in\IZ^2}
z^{\kappa^{~}_k}\,\sum_{s=1}^{q_{2n}}\,(\omega_n)^{s(i-1)}~
\exp\left[-\mbox{$\frac{\pi\,\theta}2$}\,\big((s+l\,q_{2n})^2\right.
\nn&&
{}~~~~~~~-\left.2\ii(s+l\,q_{2n})(j-i+\kappa^{~}_k\,q_{2n})+(j-i+
\kappa^{~}_k\,q_{2n})^2\,\big)\right] \ ,
\label{Gammanmpsiij}
\end{eqnarray}
where the integer $\kappa^{~}_k$ depends on the circular Fourier
momentum $k\in\IZ$ and the triangular block of the matrix as
\be
\kappa^{~}_k:=\left\{\begin{matrix}k&i<j\\\,k+1&i\geq j\end{matrix}
\right. \ .
\label{kappar}\end{equation}
We can transform the sums over $s\in\IZ_{q_{2n}}$ and $l\in\IZ$ in
(\ref{Gammanmpsiij}) into a sum over a single integer
$m=s+l\,q_{2n}\in\IZ$, and the sum over $k\in\IZ$ to one over
$r=\kappa^{~}_k\in\IZ$. Unlike the continuum case, in the matrix
regularization one cannot decouple the sums over $m$ and $r$, and
instead of factorizing into the product of two genus one theta-functions
as in (\ref{wigmpsi}), the matrix elements (\ref{Gammanmpsiij}) can
generically only be written in terms of genus {\it two} Jacobi
theta-functions
\be
\vartheta(\mnu\,|\ii\mtau)=\sum_{\boldsymbol{m}\in\IZ^2}\e^{-\pi\,
\boldsymbol{m}\,\cdot\,\mtau\boldsymbol{m}+2\pi\ii\boldsymbol{m}
\,\cdot\,\mnu} \ ,
\label{genus2Jacobi}\end{equation}
where $\boldsymbol{m}\cdot\mnu:=m_1\,\nu_1+m_2\,\nu_2$. The functions
(\ref{genus2Jacobi}) are holomorphic in $\mnu\in\IC^2$ for symmetric
$2\times 2$ period matrices $\mtau$ of positive definite real part.

With $z:=\e^{2\pi\ii\tau/r_n}$, $\tau\in[0,r_n)$, after some algebra
we may thereby write (\ref{Gammanmpsiij}) as
\be
\big(\Gamma_n(\mpsi^{~}_\theta)\big)_{ij}=\theta~
\e^{-\frac{\pi\,\theta}2\,(i-j)^2}~\vartheta\left(\left.
\mnu^{(n)}_{ij}\,\right|\ii\mtau^{(n)}\right) \ ,
\label{mpsiijTheta}\end{equation}
where
\bea
\mnu_{ij}^{(n)}&=&\begin{pmatrix}\,\ii\left(\frac{p_{2n}}{q_{2n}}\,(i-1)+
\frac\theta2\,(j-i)\right)\,\\\frac{q_{2n}\,\theta}2\,(i-j)+
\frac{\ii\tau}{r_n}\end{pmatrix} \ , \nn\mtau^{(n)}&=&\frac\theta2\,
\begin{pmatrix}1&-\ii q_{2n}\,\\\,-\ii q_{2n}&(q_{2n})^2\,\end{pmatrix} \ .
\label{mtauij}
\end{eqnarray}
The computation of the matrix elements corresponding to the second
tower is completely analogous with the replacements
$(p_{2n},q_{2n})\to(p_{2n-1},q_{2n-1})$, $\tau\to\tau'$ and $r_n\to
r_n'$ in the above. In this way we arrive at the matrix
approximation to the Boca projection $\B_\theta$ in the limit
$\theta\to0$ in the form
\be
\Gamma_n(\mpsi^{~}_\theta)=\theta\,\begin{pmatrix}\,\left(
\e^{-\frac{\pi\,\theta}2\,(i-j)^2}~\vartheta\left(\left.
\mnu^{(n)}_{ij}\,\right|\ii\mtau^{(n)}\right)\right)&(0)_{q_{2n}
\times q_{2n-1}}\\(0)_{q_{2n-1}\times q_{2n}}&
\left(\e^{-\frac{\pi\,\theta}2\,(i'-j')^2}~\vartheta\left(\left.
\mnu^{\prime\,(n)}_{i'j'}\,\right|\ii\mtau^{\prime\,(n)}\right)\right)\,
\end{pmatrix} \ ,
\label{GammanmpsiTheta}\end{equation}
and thus the matrix regularization on the torus naturally
involves hyperelliptic functions.

To transform to an appropriate small $\theta$ limit, as before we
apply the Jacobi inversion formula for the genus two theta functions
(\ref{genus2Jacobi}),
\be
\vartheta(\mnu\,|\ii\mtau)=\frac{\e^{-\pi\,\mnu\,\cdot\,\mtau^{-1}\mnu}}
{\sqrt{\det\mtau}}~\vartheta\left(\left.\ii\mtau^{-1}\mnu\,\right|
\ii\mtau^{-1}\right) \ ,
\label{Jacobiinv2}\end{equation}
which now holds up to an irrelevant phase factor. By substituting
(\ref{mtauij}) into (\ref{genus2Jacobi}) it is
straightforward to show that the $\theta\to0$ limit of the right-hand
side of (\ref{Jacobiinv2}) also picks out the zero mode
$\boldsymbol{m}=\boldsymbol{0}$. Then, from (\ref{mpsiijTheta}),
(\ref{mtauij}) and (\ref{Jacobiinv2}), after some algebra we have in
the first tower
\bea
\big(\Gamma_n(\mpsi^{~}_\theta)\big)_{ij}&=&\frac1{q_{2n}}~
\e^{-\frac{\pi\,\theta}2\,(i-j)^2}~\exp\left\{\mbox{$-\frac\pi
{(q_{2n})^2\,\theta}\,\left[(p_{2n})^2\,(i-1)^2+2\ii p_{2n}\,(i-1)\,
\frac\tau{r_n}\right.$}\right.\nn&&
{}~~~~~~~~~~~~~~~~~~~~~~~~~~~~~~~~~~~-\left.\left.\left(\mbox{$
q_{2n}\,\theta\,(i-j)+\frac{\ii\tau}{r_n}$}\right)^2\right]\right\} \ .
\label{Gammanijsmall}
\end{eqnarray}
As before, we now substitute into (\ref{Gammanijsmall}) the rescaling
(\ref{thetarescale}) and the circular coordinates
\be
\frac\tau{r_n}=\frac t{2\pi\,R}
\label{taurescale}\end{equation}
with $t\in\IR$, and take the large area limit $R\to\infty$. Since
$\theta\to0$, it follows from (\ref{thetarescale}) and appendix~\ref{appa},
eq.~(\ref{seqincrease}) that $p_{2n}\to0$ as $p_{2n}\sim
q_{2n}\,\theta\sim\frac1{R^2}$ also in this limit. A completely analogous
analysis carries through for the second tower, and in this way we arrive
finally at the matrix version of the basic GMS soliton (\ref{mpsi00}) (or
(\ref{wigmpsiGMS})) in the form
\be
\Gamma_n(\mpsi^{~}_\Theta)(t,t'\,)=\begin{pmatrix}\frac1{q_{2n}}~
\e^{-t^2/(2q_{2n})^2\,\Theta}~(1)_{q_{2n}\times q_{2n}}&
(0)_{q_{2n}\times q_{2n-1}}\\
(0)_{q_{2n-1}\times q_{2n}}&\frac1{q_{2n-1}}~
\e^{-t^{\prime\,2}/(2q_{2n-1})^2\,\Theta}~(1)_{q_{2n-1}
\times q_{2n-1}}\end{pmatrix} \ .
\label{Gammansmallmatrix}\end{equation}

The matrix regularization (\ref{Gammansmallmatrix}) determines
solitons on noncommutative $\IR^2$ as approximate projections in
the 
algebra
$\mat_{q_{2n}}(C^\infty(\IR))\oplus\mat_{q_{2n-1}}(C^\infty(\IR))$
of 
matrix-valued functions on two copies of the real line $\IR$.
They 
are approximate in the sense that while the 
matrices
$\frac1{q_{2n}}\,(1)_{q_{2n}\times q_{2n}}$ 
and
$\frac1{q_{2n-1}}\,(1)_{q_{2n-1}\times q_{2n-1}}$ are 
projection
operators in $\mat_{q_{2n}}(\IC)$ and 
$\mat_{q_{2n-1}}(\IC)$,
respectively, the Gaussian prefactors in 
$S(\IR)$ combine to
projections only in the $n\to\infty$ limit. After 
an appropriate
rescaling of the coordinates $t,t'\in\IR$, the matrix 
soliton
(\ref{Gammansmallmatrix}) evidently converges to the 
GMS
one-soliton configuration (\ref{psi00}), and in particular, 
by
using translational symmetry, its moduli space is 
naturally
isomorphic to $(\IR\times\IN)\times(\IR\times\IN)$, where 
the
extra factors of $\IN$ come from the freedom in replacing 
the
matrices $(1)_{q\times q}$ by $\frac1k\,(k)_{q\times q}$ and 
rescaling
$q_{2n}\to
k\,q_{2n}$, $q_{2n-1}\to k\,q_{2n-1}$ for any 
$k\in\IN$ in the
limit $n\to\infty$. Note, however, that since the 
two towers are
independent, the soliton moduli space is determined by 
a pair of
one-dimensional, localized matrix-valued functions on $\IR$ 
and it
is no longer a complex manifold at the finite level. Higher 
Landau
levels can be approximated by constructing appropriate
finite 
versions of the differential operators~(\ref{abdag}),
similarly to 
section~\ref{NCFTMQM}, and applying them to the basic
soliton fields 
(\ref{Gammansmallmatrix}), as in (\ref{psinm}). The
corresponding 
$k$-soliton moduli space will then be a symmetric
orbifold of the 
single soliton one. We shall not pursue this
construction any further 
here, but in any case this gives a
precise way to regulate field 
theories on the noncommutative plane
by means of matrix quantum 
mechanics.

\subsection*{Acknowledgments}

We thank G.~Elliott, 
D.~Evans, L.~Griguolo, J.-H.~Park and J.~Varilly
for helpful 
discussions and correspondence. The work of G.L. and
F.L. is 
supported in part by the {\sl Progetto di Ricerca di Interesse
 Nazionale SINTESI}. The work of R.J.S. is supported in part by 
an
Advanced Fellowship from the {\sl Particle Physics and Astronomy
 Research Council}~(U.K.). F.L. would like to thank the Department 
of
Mathematics at
Heriot-Watt University for the hospitality 
presented to him during
several extended visits in his sabbatical 
year.

\setcounter{section}{0}
\renewcommand{\thesection}{\Alph{section}}
\setcounter{equation}{0}

\section{Continued 
Fraction Expansions\label{appa}}

A well known result of number 
theory~\cite{hw} states that
any irrational number $\theta$ can be 
uniquely represented as a simple
continued fraction expansion
\be
\theta=\lim_{n\to\infty}\,\theta_n \ , ~~ \theta_n:=\frac{p_n}{q_n}
\label{thetalimbis}
\end{equation}
involving positive integers $c_k>  0$, $k\geq 1$ and $c_0\in\zed$. The
continued fraction is the definition of a sequence of rational numbers
$\{\theta_n\}$ (the approximants), which converge to $\theta$. The
$n^{\rm th}$ approximant $\theta_n$ of the expansion is given by
\be
\theta_n=c_0 + {1\over\displaystyle c_1+
{\strut 1\over \displaystyle c_2+ {\strut
1\over\displaystyle\ddots {}~ c_{n-1}+{\strut 1\over c_n}}}} \ \ \ .
\label{thetandef}
\end{equation}
A short-hand notation for the expansion is
\be
\theta = [c_0, c_1, c_2, \dots ~]~.
\end{equation}
The relatively prime integers $p_n$ and $q_n$ in (\ref{thetalimbis}) may
be computed recursively from (\ref{thetandef}) by using the formul{\ae}
\bea
p_n&=&c_n\,p_{n-1}+p_{n-2}~~~~~~,~~~~~~p_0~=~c_0~~,~~p_1~=~c_0c_1+1 \ , \nn
q_n&=&c_n\,q_{n-1}+q_{n-2}~~~~~~,~~~~~~q_0~=~1~~,~~q_1~=~c_1
\label{pqrec}
\end{eqnarray}
for $n\geq2$. Note that all $q_n>0$ while
$p_n\in\zed$, and that both $q_n$ and $|p_n|$ are strictly increasing
sequences which therefore diverge as $n\to\infty$. The sequence of
convergents (\ref{thetandef}) can be shown to satisfy the bound
\be
|\theta-\theta_n|\leq\frac1{(q_n)^2} \ ,
\label{convbound}\end{equation}
showing how fast the limit in (\ref{thetalimbis}) converges.

When $0<\theta<1$ (so that $p_0=c_0=0$ and all $p_n>0$), the even
order convergents are always smaller than $\theta$, while the odd order
ones are larger. Thus the even (resp. odd) order convergents form
an increasing (resp. decreasing) sequence which converges to $\theta$ as
\be
\theta_{2n-2}<\theta_{2n}<\theta<\theta_{2n+1}<\theta_{2n-1} \ .
\label{seqincrease}\end{equation}
Furthermore, the
approximants fulfill Diophantine properties
\be
\begin{pmatrix}\,p_{2n\pm 1}&p_{2n}\,\cr\,q_{2n\pm 1}&q_{2n}\,\cr
\end{pmatrix}\in SL(2,\zed) \ ,
\label{Dioidpq}
\end{equation}
which follow from the recursion relations (\ref{pqrec}) by
induction 
on $n$.
We also define the decreasing 
sequence
\bea
\beta_{2n}&=&p_{2n-1}-q_{2n-1}\,\theta~=~q_{2n-1}\,(\theta_{2n-1}-\theta) 
\ ,
\label{defbetaeven} 
\\
\beta_{2n-1}&=&q_{2n}\,\theta-p_{2n}~=~q_{2n}\,(\theta-\theta_{2n})
{}~, 
\label{defbetaodd}
\end{eqnarray}
with $\beta_k>0$ and 
$\lim_{k\to\infty}\beta_k=0$.
  For each $n$ the properties 
\eqn{Dioidpq} imply 
that
\be
q_{2n}\,\beta_{2n}+q_{2n-1}\,\beta_{2n-1}=1 \ , 
~~~
q_{2n}\,\beta_{2n+2}+q_{2n+1}\,\beta_{2n-1}=1 \ 
.
\label{qbetaid1}
\end{equation}

\app{Matrix Unit 
Relations\label{appb}}
\setcounter{equation}{0}

To prove that the 
collections of operators $\{ \P^{ii} \}$ and
$\{\P^{i+2,i+1}\}$ 
in~(\ref{transe}) and~(\ref{ejkdef}) satisfy
the 
relations
\be
\P^{ij}\,\P^{kl}=\delta^{jk}\,\P^{il} \ , 
\label{matrixmultbis}
\end{equation}
we will first prove 
that
\be
\P^{21}\,\P^{ii}=0 ~~~~ \forall i>1 \ 
.\label{e21ii}
\end{equation}
For this, we will show that 
$\P^{21}\,\P^{ii}|\psi\rangle=0$ for all
vectors $|\psi\rangle\in\cal 
H$ of the underlying Hilbert space on which
the algebra $\atheta$ is 
represented. If $|\psi\rangle$ has no
component in the subspace 
${\cal H}_i={\rm im}(\P^{ii})$, then this is
trivially true, so we 
suppose that $|\psi\rangle\in{\cal H}_i$. Note
that the operator 
$\Pi^{21}$ in (\ref{Pi21}) contains the projection
$\P^{11}$ to its 
extreme right. With this observation, we can now
exploit a standard 
result of functional
analysis (see for 
instance~\cite[Theorem~2.3.4]{mu}) which states
that the kernel of
a 
bounded linear operator coincides with the kernel of the partial 
isometry
in its polar decomposition. Since $i>1$, we 
have
$\P^{11}|\psi\rangle=0$, and so
\be
{\cal 
H}_i\subset\ker\left(\Pi^{21}\right)=\ker\left(\P^{21}\right) \ 
.
\end{equation}
It follows that for any vector $|\psi\rangle\in\cal 
H$ we 
have
\be
\P^{ii}|\psi\rangle\in\ker\left(\P^{21}\right)
\end{equation}
for 
$i>1$, which establishes \eqn{e21ii}. By repeating this argument 
for
the adjoint operator $(\Pi^{21})^\dagger$ and using the 
definition
(\ref{Pjidagdef}) one similarly 
proves
\be
\P^{12}\,\P^{11}=0 \ .
\end{equation}

We will now prove 
the identity
\be
\P^{21}\,\P^{11}=\P^{21} \ .
\end{equation}
For 
this, we decompose a generic vector $|\psi\rangle\in\cal H$ 
as
\be
|\psi\rangle=|\psi^{~}_1\rangle\oplus|\psi_1^\perp\rangle
\end{equation}
with 
$|\psi^{~}_1\rangle\in{\cal H}_1$ and $|\psi^\perp_1\rangle\in{\cal 
H}_1^\perp$. 
Then
\be
\P^{21}\,\P^{11}|\psi\rangle=\P^{21}|\psi^{~}_1\rangle \ 
,
\end{equation}
and the desired result now follows from the fact 
that
\be
|\psi_1^\perp\rangle\in\ker\left(\P^{11}\right)\subset\ker\left(\Pi^{21}
\right)=\ker\left(\P^{21}\right) 
\ .
\end{equation}

Finally, to establish the 
expression
\be
\P^{22}\,\P^{21}=\P^{21} \ , \label{e2221}
\end{equation}
we 
note first of all that since the operator $\Pi^{21}$ contains 
an
orthogonal projection on its right, it has 
closed
range\footnote{\baselineskip=12pt The statement that a bounded 
linear
    operator $T$ is closed is equivalent to the following
 statements~\cite{w-o}: (a) 0 is an isolated point in the
    spectrum 
of the self-adjoint operators $T^\dag T$ and $TT^\dag$; (b)
    the 
right and left ideals $T\,\atheta$ and $\atheta\,T$ are closed in
the norm topology on $\atheta$; and (c) there exist projection
operators $\sf P$ and $\sf Q$ with $\atheta\,T=\atheta\,{\sf P}$ and
$T\,\atheta={\sf Q}\,\atheta$.}. With this observation, we can now 
exploit
another standard functional analytic result
(see for 
instance~\cite[Theorem~15.3.8]{w-o}) which states that the range
of a 
closed operator is the same as the range of the partial isometry 
in
its polar decomposition. It follows from this, and from the fact 
that the
operator $\Pi^{21}$ contains the projection $\P^{22}$ to its 
extreme
left, that
\be
\P^{21}({\cal H})=\Pi^{21}({\cal 
H})\subset{\cal H}_2 \ .
\end{equation}
Since $\P^{22}$ acts as the 
identity operator on the subspace ${\cal
    H}_2$, the relation 
\eqn{e2221} follows. The corresponding
identities for the projections 
and partial isometries with index labels
larger than 2 can then be 
constructed either by direct multiplication
or by use of the 
automorphism $\alpha$ defined in 
(\ref{rhoautodef}).

\app{Approximating the Torus 
Algebra\label{appc}}
\setcounter{equation}{0}

The proof of the fact 
that
\be
\lim_{n\to\infty} ~\big\|a-\Gamma_n(a)\big\|^{~}_0 ~ = ~0 
~,
\end{equation}
with $a\in\atheta$ and $\Gamma_n$ the projection 
onto the finite
level subalgebra $\athetan$ in \eqn{An}, comes from 
repeated
applications of the
triangle and product inequalities for 
the $C^*$-norm.
For any $m,r\in\zed$, by using the fact 
that
$\|(\U_n)^m\|^{~}_0=\|(\V_n)^r\|^{~}_0=1$ 
and
$\|U^m\|^{~}_0=\|V^r\|^{~}_0=1$, we 
have
\be
\big\|(\U_n)^m\,(\V_n)^r-U^m\,V^r\big\|^{~}_0\leq
\big\|(\U_n)^m-U^m\big\|^{~}_0+\big\|(\V_n)^r-V^r\big\|^{~}_0 
\ .
\label{triprodcomp}\end{equation}
Using (\ref{UVconvn}), we now 
define $\U_n=U+\Delta_n$ with
$[U,\Delta_n]=0$ and 
$\|\Delta_n\|^{~}_0\leq\varepsilon_n$. 
Then
\bea
\big\|(\U_n)^m-U^m\big\|^{~}_0&=&\left\|\,\sum_{p=1}^m{m 
\choose 
p}\,
U^{m-p}\,(\Delta_n)^{p}\,\right\|^{~}_0\nn&\leq&\sum_{p=1}^m{m\choose 
p}\,
\big(\left\|\Delta_n\right\|^{~}_0\big)^p\nn&\leq&1-(1-\varepsilon_n)^m~<~
m\,\varepsilon_n 
\ .
\end{eqnarray}
A completely analogous calculation 
gives
\be
\big\|(\V_n)^r-V^r\big\|^{~}_0<r\,\varepsilon_n \ 
.
\label{Vbounds}\end{equation}
{}From (\ref{Gammanhomo}) we 
have
\be
\big\|a-\Gamma_n(a)\big\|^{~}_0
\leq\sum_{(m,r) \in 
\IZ^2}|a_{m,r}|~\big\|(\U_n)^m\,(\V_n)^r-U^m
\,V^r\big\|^{~}_0 \ 
,
\label{aGammaannorm}
\end{equation}
and so from 
(\ref{triprodcomp})--(\ref{Vbounds}) it follows 
that
\be
\big\|a-\Gamma_n(a)\big\|^{~}_0
<\varepsilon_n\,\sum_{(m,r) 
\in \IZ^2}(m+r)\,|a_{m,r}| \ ,
\end{equation}
which vanishes as 
$n\to\infty$ for Schwartz sequences $\{a_{m,r}\}$.
Therefore, to each 
element of $\atheta$ there always corresponds an element
of the 
subalgebra $\athetan$ to within an arbitrarily small radius in 
norm.

\app{Inductive Limit\label{appd}}
\setcounter{equation}{0}

In 
this appendix we will show how to obtain the noncommutative torus
as 
an inductive limit\footnote{\baselineskip=12pt Strictly speaking, 
the
following
   discussion is only rigorously valid in the
 continuous category. However, on the noncommutative torus
   the 
proof of~\cite{ee} should go through also for smooth
   functions, 
with the appropriate technical
    modifications.}
\be
\atheta= 
\bigcup_{n=0}^\infty\bthetan
\end{equation}
of an appropriate 
inductive system of algebras
$\{\bthetan,\iota_n\}_{n\geq0}$, 
together with injective unital $*$-morphisms
$ \iota_{n} : \bthetan 
\hookrightarrow \cb_{{n+1}}$~\cite{ee}.
It turns out that for 
K-theoretical reasons one needs to take
$\bthetan = 
\ca_{2n+1}=\mat_{q_{4n+2}}\left(C^\infty(\circles^1)\right)\oplus
\mat_{q_{4n+1}}\left(C^\infty(\circles^1)\right)$. 
The crucial point is to
explicitly construct the embeddings $\iota_n$ 
in such a way that the
K-theory groups (\ref{K0Atheta}) and 
(\ref{K1Atheta}) of the
noncommutative torus are obtained in the 
limit out of the ``finite
level'' counterparts \eqn{K01Athetan}. For 
this, one needs to exploit
the continued fraction expansion of 
$\theta$ and
the recursion relations (\ref{pqrec}) in a very careful 
manner.

It is well known \cite{es} that on the ``matrix part'', the 
continued
fraction expansion of $\theta$ directly gives the required 
dimension group
of the torus $\atheta$ while leading to a trivial 
$\K_1$-group, this
being the appropriate setting for immersions into 
an AF-algebra \cite{pvb}. In
the present case, in order to get the 
correct $\K_1$-group (\ref{K1Atheta}),
which is generated by the two 
independent classes $[U]$ and $[V]$,
one needs to modify the 
construction somewhat. The main point is that
the clock operators 
appearing in (\ref{defUVn}) are elements of
finite-dimensional matrix 
algebras and therefore have trivial
$\K_1$-class. The non-trivial 
group is thereby generated by the
generalized shift operators in 
(\ref{defUVn}), and this must be kept in
mind when embedding from one 
level to the next. To this end, one
``skips'' a step~\cite{ee}, by 
going from level $n$ to level $n+1$
(so as to send $q_{n}$ 
to
$q_{n+4}$) . With
the skipping of steps, the roles of the clock 
and shift operators in
(\ref{defUVn}) change from one level to the 
next.

We thereby use the recursion relations (\ref{pqrec}) to 
define
integer valued 
matrices
\be
P_n=\begin{pmatrix}
\,r_n&s_n\,\cr\,t_n&u_n\,\cr\end{pmatrix} 
\ , ~~ n=0,1,2, \dots
\label{pdef}
\end{equation}
with the property 
that
\be
P_n 
\begin{pmatrix}\,q_{4n+2}\,\cr\,q_{4n+1}\,\cr\end{pmatrix}=
\begin{pmatrix}\,q_{4(n+1)+2}\,\cr
\,q_{4(n+1)+1}\,\cr\end{pmatrix} 
\ .
\label{rstudef}
\end{equation}
To simplify notation for the rest 
of this appendix, let us denote
$d_n=q_{4n+2}$ and 
$d'_n=q_{4n+1}$.
As mentioned before, we shall take the algebra in 
the inductive limit to 
be
\be\label{Bn}
\bthetan~=~\ca_{2n+1}~=~\mat_{d_n}\left(C^\infty(\circles^1)\right)\oplus
\mat_{d'_n}\left(C^\infty(\circles^1)\right)\, 
, ~~ n = 0, 1, 2, \dots \ ,
\end{equation}
and we are now ready to 
describe the embedding $ \iota_{n} : 
\bthetan
\hookrightarrow\cb_{{n+1}}$.

Since a generic element of the 
matrix algebra (\ref{Bn}) is of the 
form
\be
\a_n=\sum_{k\in\IZ}~\sum_{i,j=1}^{d_{n}}a_{ij;k}^{(n)}~
z^k\,\P_n^{ij}~\oplus~\sum_{k'\in\IZ}~
\sum_{i',j'=1}^{d'_{n}}a_{i'j';k'}^{\prime\,(n)}~z^{\prime\,k'}
\,\P_n^{\prime\,i'j'} 
\ ,
\label{Angeneric}
\end{equation}
it suffices to give the immersions of the $d_{n}\times d_{n}$ and
$d'_{n}\times d'_{n}$ matrices $\big(a_{ij;k}^{(n)}\big)$ and
$\big(a_{i'j';k'}^{\prime\,(n)}\big)$ for given fixed values of the circular
Fourier modes $k$ and $k'$, and of the two unitaries $z$ and $z'$
which generate the center of the algebra $\athetan$. Denoting the mode
restrictions by $\a_n\!\!\upharpoonright_k$, the
embedding of the matrix degrees of freedom is given by
\bea
\iota_n\left[\begin{pmatrix}\,\a_n\!\!\upharpoonright_k& \cr &
(0)_{d_n'\times d'_{n}}\,\cr\end{pmatrix}\right]&=&
\begin{pmatrix}\scriptstyle\,\id_{r_n}\otimes\left(a^{(n)}_{ij;k}\right)& & &
\cr
   &\scriptstyle (0)_{s_n\,d'_{n}\times s_n\,d'_{n}} & & \cr & &
\scriptstyle\id_{t_n}\otimes\left(a_{ij;k}^{(n)}\right)& \cr & & &
\scriptstyle(0)_{u_n\,d'_{n}\times u_n\,d'_{n}}\,\cr\end{pmatrix} \ ,
\nn&&{~~}^{~~}_{~~}\nn
\iota_n\left[\begin{pmatrix}\,(0)_{d_n\times d_{n}}& \cr &\a_n\!\!
\upharpoonright_{k'}\,\cr\end{pmatrix}\right]&=&
\begin{pmatrix}\,\scriptstyle(0)_{r_n\,d_{n}\times r_n\,d_{n}}& & &
\cr &\scriptstyle\id_{s_n}\otimes
\left(a_{i'j';k'}^{\prime\,(n)}\right)& & \cr & &
\scriptstyle(0)_{t_n\,d_{n}\times t_n\,d_{n}}& \cr & &
&\scriptstyle\id_{u_n}\otimes
\left(a_{i'j';k'}^{\prime\,(n)}\right)\,\cr\end{pmatrix} \ , \nn&&
\label{iotanmatrix}
\end{eqnarray}
while $z$ and $z'$ are embedded as
\bea
\iota_n\left[\begin{pmatrix}\,z\,\id_{d_{n}}& \cr &
(0)_{d_n'\times d'_{n}}\,\cr\end{pmatrix}\right]&=&
\begin{pmatrix}\,\scriptstyle{\cal S}_{r_n}(z)\otimes\id_{d_{n}}& & &\cr
    &\scriptstyle(0)_{s_n\,d'_{n}\times s_n\,d'_{n}} & & \cr & &
\scriptstyle{\cal S}_{t_n}(1)\otimes\id_{d_{n}}&
\cr & & &\scriptstyle(0)_{u_n\,d'_{n}\times u_n\,d'_{n}}\,
\cr\end{pmatrix} \ , \nn&&{~~}^{~~}_{~~}
\nn\iota_n\left[\begin{pmatrix}\,(0)_{d_n\times d_{n}}
& \cr &z'\,\id_{d'_{n}}\,\cr\end{pmatrix}\right]&=&
\begin{pmatrix}\,\scriptstyle (0)_{r_n\,d_{n}\times r_n\,d_{n}}& & & \cr
    &\scriptstyle {\cal S}_{s_n}(1)\otimes\id_{d'_{n}}& & \cr & &
\scriptstyle (0)_{t_n\,d_{n}\times t_n\,d_{n}}& \cr & & &
\scriptstyle {\cal S}_{u_n}(z'\,)\otimes
\id_{d'_{n}}\,\cr\end{pmatrix} \ . \nn&&
\label{iotanz}
\end{eqnarray}
When lifted to the K-theory groups, the homomorphism $\iota_n$
acts as the matrix $P_n$ on $\K_0(\bthetan)=\zed\oplus\zed$ and as the
identity on  $\K_1(\bthetan)=\zed\oplus\zed $, so
that the inductive limit algebra has the appropriate K-theory groups
(\ref{K0Atheta})--(\ref{K1Atheta}). Furthermore, in \cite{ee} it is
shown that the limit algebra is a simple unital algebra that has a
unique trace state. All of these properties select the noncommutative
torus algebra $\atheta$ up to isomorphism.

\app{Approximating the Leibniz Rule\label{appf}}
\setcounter{equation}{0}

In this appendix we will show that the two ``derivatives'' defined in
\eqn{derin} satisfy an approximate Leibniz rule, which becomes the
usual one in the $n\to\infty$ limit. To keep formul{\ae} from becoming
overly cumbersome, we will only indicate explicitly terms appearing in
the first tower. Analogous expressions are always understood to
appear in the second tower. We will denote by $[a]^{~}_q$
the integer part of a real number $a$ modulo $q$, with the convention that
$[a]^{~}_0$ is its integer part in $\zed$.

The product of two elements of the finite level algebra $\athetan$
with expansion (\ref{Aexp2}) is given by
\bea
\a_n\b_n&=&\sum_{i,j,s,t=1}^{q_{2n}}~\sum_{k,l\in\IZ}\,\alpha_{i+
\big[\frac{q_{2n}}2\big]_0,j;k}^{(n)}
{}~\beta_{s+\big[\frac{q_{2n}}2\big]_0,t;l}^{(n)}~(\omega_n)^{js-it}~
z^{k+l+\big[\frac{j+t}{q_{2n}}\big]_0}\nn&&~~~~~~~~~~~~~~~~~~~~~~~~~\times
\,\left({\cal C}_{q_{2n}}\right)^{[i+s]_{q_{2n}}}\,\big( {\cal
S}_{q_{2n}}(z)\big)^{[j+t]_{q_{2n}}}\nn&&\oplus~({\rm
second}~{\rm tower}) \ . \label{anbnprod}
\end{eqnarray} By using
(\ref{derin}) one may calculate the derivative of the product
(\ref{anbnprod}) to be
\bea
\mnabla_1(\a_n\b_n)&=&2\pi\ii\,\sum_{i,j,s,t=1}^{q_{2n}}~\sum_{k,l\in\IZ}\,
[i+s]_{q_{2n}}~\alpha_{i+\big[\frac{q_{2n}}2\big]_0,j;k}^{(n)}~
\beta_{s+\big[\frac{q_{2n}}2\big]_0,t;l}^{(n)}~
(\omega_n)^{js-it}~z^{k+l+\big[\frac{j+t}{q_{2n}}\big]_0} \nn&&
{}~~~~~~~~~~~~~~~~~~~~~~~~~~~~~~\times\,\left({\cal
C}_{q_{2n}}\right)^{[i+s]_{q_{2n}}}\,\big( {\cal
S}_{q_{2n}}(z)\big)^{[j+t]_{q_{2n}}}~ \nn&& \oplus~({\rm
second}~{\rm tower})\ ,
\end{eqnarray}
while a direct calculation using the definition (\ref{derin}) and the
product formula (\ref{anbnprod}) gives
\bea
(\mnabla_1\a_n)\b_n+\a_n(\mnabla_1\b_n)&=&2\pi\ii\,
\sum_{i,j,s,t=1}^{q_{2n}}~\sum_{k,l\in\IZ}\,
(i+s)~\alpha_{i+\big[\frac{q_{2n}}2\big]_0,j;k}^{(n)}~
\beta_{s+\big[\frac{q_{2n}}2\big]_0,t;l}^{(n)}\nn&&~~~~~~~~~\times\,
(\omega_n)^{js-it}~z^{k+l+\big[\frac{j+t}{q_{2n}}\big]_0}~\left({\cal
C}_{q_{2n}}\right)^{[i+s]_{q_{2n}}}\,\big( {\cal
S}_{q_{2n}}(z)\big)^{[j+t]_{q_{2n}}}~ \nn&& \oplus~({\rm
second}~{\rm tower}) \ .
\end{eqnarray}
The difference between these two expressions occurs 
when at least
one of the integers $i$ or $s$ is of 
order
$\big[\frac{q_{2n}}2\big]_0$, in which case the 
corresponding
coefficient of $\a_n$ or $\b_n$ is exponentially 
small
in the limit. The two expressions thereby coincide at 
$n\to\infty$. A
completely analogous computation 
gives
\bea
\mnabla_2(\a_n\b_n)&=&2\pi\ii\,\sum_{i,j,s,t=1}^{q_{2n}}~\sum_{k,l\in\IZ}\,
\left\{[j+t]_{q_{2n}}+q_{2n}\left(k+l+\left[\mbox{$\frac{j+t}{q_{2n}}$}
\right]_0\right)\right\} 
\nn&&~~~~~~~~~~~~~~~~~~~~~~~~~\times
\,\alpha_{i+\big[\frac{q_{2n}}2\big]_0,j;k}^{(n)}~\beta_{s+
\big[\frac{q_{2n}}2\big]_0,t;l}^{(n)}~
(\omega_n)^{js-it}~z^{k+l+\big[\frac{j+t}{q_{2n}}\big]_0}\nn&&
{}~~~~~~~~~~~~~~~~~~~~~~~~~\times\,
\left({\cal C}_{q_{2n}}\right)^{[i+s]_{q_{2n}}}\,\big( {\cal
S}_{q_{2n}}(z)\big)^{[j+t]_{q_{2n}}}\nn&& \oplus~({\rm
second}~{\rm tower})
\end{eqnarray}
and
\bea
(\mnabla_2\a_n)\b_n+\a_n(\mnabla_2\b_n)&=&2\pi\ii\,\sum_{i,j,s,t=1}^{q_{2n}}~
\sum_{k,l\in\IZ}\,\big(j+t+q_{2n}(\,k+l\,)\big)\nn&&~~~~~~~~~~~~\times
\,\alpha_{i+\big[\frac{q_{2n}}2\big]_0,j;k}^{(n)}~\beta_{s+
\big[\frac{q_{2n}}2\big]_0,t;l}^{(n)}~
(\omega_n)^{js-it}~z^{k+l+\big[\frac{j+t}{q_{2n}}\big]_0}
\nn&&~~~~~~~~~~~~\times
\,\left({\cal C}_{q_{2n}}\right)^{[i+s]_{q_{2n}}}\,\big( {\cal
S}_{q_{2n}}(z)\big)^{[j+t]_{q_{2n}}}\nn&& \oplus~({\rm
second}~{\rm tower}) \ .
\end{eqnarray}
Again the two expressions differ only for large momenta $j$ and
$t$.

\vfill\eject
\providecommand{\href}[2]{#2}

\end{document}